\documentclass[preprint,floats,nofootinbib,superscriptaddress,color]{revtex4}
\usepackage{bm}
\usepackage{amsmath}
\usepackage{graphicx}
\usepackage{graphics}
\usepackage[dvips]{color}

\newcommand{\be}{\begin{equation}}
\newcommand{\ee}{\end{equation}}
\newcommand{\ba}{\begin{eqnarray}}
\newcommand{\ea}{\end{eqnarray}}

\newcommand{\nk}{{\bf      k}}
\newcommand{\np}{{\bf      p}}
\newcommand{\nq}{{\bf      q}}

\begin{document}

\title{Parity Violation in Elastic Electron-Nucleon Scattering: Strangeness Content in the Nucleon}
\author{R. Gonz\'alez-Jim\'enez}
\affiliation{Departamento de F\'{\i}sica At\'{o}mica, Molecular y Nuclear,
Universidad de Sevilla,
  41080 Sevilla, SPAIN}
\author{J.A. Caballero}
\affiliation{Departamento de F\'{\i}sica At\'{o}mica, Molecular y Nuclear,
Universidad de Sevilla,
  41080 Sevilla, SPAIN}
\author{T.W. Donnelly}
\affiliation{Center for Theoretical Physics, Laboratory for Nuclear
  Science and Department of Physics, Massachusetts Institute of Technology,
  Cambridge, MA 02139, USA}

\begin{abstract}

Parity violation in elastic electron-nucleon scattering is studied
with the basic goal of improving the understanding of electroweak
hadronic structure with special emphasis on the strangeness content
in the nucleon. Models for the parity-violating (PV) asymmetry are
provided and compared with the world data measured at very different
kinematics. The effects introduced in the PV asymmetry due to
alternative descriptions of the hadronic structure are analyzed in
detail. In particular, a wide selection of prescriptions for dealing
with the electromagnetic and neutral current weak interaction
nucleon form factors, including the most recent ones used in the
literature, is considered.

\end{abstract}


\maketitle

\tableofcontents

\section{Introduction}\label{sec:intro}


Over the years electron scattering has provided much of the precise
information on hadronic and nuclear structure. Most such studies
have considered only the purely electromagnetic (EM) interaction,
that is, parity-conserving (PC) electron scattering processes. The
analysis of inclusive and semi-inclusive reactions, as well as the
measurement of polarization observables in very different
kinematical regimes, has allowed us to deepen very significantly our
knowledge of the internal structure of hadrons and nuclei: the EM
nucleon form factors, the nuclear ground-state density, the momentum
and energy distributions of nucleons inside nuclei, the role of
meson-exchange currents, {\it etc.,} have been investigated in depth
previously~\cite{Boffi93,Boffi96,Kelly,Walecka,RaskDon,DonRask}.

The weak interaction, although orders of magnitude weaker than the
EM one, also plays a role in electron scattering processes leading
to effects that can shed some light on ingredients of the reaction
mechanism that are not available through the purely EM interaction.
Studies of parity-violating (PV) observables through analyses of
electron scattering --- denoted simply as PV electron scattering to
contrast it with PC scattering where only the EM interaction enters
--- has three basic motivations: i) to serve in testing the Standard
Model of electroweak interactions, ii) to provide a tool for
determining the electroweak form factors of the nucleon, and iii) to
employ the semi-leptonic electroweak interaction as a probe of
nuclear structure, much as PC (purely EM) electron scattering has
been used for decades now. In this work, our interest is restricted
to the second point. It is important to point out that the smallness
of the weak coupling, compared with the EM one, forces one to
analyze observables that are strictly linked to parity-violating
effects, requiring at the same time, excellent control of the EM
ingredients that enter into the description of the scattering
reaction. Furthermore, from the combined study of electron
scattering from the proton and from nuclei, involving elastic,
inelastic and quasielastic regimes, additional useful constraints on
the form factors can emerge. In this work we restrict ourselves to
elastic PV electron-proton scattering. The extension to studies of
nuclear structure effects will be considered in a forthcoming paper.

The first PV electron scattering experiment on deuterium was
performed at high energies at SLAC in 1976~\cite{Prescott}, where a
powerful new experimental technique was introduced, the measurement
of helicity-dependent electron scattering cross sections. Since
then, considerable interest has emerged in the use of such
experiments to study electroweak physics at intermediate energies.
In particular, considerable attention has been paid to exploring the
strange-quark content in the structure of the nucleon.

The existence of the quark sea and its influence on some basic
properties of the nucleon (mass, spin, magnetic moment) has been
firmly established in several experimental studies. However, the
specific role of $s\overline{s}$ pairs in the static EM properties
of the nucleon still remains elusive. Polarization measurements in
deep inelastic scattering (DIS) provides access to the contribution
of strange quarks to the nucleon spin. However, the analysis of
several experiments has led to rather different conclusions: while
the original EMC experiment~\cite{EMC} suggested a contribution of
the order of $\sim$10$\%$, more recent
experiments~\cite{COMPASS,HERMES} are compatible with a null
contribution (see \cite{BeckMcKeown} for details). Other
discussions, both from theoretical and experimental points of view,
also address the role of the $s$-quark with respect to the hadron
mass~\cite{Lhuillier,Harrach,GassLeut}. In this work our interest is
focused on the contribution of the strange quarks in the electroweak
current of the nucleon. As already suggested by several
authors~\cite{Pollock,KaplanManohar}, knowledge of neutral current
form factors, combined with the EM ones, provides access to specific
strangeness content in the nucleon through analysis of PV electron
scattering.

\begin{figure}[htbp]
    \centering
        \includegraphics[width=.8\textwidth,angle=0]{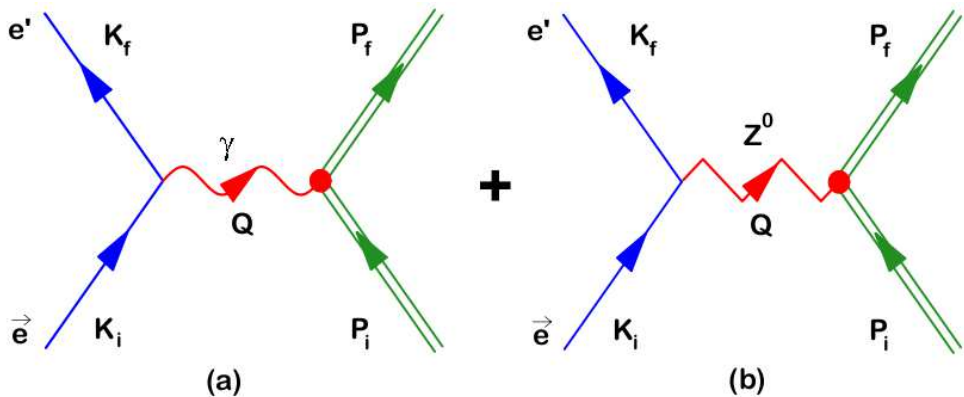}
    \caption{One-boson-exchange diagrams for electron-proton scattering: (a) EM
          interaction and (b) WNC interaction.}
    \label{fig:feynman}
\end{figure}

A complete description of the scattering process between electrons
and hadrons and/or nuclei requires not only the dominant EM
interaction but also the much smaller weak interaction, the latter
being responsible for parity violation. Assuming the Born
Approximation, {\it i.e.,} only one virtual boson exchanged in the
process (photon $\gamma$ for the EM interaction and neutral $Z$
boson for the weak neutral current (WNC) process), the corresponding
Feynman diagrams are depicted in Fig.~\ref{fig:feynman}. Because of
the smallness of the weak coupling constant compared with the EM
one, the leading-order PV contribution arises from the interference
between the two processes shown diagrammatically in Fig.~1, namely,
between the $\gamma$-exchange amplitude (diagram a: purely EM
interaction) and the $Z$-exchange (diagram b: WNC process). In what
follows we denote the scattering amplitudes associated with
$\gamma$-exchange (Fig.~\ref{fig:feynman}~(a)) and $Z$-exchange
(Fig.~\ref{fig:feynman}~(b)) as ${\cal M}_\gamma$ and ${\cal M}_Z$,
respectively. The contribution of the purely weak term $|{\cal
M}_Z|^2$ is typically negligible. It is important to stress that the
PV interference contribution, $Re({\cal M}^\ast_\gamma {\cal M}_Z)$
is $\sim$4--5 orders of magnitude smaller than the purely EM one,
$|{\cal M}_\gamma|^2$. Hence, the determination of parity violation
through electron scattering requires measurements of observables
that only exist if the weak interaction comes into play.
Furthermore, the investigation of the WNC nucleon form factors using
PV electron scattering depends strongly on the depth of our
knowledge of the purely EM form factors. Notice that higher-order
corrections beyond the one-photon-exchange approximation in the EM
sector, such as two-photon effects~\cite{YuChen}, can give
contributions of the same order or higher than the $\gamma$--$Z$
interference term. Hence, it is important to explore how sensitive
the EM nucleon form factors are to these higher-order corrections.
In~\cite{Guichon} it is argued that these corrections may affect the
Rosenbluth separation, but are relatively much less important for
the extraction of the form factor ratio using polarization
observables, and that, accordingly, using polarization degrees of
freedom in elastic $ep$ scattering can provide a clean separation of
the form factors. In this work we will use the most recent models or
prescriptions describing the EM nucleon form factors which account
for effects coming from two-photon exchange (TPE) and other
higher-order corrections in the EM interaction. Another topic of
interest that has recently been investigated is the potential impact
of isospin violations on the extraction of the nucleon strange
vector form factors~\cite{Kubis2006,Viviani2009}.

Parity violation in electron scattering emerges from the measurement
of the helicity asymmetry, also denoted as the PV asymmetry, that is
given as the ratio between the difference and the sum of the cross
sections corresponding to right and left-handed incident polarized
electrons,
\be {\cal
A}^{PV}=\frac{d\sigma^+-d\sigma^-}{d\sigma^++d\sigma^-}=\frac{d\sigma^{PV}}{d\sigma^{PC}}
\, , \label{eq1}
\ee
where the index $\pm$ indicates the helicity of
the incident electron beam. It is important to point out that the
above cross sections refer to single-arm (inclusive) scattering of
longitudinally polarized electrons with no hadronic/nuclear
polarizations. Otherwise, parity-conserving effects that are
generally much larger than the effects coming solely from PV may
also contribute to the asymmetry. As a basic example, in the case of
exclusive $A(\vec{e},e'N)$ reactions, that is, polarized electrons
and unpolarized nuclear targets, a purely EM response (called the
fifth response) which is linked to the electron helicity and
final-state interactions, contributes to the asymmetry unless
coplanar kinematics are selected~\cite{RaskDon}. Similar comments
also apply to inclusive $(\vec{e},e')$ processes but with nuclear
polarizations~\cite{DonRask}. In contrast, in the particular case of
single-arm electron scattering with no hadronic polarization, the
purely EM cross section does not depend on the electron helicity;
thus the difference in the numerator in Eq.~(\ref{eq1}) only enters
because of the weak interaction (given through the interference
between the amplitudes ${\cal M}_\gamma$ and ${\cal M}_Z$), and
therefore, a value of ${\cal A}^{PV}\neq 0$ is a clear signature of
PV effects (exchange of the $Z$-boson). On the other hand, the sum
in the denominator is dominated by the EM interaction (basically
twice the unpolarized cross section), since residual contributions
from the interference terms are negligible.

In recent years a great deal of effort has been devoted to the
measurement and determination of the PV asymmetry. The analysis of
data requires increasingly precise knowledge of the EM and WNC
nucleon form factors. In what follows we summarize the main
experimental programs that are focused on PV electron scattering
from the proton, and in some cases using deuterium or helium.

\begin{enumerate}
\item
  {\bf SAMPLE~\cite{SAMPLE1,SAMPLE2}}
  These experiments were run at MIT-Bates and involved PV electron scattering from hydrogen and
  deuterium targets.  Longitudinally polarized electrons with an energy of about 200 MeV were employed with
  scattering at large angles.
\item
  {\bf HAPPEX~\cite{Happex-99,Happex-a,Happex-b,Happex-III}} This sequence of experiments employs the facilities of Jefferson
  Lab (Hall A).
  The energy of the polarized electron beam is typically around 3 GeV and both hydrogen and helium have been used as targets;
  forward-angle scattering is involved in all cases.
  \item
  {\bf PVA4~\cite{PVA4,PVA4-2,PVA4back}} This experiment was undertaken at MAMI (Mainz). To date
  longitudinally polarized electrons were
  scattered from hydrogen and various electron beam energies and scattering angles were employed.
  Future measurements are planned for hydrogen and deuterium.
\item
  {\bf G0~\cite{G0,G0back}} This experiment was undertaken in Hall C (Jefferson Lab). Polarized
  electrons were scattered from hydrogen and deuterium.
  Forward- and backward-angle scattering measurements have been made, where in the forward configuration recoil protons
  were detected at angles corresponding to electrons in the forward direction, while in the backward configuration the apparatus
  was reversed and scattered electrons were detected.
\item
  {\bf Q-weak~\cite{Qweak1,Qweak2,Qweak3}}
  Experiment performed at JLab. Polarized electrons scattered from hydrogen at very low
  momentum transfer serve as a Standard Model test.
  Data acquisition was completed in May, 2012 and their analysis is now in progress.
 \item {\bf PVDIS at 6 GeV~\cite{PVDIS}} This experiment aims to measure the parity-violating
asymmetry for deep inelastic scattering (DIS) of polarized electrons from $^2$H
at $|Q^2|$ = 1.1 and 1.9 (GeV/c)$^2$. The combination of the two measurements (data analysis is presently
in progress) will provide a significant constraint on non-perturbative QCD effects. This may shed light
on the knowledge of the neutral effective
weak coupling constant combination 2$C_{2u}$ - $C_{2d}$. This experiment is also sensitive to the
weak effective couplings, $C_{1u}$ and $C_{1d}$, and it will provide
a baseline measurement for the future 12 GeV program~\cite{12GeV} at JLab.

\end{enumerate}

In this work we compare our theoretical predictions evaluated with
several recent descriptions of the hadronic structure with all
available ${\vec e}p$ data. Electroweak radiative corrections as
well as effects from higher-order terms in the description of the EM
interaction have been incorporated in the analysis. Special
attention is devoted to the strangeness content in the electric,
magnetic and axial-vector nucleon form factors.

Several different approaches to studies of PV ${\vec e}p$ scattering
may be followed, including
\begin{itemize}
\item Performing measurements at specific values of momentum transfer
but different values of the electron scattering angle and thereby extracting
the hadronic responses without having to resort to model assumptions for the underlying form factors.
\item Using PV ${\vec e}p$ data from different measurements at different values
of momentum transfer and scattering angle, although now invoking
some representation for the underlying form factors. For very low
momentum transfers it may be possible to make low-$Q^2$ expansions;
however, for extended ranges of momentum transfer one has to resort
to parametrizations of the form factors.
\end{itemize}
The first option is clearly preferable, although this has been
possible only for a subset of the world data, and the second option
at very low momentum transfers is only applicable for that kinematic
region. In the present study we follow the second option invoking
specific ``reasonable" models and/or parametrizations for the
nucleon form factors. In particular, for the electromagnetic form
factors we use all of high-quality the world data and consider several specific
vector meson dominance based models, as well as several popular
parametrizations of these quantities. When parity-conserving
${\vec e}p$ scattering measurements differ significantly, we investigate the
impact this has on the PV asymmetry. For the weak neutral current
form factors we consider several parametrizations. The spirit of our
approach follows that of \cite{MusDon92} where the axial-vector and
strangeness form factors were characterized by simple ``reasonable"
expressions containing a few parameters to be varied. As discussed
later in the paper, we explore the consequences of having a
non-traditional axial-vector form factor as suggested by some recent
neutrino reaction studies.

In recent years, there has been a great deal of progress in studying
the structure of the nucleon from a theoretical point of view
(see~\cite{Musolf_EPJA04} and refs. therein). In particular,
computation of strangeness form factors using 
microscopic calculations based on lattice-QCD has been presented in
previous work, for instance, results from the quenched calculations
in~\cite{Dong1998,Lewis2003}. More recently, the nucleon's strange
form factors have been simulated~\cite{Doi2009}, 
and preliminary determinations of $\Delta s$ presented
in~\cite{Babich2008,Bali2010,Collins2010,Toussaint2009,Takeda2011}.
A general study, combining lattice with chiral perturbation theory,
can be found in~\cite{Leinweber0506,Wang2009}.

Here we summarize some of our key findings --- these are discussed at length later.
We show that the uncertainty associated with the description of the EM nucleon form factors 
can be comparable to
that due to the particular choice for the nucleon�s axial-vector structure. Concerning
the specific strangeness content in the nucleon, our analysis of all available data
is basically consistent with zero magnetic strangeness. However, present data
do not allow us to select a specific strangeness content value that provides
a successful description of data at all values of $Q^2$. On the contrary, our study shows consistency
with positive electric strangeness. Although some caution should be drawn from these
general conclusions because of the present uncertainties in the evaluation of the
PV asymmetry, our results emerge from a global analysis of world data showing
the $1\sigma$ and $2\sigma$ confidence ellipses in the magnetic-electric strangeness plane.
We have also applied our study to the Q-weak experiment~\cite{Qweak1,Qweak2,Qweak3}, and
have shown how the confidence region may change due to the hypothetical Q-weak measurement,
and moreover, how the various descriptions of the EM and WNC form factors can importantly affect the
interpretation of this experiment.

The general organization of the paper is as follows: in
Sect.~\ref{sec:form} we present the general formalism needed in the
description of elastic electron-nucleon scattering processes with
parity violation. We show how to evaluate the differential cross
section and PV asymmetry specifying the different approaches
considered in this study. The study of the hadronic structure is
presented in Sect.~\ref{sec:ff}. The electroweak nucleon form
factors that enter in the general expressions for the EM and WNC
contributions are analyzed in detail. Very different prescriptions,
widely used in the literature, for the EM nucleon form factors are
considered, including some of the most recent descriptions. In
Sect.~\ref{sec:analysis} we present the results obtained for the PV
asymmetry. Various kinematical regimes are considered and we analyze
the sensitivity of ${\cal A}^{PV}$  to the specific choices made for
the nucleon's EM form factors (Sect.~\ref{sec:EMdep}) and
axial-vector form factor (Sect.~\ref{sec:axialdep}). The influence
of the $s\overline{s}$ sea quark in the electric and magnetic
strangeness form factors is also investigated in detail
(Sect.~\ref{sec:strangedep}). A systematic comparison between the
theoretical results and experimental data is provided. As discussed
later, for these kinds of study having excellent control over the EM
structure of the nucleon is required before definitive conclusions
on the strangeness content in the nucleon can be reached. The impact
of radiative corrections is also briefly addressed
(Sect.~\ref{sec:radcorr}). In Sect.~\ref{sec:fit} a global analysis
of all data is presented, while in Sect.~\ref{sec:qweak}
implications for the interpretation of the Q-weak experiment and for
potential future experiments at higher energies
(Sect.~\ref{sec:expt}) are discussed. Finally, in
Sect.~\ref{sec:concl} we summarize our basic results and present our
conclusions.


\section{General formalism for PV elastic $(e,N)$ scattering}\label{sec:form}

In this section we summarize the general formalism involved in the
description of elastic parity-violating electron-nucleon scattering.
Although, for simplicity, we restrict ourselves to the plane-wave
Born approximation (PWBA) with a single-boson ($\gamma$ or $Z$)
exchange, higher-order electroweak corrections will be discussed and
incorporated in the analysis of the results. The processes we
consider are represented in Fig.~1. Here, an electron with
four-momentum $K_i^\mu=(\epsilon_i,\nk_i)$ and helicity $h$ is
scattered through an angle $\theta_e$ to four-momentum
$K_f^\mu=(\epsilon_f,\nk_f)$, exchanging a photon (EM interaction)
or a neutral boson $Z$ (WNC interaction). The hadronic variables are
$P_i^\mu=(E_i,\np_i)$ the incident nucleon four-momentum and
$P_f^\mu=(E_f,\np_f)$ the final one. The transferred four-momentum
in the process is given by
$Q^\mu=(K_i-K_f)^\mu=(P_f-P_i)^\mu=(\omega,\nq)$. We use the
conventions and metric of \cite{Bjorken} and accordingly, with the
notation employed in previous work, $Q^2=\omega^2-q^2\leq 0$.

Given the restriction to the PWBA the transition matrix amplitude for the
scattering process can be written as \be S_{fi}=-i\int d^4X\left[
j^{\mu}_{EM}(X)A_{\mu}(X) + j^{\mu}_Z(X) Z_{\mu}(X)\right] \, , \ee
where we have introduced the EM and WNC leptonic currents,
$j^\mu_{EM},\, j^\mu_Z$, and the corresponding quantum fields,
$A^\mu$ and $Z^\mu$, attached to the hadronic vertex. The explicit
expressions for the leptonic currents are: \ba
j_{EM}^{\mu}(X)&=& - e\bar{\Psi}(X)\ \gamma^{\mu}\ \Psi(X)\\
j^{\mu}_Z(X)&=& \frac{g}{4\cos\theta_W}
      \bar{\Psi}(X)\ \left(a_V\gamma^{\mu}+a_A\gamma^5\gamma^{\mu}\right)\ \Psi(X) \, ,
\ea
where the vector and axial-vector electron couplings at tree level are assumed~\cite{DonnellyPeccei,DonDubSick}:
\be
a_V\equiv g_V=-1+4\sin^2\theta_W \,\, , \,\,\,\,\, a_A\equiv -g_A=-1\label{aAaV} \, .
\ee

After some algebra the scattering amplitude can be finally cast in
the form \be
S_{fi}=-\frac{(2\pi)^4}{V^2}\delta^4(P_f-P_i+K_f-K_i)\frac{mM}
{\sqrt{\epsilon_i\epsilon_f E_i E_f}}[{\cal M}_{\gamma}+{\cal
M}_{Z}] \label{Sfi} \, , \ee where $m$ and $M$ represent the
electron and hadron masses, respectively, and where we have
introduced the amplitudes
\ba
 {\cal M}_{\gamma}&=&j_{EM}^{\mu}\left(\frac{-ig_{\mu\nu}}{Q^2}\right)J^{\nu}_{EM}\label{Ag}\\
 {\cal M}_{Z}&=&j_{Z}^{\mu}\left(\frac{ig_{\mu\nu}}{M_Z^2}\right)J^{\nu}_{Z}\label{Az}
\end{eqnarray}
with $M_Z$ the neutral boson mass.

In the extreme relativistic limit (ERL) for the electrons,
$k_{i,f}\approx \epsilon_{i,f}$, the differential cross section is
given as \ba  \frac{d\sigma}{d\Omega_f}&=& \frac{m^2}{(2\pi)^2}
\left(\frac{\epsilon_f}{\epsilon_i}\right)^2 \overline{\sum}
\left[j^{EM}_{\mu}\left(\frac{-i}{Q^2}\right)J^{\mu}_{EM} +
j^{Z}_{\mu}
\left(\frac{i}{M_Z^2}\right)J^{\mu}_{Z}\right]^* \nonumber \\
&\times & \left[j^{EM}_{\nu}\left(\frac{-i}{Q^2}\right)J^{\nu}_{EM}
+ j^{Z}_{\nu}\left(\frac{i}{M_Z^2}\right)J^{\nu}_{Z}\right] \, , \ea
where the lab frame has been chosen, {\it i.e.,} the proton target
is taken to be at rest. The symbol $\overline{\sum}$ refers to
sum/average over spins of final/initial particles other than the
incident electron which is assumed to be longitudinally polarized
(see below).
We may isolate the contribution
coming solely from the EM interaction and the EM-WNC interference
term: \ba
\overline{\sum}\left[\left(j_{\mu}^{EM}\right)^*\left(j_{\nu}^{EM}\right)
\
\left(J^{\mu}_{EM}\right)^*\left(J^{\nu}_{EM}\right)\right]&=&l_{\mu\nu}W^{\mu\nu}
 \label{aem} \\
\overline{\sum}\left[\left(j_{\mu}^{EM}\right)^*
\left(j_{\nu}^{Z}\right) \
\left(J^{\mu}_{EM}\right)^*\left(J^{\nu}_{Z}\right)\right] &=&
\widetilde{l}_{\mu\nu}\widetilde{W}^{\mu\nu} \, , \label{aint} \ea
where we have neglected the contribution linked to the purely WNC
terms that are several orders of magnitude smaller than the
interference term, and have introduced the leptonic ($l_{\mu\nu},\,
\widetilde{l}_{\mu\nu}$) and hadronic ($W^{\mu\nu},\,
\widetilde{W}^{\mu\nu}$) tensors that can be evaluated from the
general expressions for the EM and WNC current operators.

In the case of longitudinally polarized incident electrons, as
assumed in this work, the leptonic tensors can be written as
\begin{eqnarray}
 l_{\mu\nu}&=&\frac{e^2}{8m^2}\bigl(s_{\mu\nu}+ha_{\mu\nu}\bigr)\label{lmunuFinal}\\
 \widetilde{l}_{\mu\nu}&=&\frac{-eg}{4\cos\theta_W}\frac{1}{8m^2}\biggl[(a_V-ha_A)s_{\mu\nu}+
   (ha_V-a_A)a_{\mu\nu}\biggr]\label{WlmunuFinal} \, ,
\end{eqnarray}
where $h$ refers to the electron helicity and where we have
separated the overall tensor into symmetric ($s_{\mu\nu}$) and
antisymmetric ($a_{\mu\nu}$) contributions: \be
 s_{\mu\nu}=4\left(K_{\mu}^iK_{\nu}^f+K_{\nu}^iK_{\mu}^f+\frac{Q^2}{2}g_{\mu\nu}\right) \,\, ,
\,\,\,
 a_{\mu\nu}=4i\epsilon_{\mu\nu\alpha\beta}K_i^{\alpha}K_f^{\beta} \, .
\label{amunu}
\ee

The hadronic tensors in Eqs.~(\ref{aem},\ref{aint}) are constructed from the general
expressions for the hadronic EM/WNC current operators,
\begin{eqnarray}
 W^{\mu\nu}&=&e^2 S^{\mu\nu}=
e^2\frac{1}{2} \text{Tr}\left[\frac{(\displaystyle{\not}P_i+M)}{2M}
\overline{\Gamma}^{\mu}_{EM} \frac{(\displaystyle{\not}P_f+M)}{2M}\Gamma^{\nu}_{EM}\right] \label{eq:17}
\\
\widetilde{W}^{\mu\nu}&=&\frac{eg}{4\cos\theta_W}
\left[\widetilde{S}^{\mu\nu}+\widetilde{A}^{\mu\nu}\right]=
\frac{eg}{4\cos\theta_{W}}\frac{1}{2} \text{Tr}
\left[\frac{(\displaystyle{\not}P_i+M)}{2M}\overline{\Gamma}^{\mu}_Z
\frac{(\displaystyle{\not}P_f+M)}{2M}\Gamma^{\nu}_{EM}\right] \, , \label{eq:18}
\end{eqnarray}
where we have taken into account that the hadronic EM current is
purely a polar vector whereas the WNC includes an axial-vector
contribution. No hadronic polarizations are assumed.

The differential cross section corresponding to elastic
electron-nucleon scattering can then be cast as
\begin{eqnarray}
 \frac{d\sigma^{(h)}}{d\Omega_f}&=&\frac{1}{(2\pi)^2}\left(\frac{\epsilon_f}{\epsilon_i}\right)^2
\frac{e^2}{8Q^2} \left\{\frac{e^2}{Q^2}s_{\mu\nu}S^{\mu\nu}  \right.\nonumber \\
 & +& \left.\left(\frac{\sqrt{2}g} {4M_Z\cos\theta_W}\right)^2
\left[(a_V-ha_A) s_{\mu\nu}\widetilde{S}^{\mu\nu}
+ (ha_V-a_A)a_{\mu\nu}\widetilde{A}^{\mu\nu}\right]\right\}
\label{SSD} \, ,
\end{eqnarray}
whose evaluation requires the explicit calculation of the hadronic
tensors that depend on the EM and WNC nucleon form factors. Hence, a
proper description of the internal structure of the nucleon is
needed in order to get a realistic description of the scattering
process. However, before entering into a detailed discussion of the
specific structure of the hadronic currents, the cross section can
also be expressed in terms of hadronic response functions obtained
by taking the appropriate components of the single-nucleon tensors
$S^{\mu\nu}$, $\widetilde{S}^{\mu\nu}$ and $\widetilde{A}^{\mu\nu}$:
\begin{eqnarray}
 \dfrac{d\sigma^{(h)}}{d\Omega_f}&=&\sigma_{M}\left(\frac{\epsilon_f}{\epsilon_i}\right)
\Biggl\{v_LR^L+v_TR^T+\Biggl. \left(\frac{Q^2}{e^2}\right)
\left(\frac{\sqrt{2}g}{4M_Z\cos\theta_W}\right)^2\Biggr.\nonumber\\
&\times&\biggl[(a_V-ha_A)(v_L\widetilde{R}^L+v_T\widetilde{R}^T) +
(ha_V-a_A)v_{T'}\widetilde{R}^{T'}\biggr]\Biggr\} \,
\end{eqnarray}
with $\sigma_{M}$ the Mott cross section and $v_k$ the lepton kinematical coefficients given in
the ERL as
\be
v_L=\left(\frac{Q^2}{q^2}\right)^2\,\, , \,\,\,
v_T=\tan^2\theta_e/2-\frac{1}{2}\left(\frac{Q^2}{q^2}\right) \,\, , \,\,\,
v_{T'}=\tan\theta_e/2\sqrt{\tan^2\theta_e/2-\frac{Q^2}{q^2}}\label{vT'} \, .
\ee

The functions $R^K$ ($\widetilde{R}^K$) are the hadronic EM (weak)
responses that contain all of the information on the structure of
the nucleon. They are given as bilinear combinations of the
corresponding tensors: $R^L = S^{00}$, $R^T = S^{xx}+S^{yy}$
(likewise for $\widetilde{R}^{L,T}$ in terms of
$\widetilde{S}^{\mu\nu}$) and
$\widetilde{R}^{T'}=2i\widetilde{A}^{xy}$ with the specific
components referred to a coordinate system in which the $z$-axis
lies along the direction of the transfer momentum $\nq$, $xz$ is the
scattering plane defined by the electron momenta $\nk_i$ and $\nk_f$
and the $y$-axis lies along $\nk_i\times\nk_f$. The indices $L$ and
$T,T'$ indicate contributions in the tensors along $\nq$ and
transverse to $\nq$, respectively.

\subsection{Hadronic Responses and PV Asymmetry}\label{sec:hadasy}

The general structure of the tensors containing the EM and WNC
contributions can be derived from general symmetry properties,
Lorentz covariance, charge conjugation and time reversal invariance
and, in the case of the EM interaction, current and parity
conservation. The EM current operator for on-shell nucleons is
simply given in the form: \be \Gamma^{\mu}_{EM} =
F_1\gamma^{\mu}+i\frac{F_2}{2M}\sigma^{\mu\nu}Q_{\nu} \,
                   \label{operador-em}
\ee with $F_1$ and $F_2$ the Pauli and Dirac nucleon form factors,
respectively. In the case of the WNC operator we have
\be
\Gamma^\mu_Z=\widetilde{F}_1\gamma^{\mu}+i\frac{\widetilde{F}_2}{2M}\sigma^{\mu\nu}Q_{\nu}
+G^e_A\gamma^{\mu}\gamma^5 +
\frac{\widetilde{G}_P}{M_N}Q^{\mu}\gamma^{5} \, ,
\label{operador-weak}
\ee
where $\widetilde{F}_{1,2}$ are the WNC
vector form factors and  $G^e_A$ ($\widetilde{G}_P$) the
axial-vector (pseudoscalar) ones.

Introducing these explicit expressions for the currents into
Eqs.~(\ref{eq:17}) and (\ref{eq:18}), the hadronic symmetric and
antisymmetric contributions to the tensors are finally given by \ba
 2M^2 S^{\mu\nu}&=&(F_1+F_2)^2\left(P_i^{\mu}P^{\nu}_f+P_i^{\nu}P_f^{\mu}+
   \frac{1}{2}Q^2g^{\mu\nu}\right) \nonumber \\
   &-&\left[F_2(F_1+F_2)-F_2^2\left(\frac{1}{2}-\frac{Q^2}{8M^2}\right)\right]
   (P_i+P_f)^{\mu}(P_i+P_f)^{\nu} \label{SMN}
\ea
\ba
2M^2\widetilde{S}^{\mu\nu}&=&(F_1+F_2)(\widetilde{F}_1+\widetilde{F}_2)
\left(P_i^{\mu}P^{\nu}_f+P_i^{\nu}P_f^{\mu}+\frac{1}{2}Q^2g^{\mu\nu}\right)\nonumber\\
&-&\left[\frac{1}{2}F_2
(\widetilde{F}_1+\widetilde{F}_2)+\frac{1}{2}\widetilde{F}_2(F_1+F_2)
-F_2\widetilde{F}_2\left(\frac{1}{2}-\frac{Q^2}{8M^2}\right)\right]
 (P_i+P_f)^{\mu}(P_i+P_f)^{\nu} \nonumber \\
\ea
\be
2M^2 \widetilde{A}^{\mu\nu}=i(F_1+F_2)G^e_A\
 \epsilon^{\mu\nu\alpha\beta}P^i_{\alpha}P^f_{\beta} \label{AMN} \, .
\ee Note that the pseudoscalar term
$\frac{\widetilde{G}_P}{M_N}Q^{\mu}\gamma^{5}$ does not contribute
to PV electron scattering at the order considered here. The explicit expressions for the EM/WNC
hadronic responses are the following:
\begin{eqnarray}
R^L&=&\left(1+\lambda\right)\left[F_1-\lambda F_2\right]^2\label{RL}\\
R^T&=&2\lambda\bigl[F_1+F_2\bigr]^2\label{RT}\\
\widetilde{R}^L&=&\left(1+\lambda\right)\left[F_1-\lambda F_2\right]
\left[\widetilde{F}_1-\lambda\widetilde{F}_2\right]\label{WRL}\\
\widetilde{R}^T&=&2\lambda(F_1+F_2)(\widetilde{F}_1+\widetilde{F}_2)\label{WRT}\\
\widetilde{R}^{T'}&=&2\kappa(F_1+F_2)G^e_A\label{WRT'} \, ,
\end{eqnarray}
where the lab frame has been assumed, {\it i.e.,}
$P_i^{\mu}=(M,\vec{0})$, $P_f^{\mu}=(E_f,0,0,q)$, and the usual
dimensionless variables $\lambda\equiv\omega/2M$ and $\kappa\equiv
q/2M$ have been introduced.

In terms of the scale of parity-violating effects, ${\cal
A}_0=G_F|Q^2|/(2\sqrt{2}\pi\alpha)$, with $G_F=g^2/(4\sqrt{2}M_Z^2\cos^2\theta_W)$
the Fermi coupling and $\alpha$ the fine structure constant, the expression for the
differential cross section that results is
\begin{eqnarray}
 \dfrac{d\sigma^{(h)}}{d\Omega_f}&=&\sigma_{M}\left(\frac{\epsilon_f}{\epsilon_i}\right)
\Biggl[v_LR^L+v_TR^T\Biggl.\Biggr.\nonumber\\
&-&\frac{{\cal A}_0}{2}\biggl((a_V-ha_A)(v_L\widetilde{R}^L+v_T\widetilde{R}^T) +
(ha_V-a_A)v_{T'}\widetilde{R}^{T'}\biggr)\Biggr] \, .
\end{eqnarray}
As noted, PV effects (linked to $\widetilde{R}^K$ responses) can be
isolated through $\vec{e}N$ measurements. In particular, by
measuring the cross sections corresponding to the two helicities
$h=\pm 1$ and taking their difference, the result depends on the
EM/WNC interference responses (denoted simply as the PV cross
section), \be
 \left(\frac{d\sigma}{d\Omega_f}\right)^{PV} = \frac{1}{2}\left(\frac{d\sigma^{(+)}}
{d\Omega_f}-\frac{d\sigma^{(-)}}{d\Omega_f}\right)\nonumber\\
= \sigma_{M}\left(\frac{\epsilon_f}{\epsilon_i}\right)\frac{{\cal
A}_0}{2}
\left[a_A(v_L\widetilde{R}^L+v_T\widetilde{R}^T)-a_Vv_{T'}\widetilde{R}^{T'}\right]
\, . \ee On the contrary, by taking the sum of the
helicity-dependent cross sections one gets the purely EM cross
section plus a minor contribution coming from the EM/WNC
interference that can be neglected. Hence, one may write \be
 \left(\frac{d\sigma}{d\Omega_f}\right)^{PC}=
\frac{1}{2}\left(\frac{d\sigma^{(+)}}{d\Omega_f}+\frac{d\sigma^{(-)}}{d\Omega_f}\right)
\approx \sigma_{M}\left(\frac{\epsilon_f}{\epsilon_i}\right) \left[v_LR^L+v_TR^T\right] \, ,
\ee
where the index PC refers to Parity-Conserving cross sections.

The helicity-difference asymmetry (also called the PV asymmetry) is defined as
\be
 {\cal A}^{PV}\equiv \frac{\left(d\sigma/d\Omega_f\right)^{PV}}
{\left(d\sigma/d\Omega_f\right)^{PC}}= \frac{{\cal A}_0}{2}
\left[\frac{a_A(v_L\widetilde{R}^L+v_T\widetilde{R}^T)-a_Vv_{T'}\widetilde{R}^{T'}}
{v_LR^L+v_TR^T}\right]\label{asimetria} \, . \ee Using the explicit
expressions for the EM and WNC hadronic responses in
Eqs.~(\ref{RL}-\ref{WRT'}) and introducing the Sachs form factors,
{\it viz.,} $G_E=F_1-\tau F_2$ and $G_M=F_1+F_2$ (likewise for
$\widetilde{G}_{E,M}$ in terms of $\widetilde{F}_{1,2}$), the PV
asymmetry can be written as
\be {\cal A}^{PV}=\frac{{\cal
A}_0}{2}\left[ \frac{a_A\left(\varepsilon
G_E^N\widetilde{G}_E^N+\tau G_M^N\widetilde{G}_M^N\right) -
a_V\sqrt{1-\varepsilon^2}\sqrt{\tau(1+\tau)}G_M^N G^{e,N}_A}
{\varepsilon(G_E^N)^2+\tau (G_M^N)^2} \right] \, , \label{scatt_angle}
\ee
with
$\tau\equiv |Q^2|/4M^2$,
$\varepsilon=\left[1+2(1+\tau)\tan^2{\theta_e}/2 \right]^{-1}$ and
the index $N$ referring to protons or neutrons.

The analysis of the PV asymmetry in different kinematical regions is
simplified by isolating the contributions linked to the electric
(longitudinal), magnetic (transverse symmetric) and axial-vector
(actually magnetic/axial-vector interference; transverse
antisymmetric) distributions. Thus, we may write \be
 {\cal A}^{PV}= {\cal A}_E + {\cal A}_M + {\cal A}_A \label{A-componentes}
\ee
with
\ba
 {\cal A}_E &=&
 \frac{{\cal A}_0}{2}\frac{a_A\varepsilon\ G_E^N\widetilde{G}_E^N}{G^2} \label{cont_e}\\
 {\cal A}_M &=& \frac{{\cal A}_0}{2}
\frac{a_A\tau\ G_M^N\widetilde{G}_M^N}{G^2} \label{cont_m}\\
 {\cal A}_A&=& -\frac{{\cal A}_0}{2}
\frac{a_V
\sqrt{1-\varepsilon^2}\sqrt{\tau(1+\tau)}G_M^N G^{e,N}_A
}{G^2} \label{cont_a}\, , \ea where we have introduced the term
$G^2\equiv\varepsilon(G_E^N)^2+\tau (G_M^N)^2=(1+\tau)\varepsilon
F^2$ that depends only on the purely EM interaction. Note that both
channels in the EM sector, {\it i.e.,} electric $E$ and magnetic
$M$, enter in the three separate PV asymmetry contributions defined
in the above equations.

In the next section we focus on the analysis of the specific
structure of the nucleon and provide various representations for the
EM and WNC nucleon current operators.


\section{Hadronic structure: EM and WNC nucleon form factors}\label{sec:ff}

As shown in the previous section, the evaluation of PV observables
(cross section, responses and helicity-asymmetry) requires knowledge
of the EM and WNC nucleon form factors. In this section we show how
to construct the electroweak hadronic currents and present a
detailed study of the nucleon structure, comparing the results of
different theoretical descriptions with experimental data.

The general Dirac structure of the vector and
axial-vector currents in the lepton channel (to leading-order
tree-level) is: $j^{\mu}_V\sim \bar{u}_\ell\gamma^{\mu} u_\ell$ and
$j^{\mu}_A\sim \bar{u}_\ell\gamma^{\mu}\gamma^5 u_\ell$, with
$u_\ell$ the lepton Dirac spinor. The EM and WNC hadronic currents,
$J_\mu^{EM}$, $J_\mu^{WNC,V}$ and $J_\mu^{WNC,A}$, are characterized
by the corresponding quark current operators:
\begin{eqnarray}
 J^{EM}_{\mu}   &=& \sum_{q}Q_q\bar{u}_q\gamma_{\mu}u_q \label{em1}\\
 J^{WNC,V}_{\mu} &=& \sum_{q}g^q_V\bar{u}_q\gamma_{\mu}u_q \label {neutv}\\
 J^{WNC,A}_{\mu} &=& \sum_{q}g^q_A\bar{u}_q\gamma_{\mu}\gamma_5u_q \label{neuta} \, ,
\end{eqnarray}
where the indices $WNC,V$ ($WNC,A$) refer to the vector
(axial-vector) contribution in the WNC, and the sum extends over all
flavors of quarks: $u$, $d$, $s$, $c$, $b$ and $t$. The term $Q_q$
represents the EM lepton charge, and $g^q_V$ ($g^q_A$) the
corresponding vector (axial-vector) charge in the weak sector.

In what follows we restrict the description of the hadronic states
to the contribution of the three lightest quarks
($u,d,s$).\footnote{The error introduced by neglecting the heavier
quarks is expected to be of the order of $10^{-4}$ ($10^{-2}$) for
the vector (axial-vector) currents~\cite{KaplanManohar,MusDon92}.}
In this case, the EM and WNC vector and axial-vector currents can be
expressed in the form~\cite{Musolf1994}
\begin{eqnarray}
J_{\mu}^{EM} &=& J^{EM}_{\mu}(T=0)\ +\ J^{EM}_{\mu}(T=1) \label{em-current}\\
J^{WNC,V}_{\mu}&=&\xi^{T=1}_V J^{EM}_{\mu}(T=1)\ +\
\sqrt{3}\,\xi_V^{T=0}
  J_{\mu}^{EM}(T=0)\ +\ \xi^{(0)}_V \hat{V}_{\mu}^{(s)}  \label{ncv-current}\\
J^{WNC,A}_{\mu}&=&\xi^{T=1}_A\hat{A}^{(3)}_{\mu}\ +\
\xi_A^{T=0}\hat{A}_{\mu}^{(8)}\ +\ \xi^{(0)}_A \hat{A}_{\mu}^{(s)}
\label{nca-current}\, ,
\end{eqnarray}
where we have separated the isoscalar and isovector EM currents and
have made use of the SU(3) octet and singlet currents. In general we
may write
\ba
 J^{EM}_{\mu}(T=0) &=& \left[\bar{u}\gamma_{\mu}u+\bar{d}
  \gamma_{\mu}d-2\bar{s}\gamma_{\mu}s\right]/6\\
 J^{EM}_{\mu}(T=1) &=&\left[\bar{u}\gamma_{\mu}u-\bar{d}\gamma_{\mu}d\right]/2\\
\hat{A}^{(3)}_{\mu}&=&\left[\bar{u}\gamma_{\mu}\gamma_5u-\bar{d}
  \gamma_{\mu}\gamma_5d\right]/2\\
\hat{A}^{(8)}_{\mu}&=&\left[\bar{u}\gamma_{\mu}\gamma_5u+
  \bar{d}\gamma_{\mu}\gamma_5d-2\bar{s}\gamma_{\mu}\gamma_5s\right]/(2\sqrt{3}) \, .
\ea
The terms $\hat{V}_{\mu}^{(s)}$ ($\hat{A}_{\mu}^{(s)}$) explicitly include
the vector (axial-vector) currents between strange
quarks: \be \hat{V}^{(s)}_{\mu}\equiv\bar{s}\gamma_{\mu}s \,\,
,\,\,\,\, \hat{A}^{(s)}_{\mu}\equiv\bar{s}\gamma_{\mu}\gamma_5s \, .
\ee Finally, the $\xi$ coefficients represent the coupling constants
that can be written as
\begin{eqnarray}
 \xi^{T=1}_V &=& g_V^u-g_V^d = 2(1- 2\sin^2\theta_W)\left[1+R_V^{T=1}\right] \label{xiv1}\\
 \sqrt{3}\,\xi^{T=0}_V &=& 3(g_V^u+g_V^d)=-4\sin^2\theta_W\left[1+R_V^{T=0}\right]\label{xiv2} \\
 \xi_V^{(0)} &=& g_V^u+g_V^d+g_V^s = -\left[1+R_V^{(0)}\right]\label{xiv3} \\
 \xi^{T=1}_A &=& g_A^u-g_A^d =-2\left[1+R_A^{T=1}\right] \label{xia1} \\
 \xi^{T=0}_A &=& \sqrt{3}(g_A^u+g_A^d)=\sqrt{3} R_A^{T=0}\label{xia2} \\
 \xi_A^{(0)} &=& g_A^u+g_A^d+g_A^s = 1+R_A^{(0)} \label{xia3} \, ,
\end{eqnarray}
where we have included the radiative corrections $R_{V,A}^{(a)}$
that are in general both $Q^2$- and process-dependent. These arise
from higher-order elementary lepton-quark amplitudes, and we note
that the effect of heavy quark current matrix elements, formally
omitted in the previous expressions, may also be included in the
$R_{V,A}^{(a)}$ functions (see~\cite{Musolf1994} for more details).

In the general expressions for the hadronic weak neutral currents in
Eqs.~(\ref{ncv-current},\ref{nca-current}), the physics associated
with the electroweak gauge theory is contained into the coupling
constants $\xi$, while hadronic effects emerge from the current
matrix elements between quarks. Information on the various current
matrix elements involved in the previous expressions can be obtained
from different sources. PC electron scattering experiments provide a
direct way to shed light on the purely EM (isoscalar and isovector)
nucleon form factors. On the other hand, $\beta$-decay and
semi-leptonic hyperon decay processes are sensitive to the
axial-vector contribution in the weak interaction current. In this
work, our interest focuses on the analysis of the strangeness
current matrix elements and their effects on PV observables, {\it
i.e.,} on the helicity asymmetry, and on the isovector axial-vector
form factor.

The single-nucleon matrix elements of the electroweak currents, that
are consistent with Lorentz covariance as well as parity and
time-reversal invariance, are given through
$\overline{u}(P_f)\Gamma^\mu_{a} u(P_i)$ where $u(P)$ are the
single-nucleon wave functions properly normalized, and
$\Gamma^\mu_{a}$ are the corresponding EM and/or WNC current
operators in Eqs.~(\ref{operador-em},\ref{operador-weak}). Making
use of the expressions given in Eqs.~(\ref{ncv-current}) and
(\ref{nca-current}) for the WNC operators, the weak interaction
nucleon form factors can be expressed in the general form
\ba
 \widetilde{G}_{a}(Q^2)&=&\xi_V^{T=1}G_{a}^{T=1}\tau_3+
 \sqrt{3}\,\xi_V^{T=0}G_{a}^{T=0}+\xi_V^{(0)}G_a^{(s)}\, , \,\,\,\, a=E,M  \\
 G^{e,N}_A(Q^2)&=&\xi_A^{T=1}G_A^{(3)}\tau_3 + \xi^{T=0}_A G_A^{(8)} + \xi_A^{(0)}G_A^{(s)}\label{gaXis}
\,\,\,\,
\ea
with $G_a^{T=0,1}$ the isoscalar and isovector
combinations of the EM Sachs form factors of the nucleon,
$G_A^{(3,8)}$ the triplet and octet axial-vector form factors, and
$G_{E,M,A}^{(s)}$ the vector and axial-vector
strange-quark form factors. At tree level, the following expressions
apply to the nucleon WNC form factors:
\begin{eqnarray}
 \widetilde{G}_{E,M}^{p,n}&=&(1-4\sin^2\theta_W)G_{E,M}^{p,n} - G_{E,M}^{n,p} - G_{E,M}^{(s)}\label{WGEMpn}\\
 G_A^{e,N}&=& - 2 G_A^{(3)}\tau_3+G_A^{(s)} = - (G_A^p-G_A^n)\tau_3 + G_A^{(s)}\label{WGApn} \, ,
\end{eqnarray}
where the nucleon has been assumed to be an eigenstate of isospin.
As shown, the WNC form factors of the nucleon are determined by the
purely EM ones $G_{E,M}^{p,n}$, the axial-vector $G_A^{e,N}$ and the
terms $G_{E,M,A}^{(s)}$ that only enter if the nucleon has non-zero
strangeness content. Hence, in order to provide reliable analyses of
PV electron scattering observables, excellent control of the EM
structure of the nucleon is needed in addition to increasingly
precise knowledge about the axial-vector form factors. In this
situation, PV data can be safely used as a basic tool to determine
how much strangeness enter in the nucleon structure. In what follows
we present a systematic study of the nucleon form factors and
discuss the large variety of prescriptions and models used in the
literature.

\subsection{EM structure of the nucleon: $G_{E,M}^{p,n}$}\label{sec:EM}

The EM structure of the nucleon is one of the basic ingredients
entering the description of lepton-nucleon scattering processes.
Quantum Chromodynamics (QCD) is the fundamental theory of strong
interactions; however, the complexity of the quark-gluon dynamics
does not allow one to obtain analytical solutions of QCD in the
energy regime relevant for low-$Q^2$ nuclear physics. Instead,
various approaches based on numerical simulations of the theory on a
lattice and/or through the use of effective hadron Lagrangians have
been used.

As discussed above, in the case of free (on-shell) nucleons and the
purely EM interaction, the hadronic structure is fully characterized
by two functions: the electric $(G_E^N$) and magnetic ($G_M^N$)
nucleon form factors (or alternatively $F_{1,2}^N$), whose dynamical
structure is given by their dependence on the only independent
scalar variable in the scattering process, {\it i.e.,} the
transferred four-momentum $Q^2$. It is important to point out that
the description of nucleons in the nuclear medium, that is,
off-shell nucleons, is much more complex: not only can the nucleon
form factors depend explicitly on new independent dynamical
variables in the process, but also the general structure of the EM
hadronic current should include additional form factors (see
\cite{CabDonPou,CrisCabDon,Naus,TieTjon} for details).

From the experimental point of view, most of the information at our
disposal on the EM nucleon form factors comes from measurements of
elastic electron-nucleon scattering. In the case of the proton, the
use of hydrogen as a target has led to excellent determinations of
the behavior of $G_{E,M}^p$ (see~\cite{Price,
Berger, Hanson, Borkowski, Walker, Murphy, Andivahis, Qattan, Gayou2002, Gayou2001, Punjabi, Christy,
Puckett, Crawford2007, Paolone, Zhan, Ron, Bosted, Sill}).
In contrast, information on the neutron
form factors is less precise because of the lack of free neutrons
and thus the requirement to use nuclei as targets. Indeed,
information on $G_{E,M}^n$ comes mostly from analyses of scattering
on light nuclear systems, such as deuterium and helium, typically
exploiting polarization degrees of freedom to isolate the form
factors (see~\cite{Warren, Riordan, Geis, Xu2003, Anderson, Lachniet, Madey,
Zhu, Schiavilla, Bermuth, Becker, Herberg, Ostrick, Passchier, Rohe, Eden, Meyerhoff, Kubon, Xu2000, Anklin1998,
Bruins, Anklin1994, Gao, Lung, Markowitz, Esaulov, Bartel1972, Bartel1969}).

\begin{figure}[htbp]
    \centering
        \includegraphics[width=0.78\textwidth,angle=270]{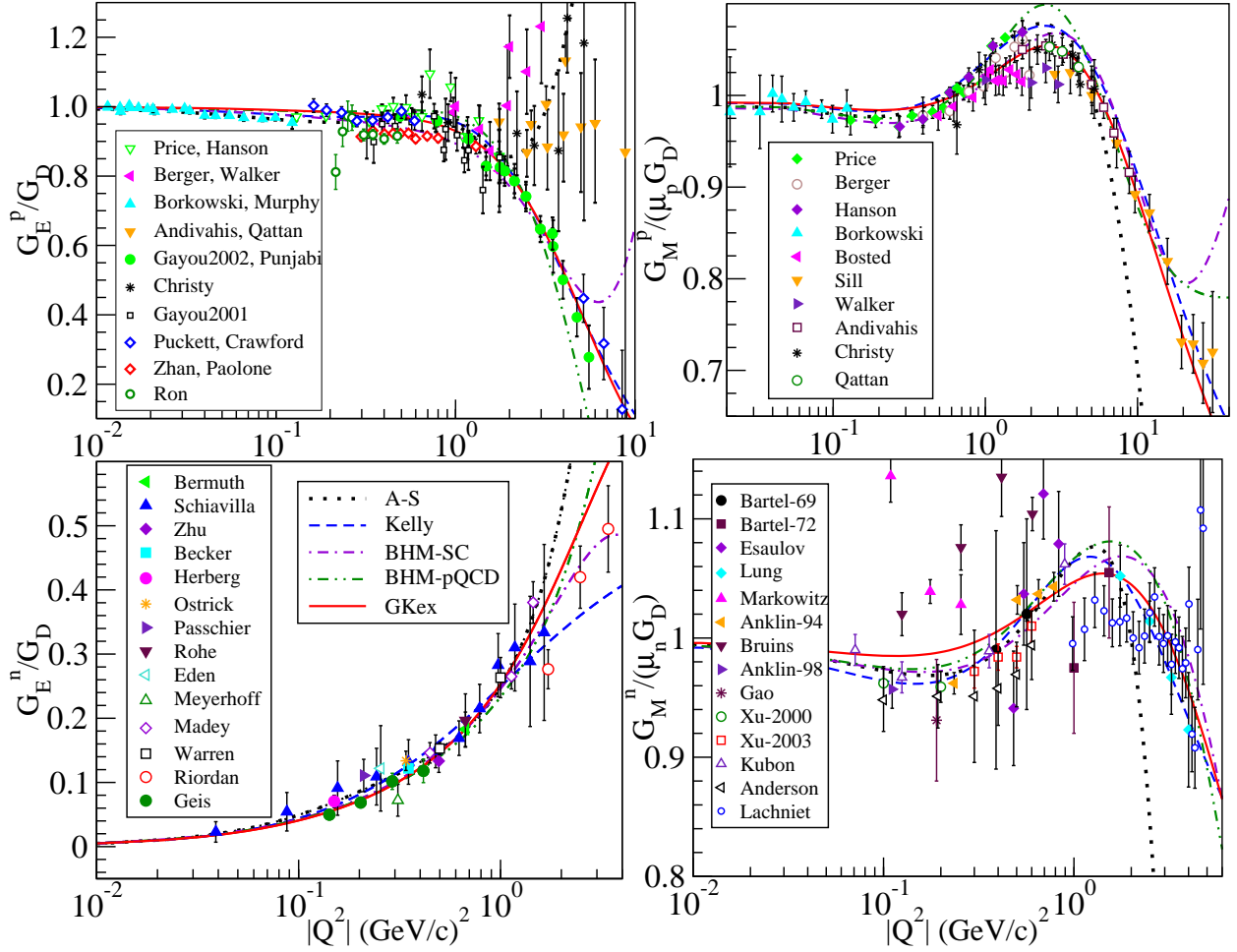}
    \caption{(Color online) EM nucleon form factors obtained using different descriptions compared
          with data. The proton electric form factor (top-left panel) corresponding to Gayou2002,
          Gayou2001, Punjabi, Puckett, Zhan, Ron, Paolone and Crawford have been obtained from
          $R_p$ data by dividing by the GKex model values of $G_M^p/\mu_p$. The same applies to Geis in the case of the electric
          neutron form factor (left-bottom), but using the GKex model $G_M^n/\mu_n$. The data are taken from references~\cite{Price,
          Berger, Hanson, Borkowski, Walker, Murphy, Andivahis, Qattan, Gayou2002, Gayou2001, Punjabi, Christy,
      Puckett, Crawford2007, Paolone, Zhan, Ron, Bosted, Sill, Warren, Riordan, Geis, Xu2003, Anderson, Lachniet, Madey,
      Zhu, Schiavilla, Bermuth, Becker, Herberg, Ostrick, Passchier, Rohe, Eden, Meyerhoff, Kubon, Xu2000, Anklin1998,
      Bruins, Anklin1994, Gao, Lung, Markowitz, Esaulov, Bartel1972, Bartel1969}.}
    \label{fig:ffem}
\end{figure}

Rosenbluth separations have been used for years to extract the
contributions from the two elastic nucleon form factors in the cross
section. However, this procedure presents important difficulties in
the region of high $|Q^2|$ because of the dominance of the $G_M^2$
term and the very small (below 1$\%$ in some experiments at high
$|Q^2|$)  contribution from $G_E^2$. More recently, the use of
nucleon polarization techniques has permitted the extraction of
interference effects that go as $G_E G_M$ and hence provide
relatively larger contributions. However, there still remain issues
to be resolved that emerge from comparison of the results of
different experiments, even in the regime of low-$Q^2$. In
Fig.~\ref{fig:ffem} we show data on nucleon form factors from a
large variety of analyses. We consider both electric (left panels)
and magnetic (right) results for the proton (top panels) and the
neutron (bottom). Data are compared with a large variety of models
that are described below.

The internal dynamics of the nucleon is governed by the constituent
quarks and the exchanged gluons that, for instance, may be simulated
using lattice-QCD. Alternatively, approaches based on phenomenology
and/or simplified models may be invoked. All of these approaches
should be consistent with the behavior of the form factors at the
limits where one can be sure of the answers. In the static limit,
{\it i.e.,} $Q^2=0$, the EM nucleon form factors should give the
correct values for the charge and magnetic moment of the nucleon:
\begin{eqnarray}
G_E^p(0)=1,&\ \ G_M^p(0)=\mu_p=2.793,\ \ \ \nonumber\\
G_E^n(0)=0,&\ \ G_M^n(0)=\mu_n=-1.913.
\end{eqnarray}
In the opposite extreme, at very high $|Q^2|$, the asymptotic
behavior of the nucleon form factors can be obtained using
perturbative QCD (pQCD). These yield $F_1$ dependent on $Q^{-4}$ and
$F_2\sim F_1/Q^2$. Once the behavior of $G_{E,M}$ in the extreme
situations is fixed, the specific dependence with the four-momentum
transfer at small-intermediate values is required. In what follows
we present different models, some of them widely used in the
literature, and compare their predictions with available data. To
make the discussion simpler, we have considered models in two basic
categories, phenomenology and Vector Meson Dominance (VMD).
\begin{itemize}
\item {\bf Phenomenological Models.} \\
Within this category, one may consider the Galster dipole parameterization that
makes use of the following functional dependence:
$G_E^p=G_D^V$, $G_E^n=-\mu_n\tau G_D^V\xi_n$, $G_M^p=\mu_p G_D^V$ and $G_M^n=\mu_n G_D^V$,
with $G_D^V=(1+\lambda_D^V\tau)^{-2}$ and $\xi_n=(1+\lambda_n\tau)^{-1}$. We consider
the standard values of the parameters: $\lambda_D^V=4.97$,
$\lambda_n=5.6$, $\mu_p=2.79$ and $\mu_n=-1.91$. This model, still
used in the literature, provides a reasonable description of proton
data at $|Q^2|\leq 1$ (GeV/c)$^2$ ($\sim$5$\%$). In the case of the
neutron, the description is significantly less precise because of
data uncertainties.

In this work we have considered two relatively new prescriptions developed by Kelly~\cite{Kelly2004} and
Arrington and Sick~\cite{A-S} (denoted A-S).
In particular, the prescription developed by Kelly
constitutes an extension of the Galster parameterization, providing a reasonable
description of recent data taken from polarization
measurements. Within this model, the electric and magnetic form factors of the proton, together with the
neutron magnetic form factor, are given by the general function:
\be
 G(Q^2)\propto\frac{\sum_{k=0}^1 a_k\tau^k}{1 + \sum_{k=1}^{3}b_k\tau^k}\label{ec:kelly} \, ,
\ee
where $a_0=1$ and the rest of parameters are given in Table~\ref{fig:tablaKelly}.

In the case of the electric form factor of the neutron, the Galster parameterization as given in
\cite{Galster} is used, {\it i.e.,}
\be
 G_E^n(Q^2)=\frac{A\tau}{1+B\tau}G_D(Q^2) \, ,
\ee
with $A$ and $B$ as given in Table~\ref{fig:tablaKelly} and
$G_D(Q^2)=(1+|Q^2|/\Lambda^2)^{-2}$ with $\Lambda^2=0.71$ (GeV/c)$^2$.
\begin{table}[htbp]
\centering
\begin{tabular}{c c c c c c c}
\hline
\hline
 F.F.      & $a_1$      & $b_1$          & $b_2$          & $b_3$          & $A$& $B$ \\
\hline
 $G_E^p$       & $-0.24\pm0.12$ & $10.82\pm0.19$ & $12.82\pm1.1$  & $21.97\pm6.8$ &    &   \\
 $G_M^p/\mu_p$ & $0.12\pm0.04$  & $10.97\pm0.11$ & $18.86\pm0.28$ & $6.55\pm1.2$  &    &   \\
 $G_M^n/\mu_n$ & $2.33\pm1.4$   & $14.72\pm1.7$  & $24.20\pm9.8$  & $84.1\pm41$   &    &   \\
 $G_E^n$       &             &                &                &               & $1.80\pm0.04$ & $3.30\pm0.32$\\
\hline
\hline
\end{tabular}
\caption{Values of the parameters in the Kelly prescription~\cite{Kelly2004}.}
\label{fig:tablaKelly}
\end{table}

The parametrizations of the form factors provided by A-S include the
effects of the two-photon exchange corrections to the extracted EM
form factors. This representation applies to momentum transfers up
to $|Q|\equiv \sqrt{|Q^2|}=1$ GeV/c, and makes use of a continued
fraction (CF) expansion in the form, \be
G_{CF}(Q)=\frac{1}{1+\frac{b_1Q^2}{1+\frac{b_2Q^2}{1+\cdots}}} \, ,
\ee where different values for the fit parameters $b_i$ are used for
the EM proton and neutron form factors. In the particular case of
$G_{E}^n$ a modified three-parameter CF expansion is considered:
$G_E^n(Q)=0.484 \times Q^2\times G_{FC}$ with $Q^2$ given in
(GeV/c)$^2$.
\item {\bf Models based on Vector Meson Dominance (VMD).}\\
A more fundamental representation of the nucleon form factors can be
obtained from models based on Vector Meson Dominance (VMD). Here,
the virtual photon is assumed to be transformed into a neutral
vector meson that couples to the corresponding hadron (see
\cite{Bhaduri} for details). Thus, the nucleon form factors are
expressed in terms of meson propagators and meson-nucleon form
factors. Within this framework a variety of descriptions of the EM
nucleon form factors have been presented in the literature. Some of
the most representative cases are: i) H\"ohler~\cite{Hohler}, based
on the use of dispersion relations to obtain the contribution of the
$\pi\pi$ continuum, fitting the width of the $\rho$ meson with a
simple function of the mass, and representing the $\omega$ and
$\phi$ mesons by simple poles, and ii)
Gari-Krumpelmann~\cite{GariKrumpelmann} that incorporates the
high-$|Q^2|$ behavior as provided by pQCD using differing
convergence rates of hadronic and quark form factors.
The use of dispersion relations in the analysis of isoscalar vector current nucleon form factors has
been considered in~\cite{PRL801998,PRC601999}.
In particular, the authors in \cite{PRC601999} include explicitly the continuum  $K\overline{K}$ contribution
in refitting the isoscalar EM form factors, and conclude that a naive VMD approach represents an effective
parametrization, but leads to erroneous valeus of the $\phi$-nucleon resonance couplings. This also has implications
for the nucleon's strange vector form factors.

Within the general framework of VMD models, in this work we present
results corresponding to two of the most recent descriptions
provided in the literature. On the one hand, we employ the
model denoted GKex~\cite{Lomon2002,Lomon2001,Crawford2010}
developed by Lomon and collaborators whose validity extends to a
wide range in the transferred momentum. This model is a generalized
description of the original GK prescription, incorporating in
addition to the asymptotic pQCD behavior, effects due to the vector
mesons $\rho,\ \rho',\ \omega,\ \omega'$ and $\phi$ and including a
width for the $\rho$. On the other hand, we also consider the model
developed by Beluskin, Hammer and Mei\ss{}ner~\cite{BHM} (BHM model)
that is an extension of the H\"ohler-type model. In addition to the
dispersion relations some constraints are also incorporated into the
model, namely, contributions to the continuum coming from $\pi\pi$,
$K\bar{K}$ and $\rho\pi$, and asymptotic convergence at high
$|Q|^2$. Two basic approaches concerning the dynamical dependence on
$|Q^2|$ at high values (pQCD behavior) are used: i) SuperConvergence
approach (denoted here as BHM-SC) and ii) explicit pQCD continuum
approach (BHM-pQCD). In the former the asymptotic behavior is
obtained by choosing the residues of the vector meson pole terms in
such a way that a spectral function consistent with asymptotic
behavior emerges (see \cite{BHM} for details). The latter approach
explicitly enforces the pQCD behavior, and it is consistent with a
nonvanishing imaginary part of the form factors in the timelike
region.
\end{itemize}


%

In Fig.~\ref{fig:ffem} we present the EM nucleon form factors versus
$|Q^2|$ for all of the models described above, and compare them with
data. As shown, all prescriptions provide reasonable descriptions of
data at low $|Q^2|$, with a relatively small dispersion between the
different curves. On the contrary, for increasing values of the
transferred momentum the differences between the models go up
significantly. This is in particular the case for the A-S
prescription whose value for $G_E^p/G_D$, with $G_D$ the standard
dipole form, starts to grow rapidly for $|Q^2|\geq 2$ (GeV/c)$^2$
whereas the other models lead to decreasing $G_E^p/G_D$.
In contrast, the other models are reasonably successful at
representing the data above 1.8 (GeV/c)$^2$ taken from polarization
measurements. Note, however, that the A-S parameterization was only
designed to be used when $|Q^2|\leq 1$ (GeV/c)$^2$.

In the case of $G_M^p$ (right-top panel), data have been measured
for a momentum transfer range that is significantly greater than for
the other form factors. As shown, the ratio $G_M^p/(\mu_pG_D)$ is
relatively close to unity until $|Q^2|\sim 1$ (GeV/c)$^2$. Then, it
increases and reaches its maximum in the region $\sim$3--5
(GeV/c)$^2$ before decreasing rapidly for $|Q^2|>7$ (GeV/c)$^2$. All
prescriptions reproduce the general behavior of data out to very
high values of $|Q^2|$, with the exception of the A-S prescription
(again, only to be used when $|Q^2|<1$ (GeV/c)$^2$). Also noteworthy
is that the BHM-pQCD model clearly overestimates data located at
$|Q^2|$-values where the maximum is reached.

As already mentioned in previous paragraphs, the extraction of the
neutron form factors from electron-deuteron and electron-$^{3}$He
scattering leads to greater uncertainties and a more restricted
momentum transfer range. This is clearly illustrated in the bottom
panels shown in Fig.~\ref{fig:ffem}. In the case of the magnetic
contribution to the neutron, the data scatter significantly in the
region below 1 (GeV/c)$^2$. The five models presented track the
average of the scattered data in this region, fitting the
higher-$|Q^2|$ behavior, except for the A-S prescription that falls
much faster. Finally, data for $G_E^n/G_D$ are presented in the
left-bottom panel compared with the five prescriptions considered.
Here, data derived from different polarization techniques as well as
values obtained from analysis of the deuteron quadrupole form factor
data~\cite{Schiavilla} are plotted. From comparison with theory we
observe that all prescriptions provide reasonable descriptions of
data up to $\sim$1 (GeV/c)$^2$. For higher $|Q^2|$ the models start
to deviate, even changing the slope of the curve for the BHM-SC
case.


\begin{figure}[htbp]
    \centering
    \includegraphics[width=.75\textwidth,angle=270]{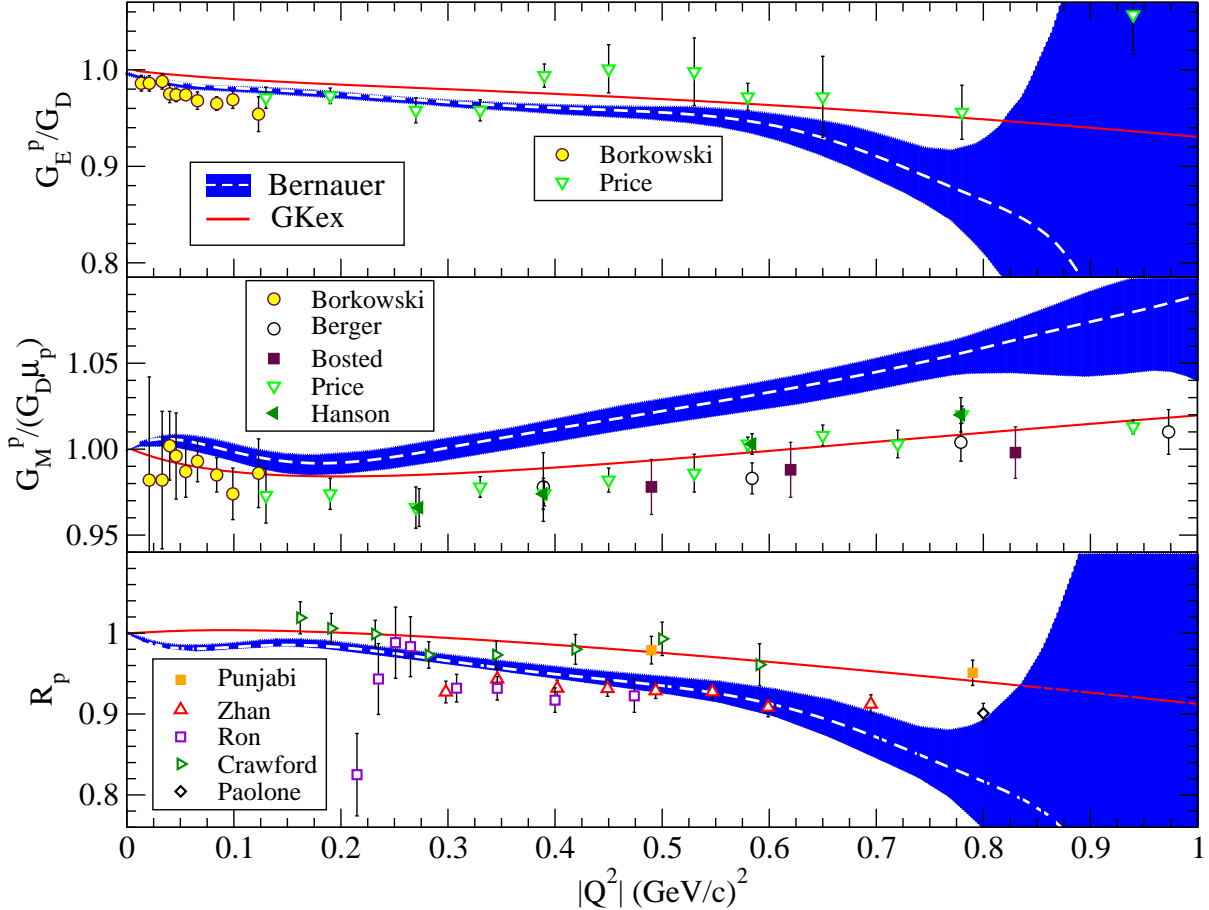}
    \caption{(Color online) EM nucleon form factors from different experiments (see Fig.~\ref{fig:ffem} for references)
    are compared with the GKex model
          and with the data of Bernauer {\it et al.}~\cite{Bernauer} (see text for details).}
    \label{fig:bernauer}
\end{figure}

To complete the discussion of the EM form factors we present in
Fig.~\ref{fig:bernauer} the analysis recently performed by Bernauer
{\it et al.}~\cite{Bernauer}, where about 1400 elastic
electron-proton cross sections were measured with four-momentum
transfers up to $|Q^2|\sim 1$ (GeV/c)$^2$. The dashed white lines in
Fig.~\ref{fig:bernauer} represent the best fits to these data,
whereas the blue shadowed areas include the statistical and
experimental systematic errors plus effects coming from Coulomb
corrections (see \cite{Bernauer} for details). We compare these data
with the results provided by the GKex model (red line). To make the
discussion that follows easier, we also include in the graph data
coming from experiments based on Rosenbluth separations in the two
upper panels and data from polarization experiments in the bottom
panel, both already shown in Fig.~\ref{fig:ffem}.

The proton electric form factor normalized to the dipole form is
presented in the top panel. We notice that GKex slightly
overestimates the behavior of the data of Bernauer {\it et al.}, but
it reproduces older Price measurements~\cite{Price}. It is important
to point out that for transfer momenta below 0.7 (GeV/c)$^2$ the
discrepancy between GKex and the data of Bernauer {\it et al.} and
Borkowski {\it et al.}~\cite{Borkowski} is on average less than
$\sim$2$\%$. The results for the proton magnetic form factor are
presented in the middle panel. In this case, the GKex model fits
nicely data coming from the older experiments, but it underestimates
the new analysis performed by Bernauer {\it et al.}; the difference
is of the order of $\sim$4$\%$ at $|Q^2|=0.7$ (GeV/c)$^2$. Finally,
the bottom panel in Fig.~\ref{fig:bernauer} shows the ratio
$R_p=\mu_pG_E^p/G_M^p$. We display the most recent data presented in
the literature, Paolone {\it et al.,}~\cite{Paolone}, Zhan {\it et
al.}~\cite{Zhan} and Ron {\it et al.}~\cite{Ron}. As shown, they are
in accord with the analysis of Bernauer {\it et al.} but differ from
previous experiments, namely those of Punjabi {\it et
al.}~\cite{Punjabi} and Crawford {\it et al.}~\cite{Crawford2007}.
At $|Q^2|=0.7$ (GeV/c)$^2$ the difference is about $\sim$6-7$\%$. In
Sect.~\ref{sec:EMdep} the consequences of using the results of
Bernauer {\it et al.} for the EM form factors, rather than the GKex
model fit to the older data, in obtaining the PV asymmetry are
examined.

Summarizing, discrepancies between data taken in different experiments and the results provided
by the GKex model are below $\sim$6-7$\%$ in the range $|Q^2|\leq 0.7-0.8$ (GeV/c)$^2$.

%
%
%
\begin{figure}[htbp]
    \centering
        \includegraphics[width=0.75\textwidth,angle=270]{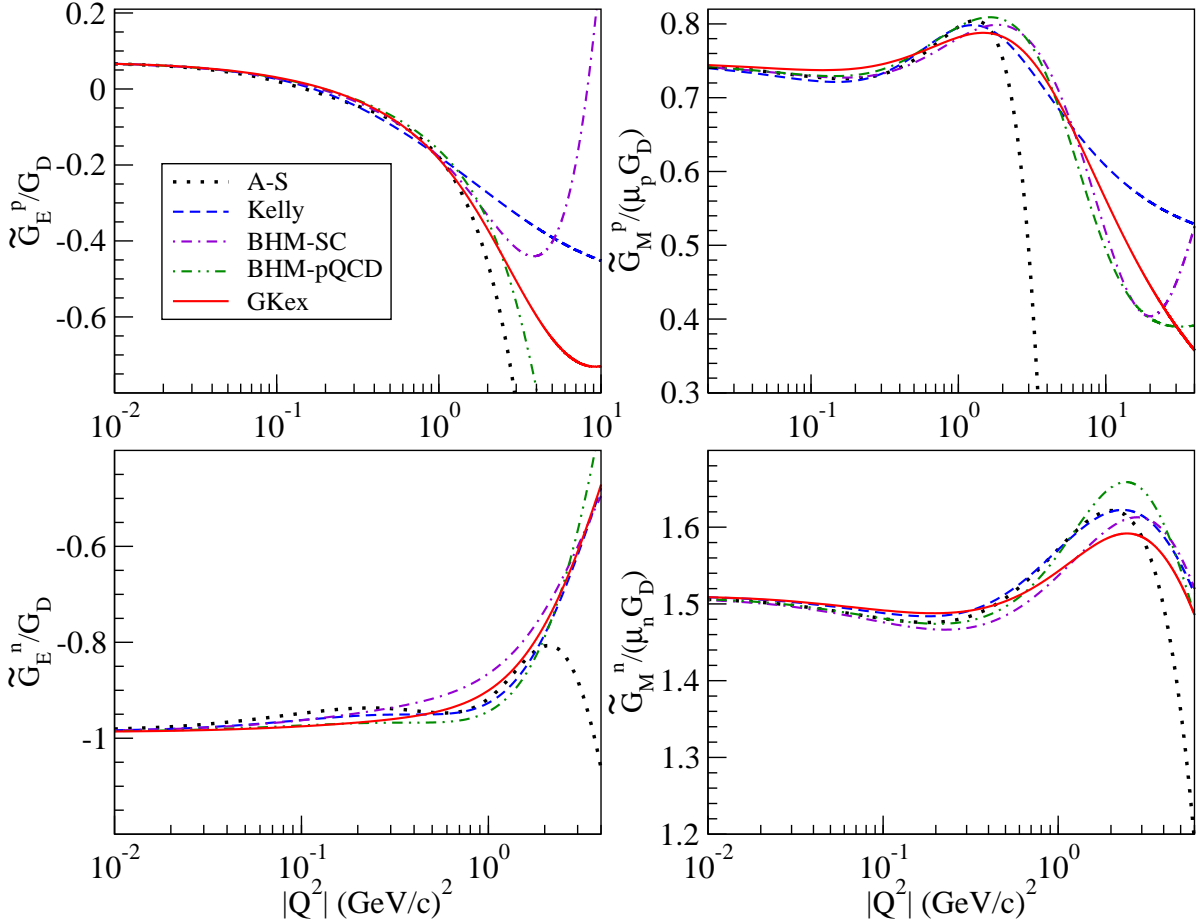}
    \caption{(Color online) Electroweak form factors obtained with the different prescriptions analyzed in this
    work. Zero strangeness has been assumed here.}
    \label{fig:ffweak}
\end{figure}

For completeness, we present in Fig.~\ref{fig:ffweak} the results
corresponding to the electroweak vector nucleon form factors,
$\widetilde{G}_{E,M}$, for the different EM descriptions considered.
In all cases strangeness content in the nucleon has not been
included, but the radiative corrections $R_V^{(a)}$ entering in the
electroweak vector coupling constants (Eqs.~(\ref{xiv1}-\ref{xiv3}))
have been incorporated assuming the general expressions
\ba
R^{T=0}_V &=& \frac{R_V^n-(1-4\sin^2\theta_W)R_V^p}{4\sin^2\theta_W}\, ,\\
R_V^{T=1} &=& \frac{(1-4\sin^2\theta_W)R_V^p +
R_V^n}{2(1-2\sin^2\theta_W)}\, \label{radiative}
\ea with
$R_V^p=-0.0520$, $R_V^n=-0.0123$ and $R_V^{(0)}=-0.0123$~\cite{Liu,PDG},
this last term not contributing to results in Fig.~\ref{fig:ffweak}
because it only enters with strangeness $G_{E,M}^{(s)}$ different
from zero.

In the present study we follow closely the arguments already
presented in~\cite{Liu} where a global analysis of experimental data
from elastic PV electron scattering at low-$Q^2$ was given.
Contributions from perturbative QCD and coherent strong interaction
effects in the radiative corrections associated with elastic nucleon
scattering have been evaluated in~\cite{Erler1,Erler2}, providing
also an improved estimate of the running of the weak mixing angle in
the $\overline{\mathrm{MS}}$ renormalization scheme. All of these
effects are included in the $R_V$-values shown above (see table~I
in~\cite{Liu}). As explained in~\cite{Liu}, the theoretical
uncertainties in $R_V^n$ and $R_V^{(0)}$ are less than $1\%$, and hence
have a negligible impact on the analysis presented in this work.
On the other hand, the theoretical error in the full expression
$(1-4\sin^2\theta_W)(1+R_V^p)$ is slightly more than $1\%$
(see~\cite{Erler2}). In this work we use the conventional
$\overline{\mathrm{MS}}$ renormalization scheme, and the weak mixing
angle, $\sin^2\theta_W$, used in the evaluation of the PV asymmetry
results that are discussed in the next section, which takes on the
value $0.23122\pm0.00015$ in accordance with the arguments presented
in~\cite{Liu,Yao}.
The use of a different $\theta_W$-value,
for instance the running $\sin^2\theta_W(0)$ given in~\cite{Erler2},
leads to differences in the PV analysis that will be considered
briefly in the next section.

As shown, results for $\widetilde{G}_{E,M}^{p,n}$ provided by the
five prescriptions are very similar for $|Q^2|\leq 1$ (GeV/c)$^2$.
This is consistent with the general behavior presented in
Fig.~\ref{fig:ffem}. On the contrary, the spread gets wider as
$|Q^2|$ increases. In general, comments already applied to the
analysis of the purely EM form factors can be also extended here,
but changing the isospin channel. Notice that the electroweak
electric and magnetic form factors for the proton (neutron) are
basically determined by the corresponding EM ones for the neutron
(proton), {\it i.e.,} $\widetilde{G}_{E,M}^{p,n}\simeq
G_{E,M}^{n,p}$.

\subsection{Axial-vector nucleon form factor: $G^e_A$}\label{sec:axial}

The neutral current axial-vector form factor for a nucleon, which is
directly linked to the
weak interaction in electron scattering, can be decomposed into its
isovector, singlet and octet axial-vector strangeness contributions,
\be
 G^{e,N}_A=\xi_A^{T=1}G_A^{(3)}\tau_3+\xi_A^{T=0}G_A^{(8)}+\xi_A^{(0)}G_A^{(s)}
\ee which may be recast into isoscalar, isovector and strangeness
contributions. In contrast to the EM and WNC vector currents, the
axial-vector current is not conserved, and hence the value of the
axial-vector nucleon form factor at $Q^2=0$ is not restricted by any
exact symmetry. However, the term $G_A^{(3)}(0)$ can be determined
from Gamow-Teller $\beta$-decay, and analogously, information on
$G_A^{(8)}$ can be obtained combining data from neutron and hyperon
$\beta$-decay measurements, assuming isospin invariance. In general,
we may write: $G_A^{(3)}(0) = (D+F)/2$ with $D+F\equiv g_A$, and
$G_A^{(8)}(0) = (3F-D)/(2\sqrt{3})$. The terms $F$ and $D$ represent
the matrix elements of the axial weak current for the states of
different hyperons belonging to the SU(3)
octet~\cite{MusDon92,Musolf1994,Alberico2002}).

Concerning the dynamical structure of $G_A^e$, {\it i.e.,}
its dependence on $Q^2$, the usual procedure is to parametrize data
making use of the standard dipole form:
$G_A^{(3,8)}(Q^2)=G_A^{(3,8)}(0)\ G_D^A (Q^2)\label{GA3}$ with
$G_D^A(q^2)=(1+|Q^2|/M_A^2)^{-2}$ and $M_A=(1.032\pm0.036)\,\,
\text{GeV}$ the axial-vector mass whose standard value comes from
the analysis of charged-current quasielastic (CCQE) neutrino
scattering processes (see \cite{Ahrens,Baker,Miller,Kitagaki}).


Some recent studies~\cite{Liu} (see also~\cite{BeckMcKeown,Musolf1994}) have addressed the importance of
including radiative corrections in the axial-vector term. An
expression for the axial-vector form factor~\cite{Liu} can be
written as follows:
\be
 G^{e,p}_A(Q^2)=-\left[g_A(1+R_A^{T=1})-\frac{3F-D}{2}R_A^{T=0}-
   \Delta s(1+R_A^{(0)})\right]\ G_D^A(Q^2)\label{GAnew} \, ,
\ee where the dependence of the axial-vector strange form factor on
$Q^2$ is also assumed to follow a dipole form. The values of the
various parameters that have been used in this work are given in
Table~\ref{fig:ctesAxial}. The ratios $R_A^{T=1}$, $R_A^{T=0}$ and
$R_A^{(0)}$ provide the effects of electroweak radiative corrections
to the isovector, isoscalar and SU(3) singlet hadronic axial vector
amplitudes, respectively. The terms $R_A^{T=1,0}$ account for {\it
one-quark} and {\it many-quark} contributions. The former correspond
to the renormalization of the effective weak couplings, $C_{2q}$,
and their values can be obtained from the SM predictions for these
couplings. The latter, {\it i.e.,} many-quark contributions, include
the {\it anapole} effects~\cite{Zhu2} as well as coherent strong
interaction contributions. Contrary to the vector corrections,
$R_V$, the importance of the many-quark effects in the $R_A$ can be
significant~\cite{Liu,PDG}. 

Studies of the anapole form factor and its potential impact on the
strangeness current have been presented in the
past~\cite{anapole1,anapole2}. In particular, the authors
in~\cite{Maekawa1,Maekawa2} evaluate the anapole form factor of the
nucleon in chiral perturbation theory to sub-leading and leading
order. The kinematic region considered is $Q\equiv\sqrt{|Q^2|}\ll
M_{QCD}$, where $M_{QCD}$ ($\sim 1$ GeV) is the typical mass scale
in QCD. In this regime a systematic expansion of the form factor in
powers of $Q/M_{QCD}$ is feasible, and the momentum dependence of
the anapole form factor was presented in~\cite{Maekawa1,Maekawa2}.
Although its inclusion in the analysis of PV electron scattering on
the nucleon should be valuable, the study presented in this work
covers a $Q^2$-range large enough where the validity of the
expansion in~\cite{Maekawa1,Maekawa2} might be questionable.
Furthermore, the uncertainty in the axial form factor, described
through the use of different values for the axial mass, could make
it difficult to isolate other residual contributions.
\begin{table}[htbp]
\centering
\begin{tabular}{c c c c}
\hline
\hline
             & $R_A^{T=1}$     & $R_A^{T=0}$       & $R_A^{(0)}$ \\
\hline
 One-quark       & $-0.172$        & $-0.253$          & $-0.551$\\
 Many-quark      & $-0.086(0.34)$  & $0.014(0.19)$     & N/A\\
 Total       & $-0.258(0.34)$  & $-0.239(0.20)$    & $-0.55(0.55)$\\
\hline
\hline
\hline
         &  Parameters     &    Values         & \\
\hline
        &  $g_A$      & $1.2695$       & \\
        & $3F-D$    & $0.58(0.12)$  & \\
        & $\Delta s$    & $-0.07(0.06)$ & \\
\hline
\hline
\end{tabular}
\caption{Values of the parameters included in Eq.~(\ref{GAnew})
(see~\cite{Liu,PDG}). The top panel includes the
one-quark~\cite{Yao} and many-quark~\cite{Zhu2} corrections to the
axial charges. In the bottom panel we give the isovector axial
form factor at zero momentum transfer, $g_A$
(see~\cite{Thomas,Thomas2,PRL99}), the SU(3) reduced matrix
elements, $3F-D$, taken from~\cite{Filippone}, and the strange quark
contribution to the nucleon spin, $\Delta s$~\cite{Adams}. }
\label{fig:ctesAxial}
\end{table}

In addition to the standard dipole form considered for the
dependence of $G^{e,p}_A$ with the transferred momentum, it is also
interesting to analyze the results when using a monopole form. This
is motivated by VMD-based analyses such as those in
\cite{Lomon2001,Lomon2002} and summarized in \cite{Crawford2010}
where one finds for the EM form factors that it is more natural to
have monopole behavior rather than dipole behavior, the latter
arising from cancelations between the contributions of the
particular vector mesons. Such cancelations cause the magnetic form
factors $G_M^p$ and $G_M^n$ to be essentially dipole-like at small
momentum transfers, but do not enter the same way for $G_E^p$ and,
in fact, the current understanding is that the electric form factors
falls faster than a standard dipole. For the axial-vector form
factor, and for the strangeness form factors discussed in the next
section, the picture may be different: the cancelations that lead to
dipole-like behavior may or may not occur and the ``standard''
assumption that a dipole form is the correct one to choose may not
be warranted. Accordingly we have investigated what might change if
monopoles rather than dipoles are assumed. In this case, the
dependence is given through the function: \be
G_M^A(Q^2)=(1+|Q^2|/\widetilde{M}^2_A)^{-1} \ee with
$\widetilde{M}_A$ being the monopole axial-vector mass introduced in
a similar way to the dipole one $M_A$.

In Sect.~\ref{sec:axialdep} we analyze in detail the effects
introduced in the PV asymmetry by the dipole versus monopole
description of the axial-vector form factor by using different
values for the axial-vector masses $M_A$ and $\widetilde{M}_A$.
Radiative corrections will also be explored in
Sect.~\ref{sec:radcorr}. Finally, concerning the strangeness content
in $G^{e,p}_A$, one finds that the PV asymmetry does not show
much sensitivity to $G_A^{(s)}$, and hence in this work all results
presented correspond to axial-vector strangeness as given in
Table~\ref{fig:ctesAxial} (see~\cite{Liu,PDG}).

\subsection{Nucleon vector form factors with strangeness: $G_{E,M}^{(s)}$}\label{sec:strange}

Our present knowledge about the strange vector nucleon form factors
is much more limited than for the form factors discussed above
although some general considerations can be made. Since the nucleon
does not present any net strangeness, the strange electric form
factor in the static limit, {\it i.e.,} $Q^2=0$, should fulfill the
constraint $G_E^{(s)}(0)=F_1^{(s)}(0)=0$. Analogously, the
strangeness magnetic moment is given by the corresponding form
factor in the limit $Q^2\rightarrow 0$: $\mu_s\equiv
F_2^{(s)}(0)=G_M^{(s)}(0)$. With regards to the functional
dependence with the transferred four-momentum, the standard
procedure is to consider the usual dipole form\footnote{In some
previous work~\cite{Musolf1994,Amaro} additional functions
$\xi_{E,M}^{(s)}$ depending on $Q^2$ were introduced; however, to
make the discussion that follows simpler here we consider only
$\xi_{E,M}^{(s)}=1$.}
\begin{eqnarray}
  G_E^{(s)}(Q^2)&=&\rho_s\tau G_D^V(Q^2)\label{GEs}\\
  G_M^{(s)}(Q^2)&=&\mu_sG_D^V(Q^2) \label{GMs}\,
\end{eqnarray}
with $G_D^V(Q^2)\equiv (1+|Q^2|/M_V^2)^{-2}$ and $M_V$ the
vector-mass parameter. The term $\mathbf{\rho_s}$  is given by the
derivative of the electric strangeness nucleon form factor with
respect to $\tau$ evaluated at $\tau=0$, {\it i.e.,}
$\mathbf{\rho_s}\equiv \left. dG_E^{(s)}/d\tau\right|_{\tau=0}$. A
detailed study of $\mathbf{\rho_s}$ and $\mathbf{\mu_s}$ and their
influence on the scattering observables will be presented in the
next section.

For completeness, following the arguments given above for the
axial-vector form factor, a functional dependence of $G_{E,M}^{(s)}$
with $Q^2$ based on a monopole form will be also explored in the
analysis of results. Thus, instead of using the function
$G_D^V(Q^2)=(1+|Q^2|/M_V^2)^{-2}$ in Eqs.~(\ref{GEs},\ref{GMs}) we
will also consider the functional dependence:
$\widetilde{G}_M^V(Q^2)=(1+|Q^2|/\widetilde{M}_V^2)^{-1}$ where
different values of the monopole vector mass $\widetilde{M}_V$ will
be considered.

\section{Parity-violating asymmetry: analysis of results}\label{sec:analysis}

In this section we present a systematic analysis of the PV asymmetry
for elastic electron-proton scattering. As already mentioned, all
results have been evaluated in the one-boson-exchange approximation.
Different kinematical regimes including backward and forward
scattering angles are considered. This study makes it possible to
compare our predictions with a large variety of data, also showing
the contribution in the asymmetry ${\cal A}^{PV}$ coming from the
separate electric, magnetic and axial-vector form factors.

The various ingredients entering the description of the process,
{\it i.e.,} the EM and WNC structure of the proton, and their
effects on ${\cal A}^{PV}$ are analyzed. In particular, the
description of the axial-vector form factor with its functional
dependence on the transferred momentum, dipole versus monopole, and
the specific value of the axial-vector mass are carefully
investigated. Finally, we discuss at length how the strangeness
content in the nucleon can modify ${\cal A}^{PV}$. This is one of
the basic objectives in the study of PV electron scattering
reactions due to the particular sensitivity shown by the PV
asymmetry to the $\overline{s}s$ content in the nucleon. In this
work we present an exhaustive analysis, isolating the contribution of
strangeness in the three channels involved in the process, electric,
magnetic and axial-vector. We compare the results obtained with all
available data, these spanning a range in $|Q^2|$ up to 1 (GeV/c)$^2$.

Before entering into a detailed discussion of the results obtained,
here we comment on effects beyond the Born approximation. In
particular, we focus on the corrections associated with two-photon
exchange (TPE), since these can be of the same order as effects from the $\gamma-Z$
interference term. We already presented some discussion on this
topic in the introduction, with particular emphasis on the
description of the EM form factors where effects due to
these higher-order corrections have been estimated.

A renewed interest in the TPE mechanism in elastic electron-proton
scattering emerged from the discrepancy between data at high-$Q^2$
for the electric and magnetic proton form factors as measured in
unpolarized (Rosenbluth separation) and polarized electron
scattering. Several studies suggested that this discrepancy could be
explained by higher-order contributions and both
theoretical and phenomenological analyses have been presented in the
literature (for a general review see~\cite{CV2007,ABM2011}). The
elastic EM form factors of the proton have been extracted using
different parameterizations that account for two-photon interference
effects. This is the case of~\cite{QAA2011} (and refs. therein)
where differences up to the order of $\sim$10\% were observed
(increasing in size as $Q^2$ goes up). These results were also
compared with phenomenological extractions and direct calculations,
showing similar results for several kinematics. A different approach
to TPE was presented in~\cite{KV2009}, where the focus was on the
large momentum transfer region, and the leading $2\gamma$
amplitude was evaluated in terms of the leading twist nucleon distribution
amplitudes; the claim made there is that the TPE contributions in fact go  as $Q^{-4}$. 

The implications of TPE for elastic PV electron scattering have been
analyzed in previous work~\cite{A-S,AC2005}. In~\cite{AC2005}
higher-order corrections are obtained within the framework of the
parton model, making use of generalized parton distributions.
TPE are shown to lead to a correction that depends on both $Q^2$ and
$\varepsilon$ (see Eq.~(\ref{scatt_angle})) reaching an increase in
the PV asymmetry of the order of $\sim$1\% compared with its Born
value. This effect results in about the same percentage decrease in
the magnitude of the weak magnetic proton form factor.

A systematic study of the EM nucleon form factors including effects
coming from TPE corrections was presented in~\cite{A-S}. This
analysis is extended up to $|Q^2|\sim 1.2$ (GeV/c)$^2$ and
incorporates the effect of the two-photon box diagrams, but not the
effect of the $\gamma-Z$ box that will be commented upon later. The
authors in~\cite{A-S} provide fits to the form factors accounting
for TPE contributions, concluding that such higher-order effects in
the PV asymmetry, due to cancelation between different terms, are
less than $1\%$ for $|Q^2|<1$ (GeV/c)$^2$. In the next section we
show results for the PV asymmetry obtained using the general
prescription for the EM form factors provided in~\cite{A-S}.

Despite the previous discussion, it is important to point out
that the role played by TPE is not yet a settled issue and three
experiments~\cite{TPEJLab,TPEOLYMPUS,Novosibirsk} are aimed at
gaining some insight by studying (PC) electron-proton versus
positron-proton scattering. Some initial results from Novosibirsk
indicate that the box and cross-box (hard) contributions may be
smaller than expected. Hence, some caution should be expressed on
how precisely we are presently able to assess the impact of the TPE effects.

To conclude these brief discussions, let us point out that isospin
breaking effects have recently been studied for PV electron
scattering on nucleons~\cite{Kubis2006} and
$^4$He~\cite{Viviani2009} (see also \cite{DonDubSick,Rama94} for discussions of isospin mixing in PC electron scattering).
The authors in~\cite{Kubis2006,Viviani2009} find isospin
violations to be smaller than strangeness uncertainties. However,
these corrections may play a role in the case of global data
analyses, particularly at increasing $|Q^2|$-values. As already
mentioned, in this work we assume isospin symmetry; however, it should be noted that the strange
form factors have sufficiently strong dependencies on the
transferred four-momentum that, within their uncertainties,  they
cover the potential impact of isospin breaking at the
kinematics of interest in this work.

\subsection{Dependence on EM nucleon structure}\label{sec:EMdep}
%
\begin{figure}[htbp]
    \centering
        \includegraphics[width=.7\textwidth,angle=270]{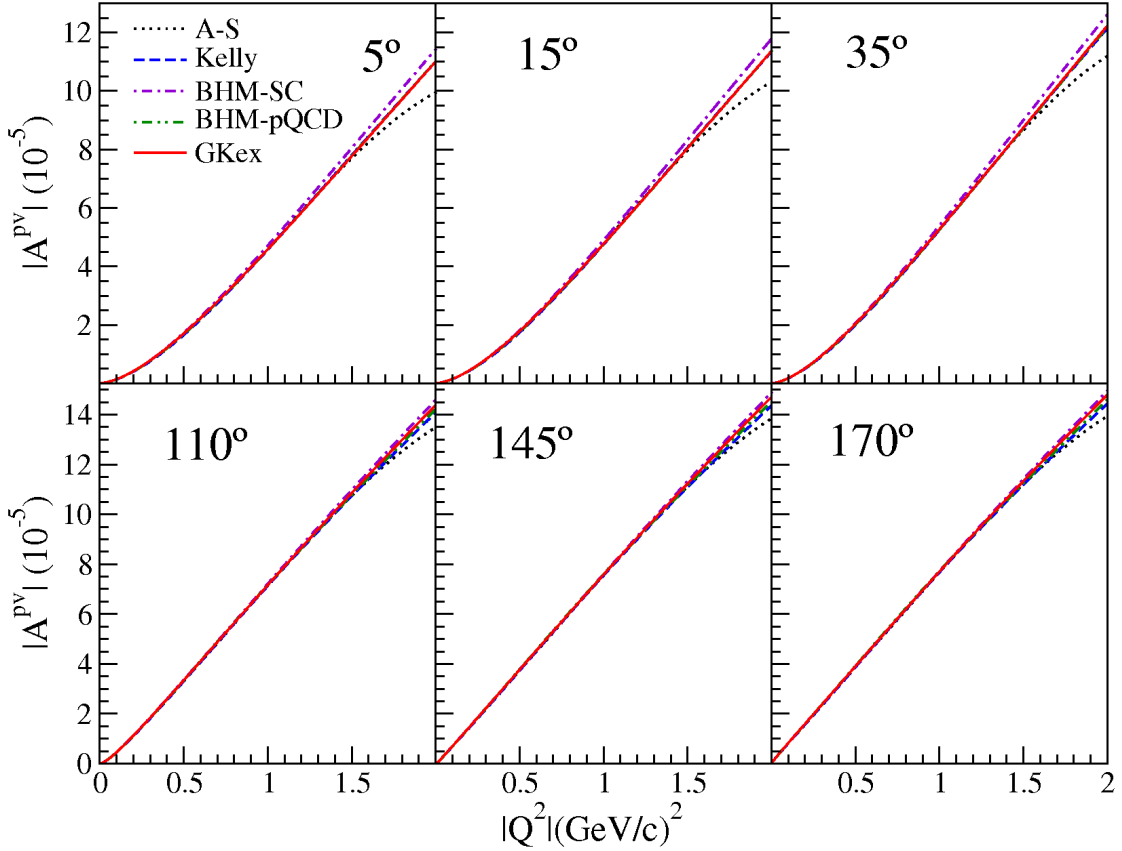}
    \caption{(Color online) Absolute value of the PV asymmetry as a function of $|Q^2|$. Results
          are presented for six scattering angles and the five prescriptions considered in the
          previous section for the EM nucleon form factors. Radiative corrections according to
          Eq.~(\ref{radiative}) have been included, but nucleon strangeness is neglected here.}
    \label{fig:AvsFF}
\end{figure}
Here our aim is to analyze the sensitivity of the PV asymmetry to
the particular description chosen for the EM form factors. Results
are shown in Fig.~\ref{fig:AvsFF} where we present $|{\cal A}^{PV}|$
versus $|Q^2|$ for six values of the scattering angle that range
from very forward ($\theta_e=5^o$) to very backward (170$^o$)
kinematics. These include the regimes where different experiments
have been performed (see discussion in next sections). For
simplicity, all results in Fig.~\ref{fig:AvsFF} have been evaluated
assuming no strangeness in the nucleon and the five prescriptions
for the EM nucleon form factors presented in Fig.~\ref{fig:ffem}
have been considered, namely, A-S (dotted line), Kelly (dashed),
BHM-SC (dot-dashed), BHM-pQCD (double-dot-dashed) and GKex (solid).
Only results based on the A-S model clearly depart from the others
for $|Q^2|$-values above $1.5$ (GeV/c)$^2$. For $|Q^2|$ above 2
(GeV/c)$^2$ (not shown in the graph) this discrepancy gets much
larger, {\it i.e.,} consistent with the behavior shown by
$G_{E,M}^{p,n}$ in Fig.~\ref{fig:ffem} (likewise for the electroweak
form factors in Fig.~\ref{fig:ffweak}). As noted earlier, A-S only applies
to $|Q^2|\leq 1$ (GeV/c)$^2$.

Concerning the four remaining prescriptions, Kelly, GKex, BHM-SC and
BHM-pQCD, they yield very similar results for all transferred
momentum values. The discrepancy is at most of the order of
$\sim$3--4$\%$ at the limit $|Q^2|=2$ (GeV/c)$^2$ and is very
similar for all scattering angles. At $|Q^2|=1$ (GeV/c)$^2$ (the limit
in the experimental data for the asymmetry) the dispersion between
the four prescriptions is about $\sim$3$\%$ in the very forward case
($\theta_e=5^o$), and gets much smaller for larger angles. In the
most backward-angle kinematics, {\it i.e.,} $\theta_e=170^o$, the
dispersion is less than $\sim$0.7$\%$. These differences are even
smaller for decreasing transferred momenta.

%
\begin{figure}[htbp]
    \centering
        \includegraphics[width=.7\textwidth,angle=270]{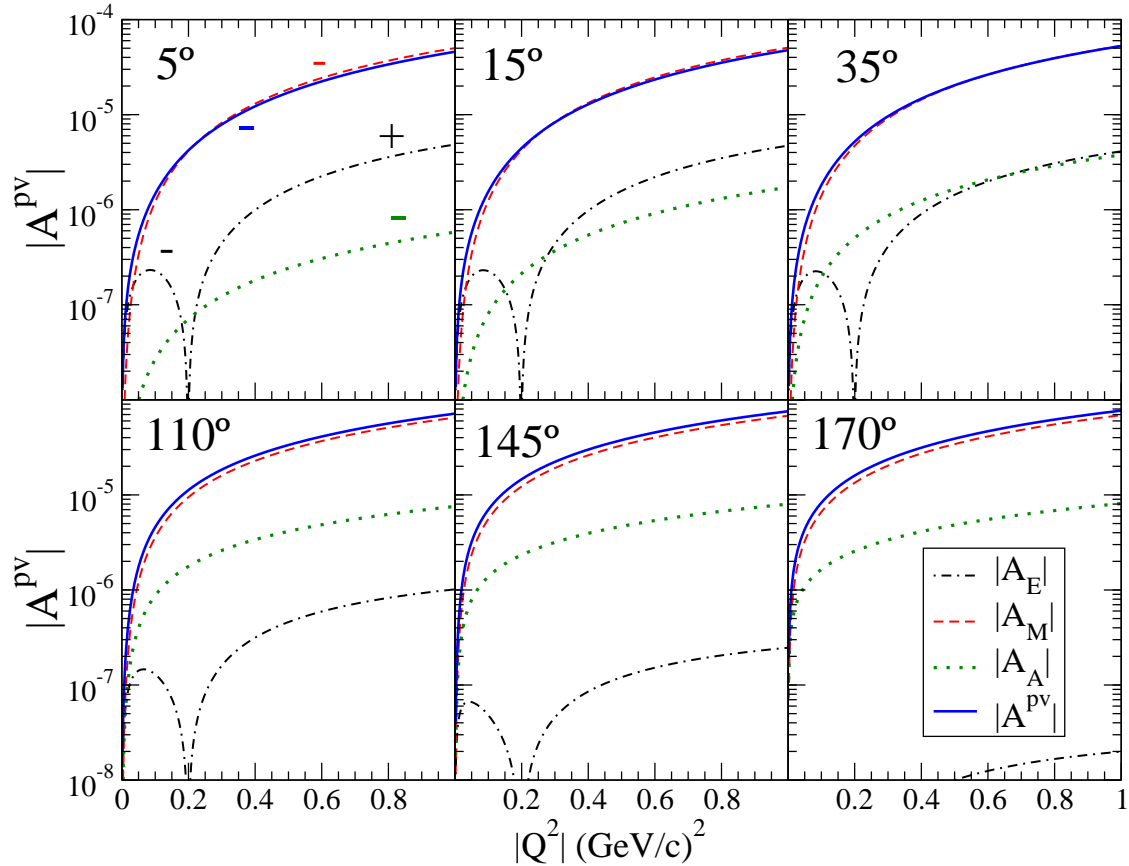}
    \caption{(Color online) Separate contributions in the PV asymmetry coming from the electric
    (black dashed-dotted line),
          magnetic (red dashed line) and axial-vector (green dotted line) distributions
          (see Eq.~(\ref{A-componentes})).
          The full asymmetry
          (blue solid line) is also shown for reference. Results correspond to the GKex
          prescription. As in the previous figure, radiative corrections are included, but
          strangeness in the nucleon is neglected.}
    \label{fig:Apv-componentes}
\end{figure}

In order to get some insight into the sensitivity of ${\cal A}^{PV}$
on the EM nucleon structure for forward and backward scattering
reactions, we isolate the contributions given by the electric,
magnetic and axial-vector distributions. These are shown in
Fig.~\ref{fig:Apv-componentes} where the three terms ${\cal
A}_{E,M,A}$ (see Eq.~(\ref{A-componentes})) are presented as
absolute values. The full asymmetry (blue solid line) is also shown
for reference. The symbol $+(-)$ indicates the positive (negative)
character of the corresponding response in the $|Q^2|$ region
selected. Note the absolute dominance of the magnetic contribution
${\cal A}_M$ (negative for all $Q^2$) in all kinematical situations,
giving rise to the full response ${\cal A}^{PV}$. Concerning the
electric and axial-vector terms, typically orders of magnitude
smaller than ${\cal A}_M$, the relative predominance of one over the
other depends on the specific kinematics. For very forward angles
${\cal A}_E$ is larger, while the reverse holds in the
backward-angle case. In fact, ${\cal A}_E$ and ${\cal A}_A$ are
similar on average for $\theta_e=35^o$.

These results can be explained easily using the general expressions
given in Eqs.~(\ref{cont_e},\ref{cont_m},\ref{cont_a}). In the limit
of very forward-angle scattering, $\theta_e\rightarrow 0^o$, we have
$\varepsilon\rightarrow 1$. Thus, the axial-vector contribution
${\cal A}_A$ approaches zero, and the ratio between the magnetic and
electric distributions simplifies to
\be
\frac{{\cal A}_M}{{\cal
A}_E}\longrightarrow \frac{\tau G_M^N\widetilde{G}_M^N}
     {G_E^N\widetilde{G}_E^N} \, .
\ee
In the backward-angle limiting case, {\it i.e.,}
$\theta_e\rightarrow 180^o$, the factor $\varepsilon\rightarrow 0$.
Hence, the electric term ${\cal A}_E$ does not enter and the two
remaining contributions are connected through
\be
\frac{{\cal
A}_M}{{\cal A}_A}\longrightarrow
\sqrt{\frac{\tau}{1+\tau}}\frac{a_A}{a_V}
      \frac{\widetilde{G}_M^N}{G_A^{e,N}}  \, .
\ee

This general discussion in the limit situations,
$\theta_e\rightarrow 0^o, \, 180^o$, is consistent with the results
presented in Fig.~\ref{fig:Apv-componentes}. For backward angles
(three bottom panels), the larger $\theta_e$ is the smaller the
contribution from ${\cal A}_E$, being several orders of magnitude
smaller than ${\cal A}_{M,A}$. Therefore, the PV asymmetry for
backward-angle kinematics is entirely determined by the magnetic and
axial-vector distributions.

In the case of forward-angle scattering, the electric term dominates
over the axial-vector one for very small angles ($\theta_e=5^o$) and
the two become similar at $\theta_e\sim 30^o$. Note the specific
signs of ${\cal A}_{E,A}$: whereas the axial-vector term is negative
for all $Q^2$, the electric one changes sign, being negative
(positive) for the smaller (higher) values of $|Q^2|$. This means
that ${\cal A}_E+{\cal A}_A$ may get canceled for some transferred
momenta at specific values of $\theta_e$. Note, however, that the
inclusion of nucleon strangeness in the analysis may introduce
significant deviations from these results (see below).

\begin{figure}[htbp]
    \centering
        \includegraphics[width=.7\textwidth,angle=270]{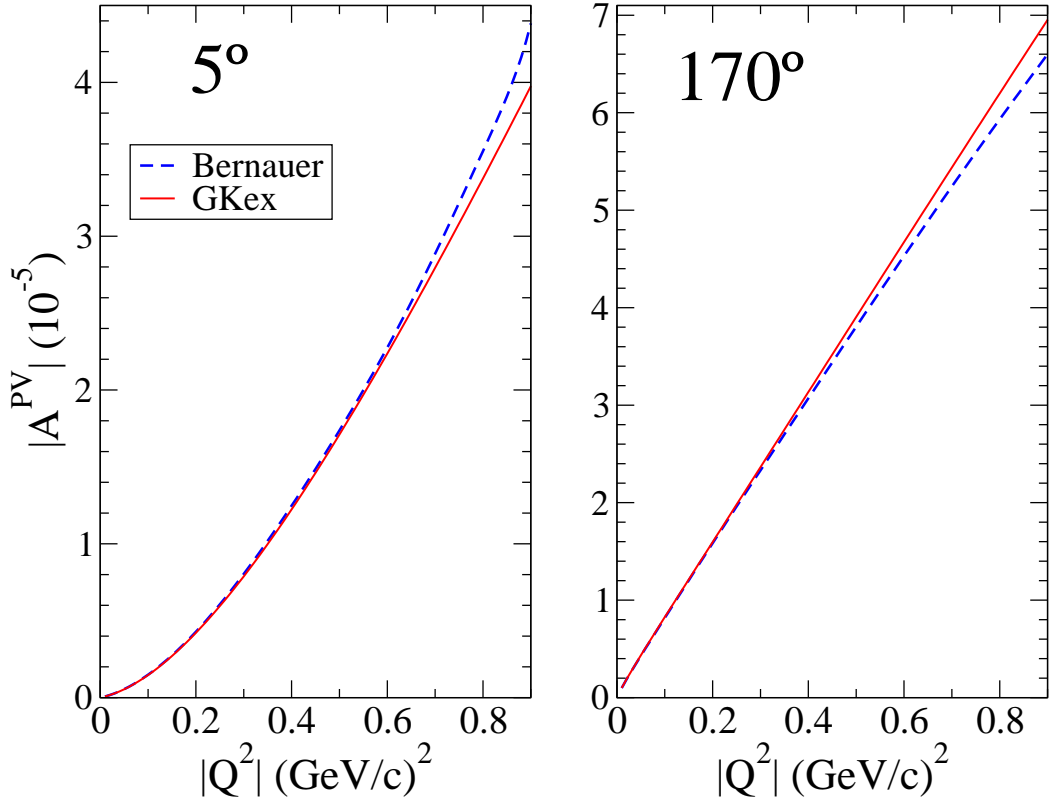}
    \caption{(Color online) PV asymmetry computed using the fit of Bernauer {\it et al.}
and the GKex description.}
    \label{fig:PVwithbernauer}
\end{figure}

Returning to
the discussion in Sect.~\ref{sec:EM} where differences in EM form
factors are seen for different experiments, we now briefly explore
the consequences of these on the PV asymmetry. In
particular, we compare results using the GKex form factors with
those obtained when $G_E^p$ and $G_M^p$ are replaced with the
results found by Bernauer {\it et al.} \cite{Bernauer} (see
Fig.~\ref{fig:bernauer}). The latter extend up to $|Q^2|\cong 0.9$ (GeV/c)$^2$.
The comparisons are shown in Fig.~\ref{fig:PVwithbernauer}
and one sees that for $\theta_e = 5^o$ the predicted asymmetries
differ by 1.6\% at $|Q^2|=0.2$ (GeV/c)$^2$ up to 5\% at $|Q^2|=0.8$
(GeV/c)$^2$, while for $\theta_e = 170^o$ they differ by 0.7\% at
$|Q^2|=0.2$ (GeV/c)$^2$ up to 4.3\% at $|Q^2|=0.8$ (GeV/c)$^2$. Note
that the magnitude of ${\cal A}^{PV}$ is larger for the $5^o$ case
when using the Bernauer {\it et al.} form factors than when using
the GKex form factors, but the opposite for the $170^o$ case.
The impact that these discrepancies may have on the analysis of the
PV asymmetry and its connection with the strangeness content
of the nucleon will be discussed in next sections.

\subsection{Dependence on axial-vector nucleon structure}\label{sec:axialdep}
%
\begin{figure}[htbp]
    \centering
        \includegraphics[width=.7\textwidth,angle=270]{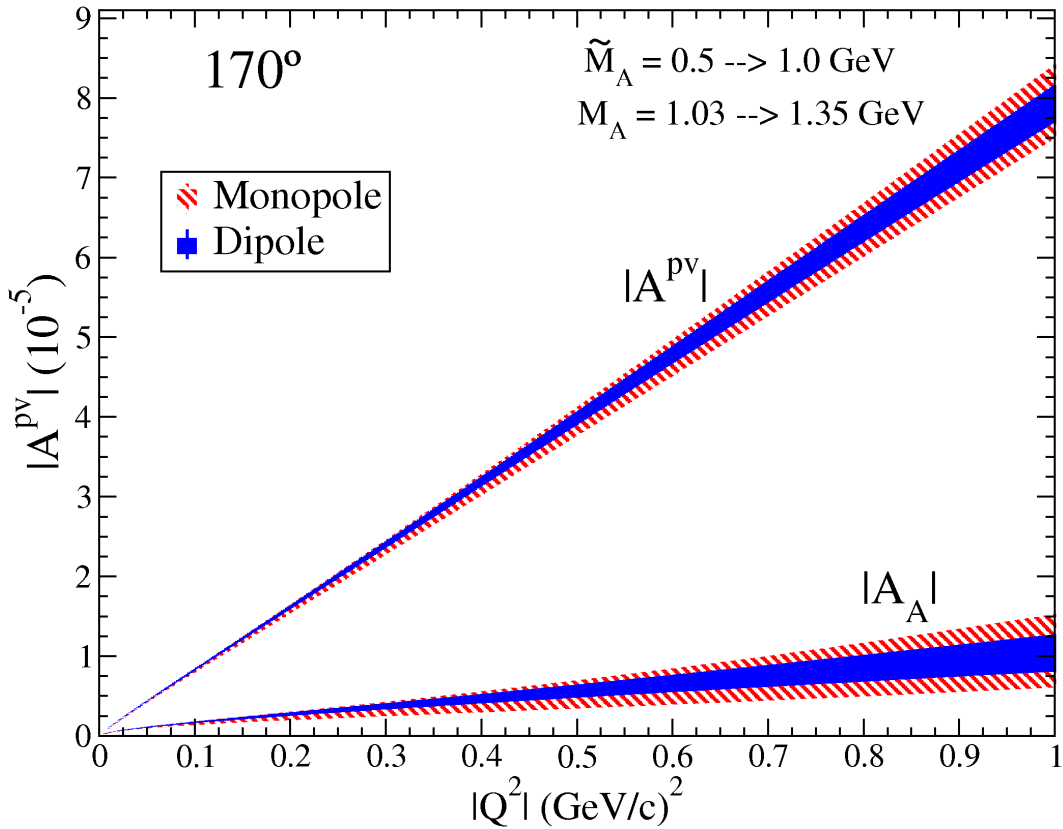}
    \caption{(Color online) Dependence of the PV asymmetry on the description of the
          nucleon axial-vector form factor.}
    \label{fig:Axial-dependence}
\end{figure}

In the previous section we already introduced the basic ideas and
expressions connected with the axial-vector nucleon form factor.
Here we analyze how the PV asymmetry depends on the specific
description of $G_A^e$. Results are presented in
Fig.~\ref{fig:Axial-dependence}, where the GKex model has been used
and no strangeness in the electric and magnetic sectors has been
considered. In contrast, radiative corrections are included and the
axial-vector strangeness content has been taken to be that in
Table~\ref{fig:ctesAxial}. Note that ${\cal A}^{PV}$ is very
insensitive to $G_A^{(s)}$. Our interest in this section is focused
on the role played by the specific functional dependence of $G_A$ on
the transferred momentum $Q^2$ and its effect on the asymmetry.
Within this context, we have evaluated the results corresponding to
the standard dipole form, $G^A_D(Q^2)=(1+|Q^2|/M_A^2)^{-2}$, as well
as to the monopole one,
$G^A_M(Q^2)=(1+|Q^2|/\widetilde{M}_A^2)^{-1}$.

Assuming either the dipole and monopole functional dependence, in
Fig.~\ref{fig:Axial-dependence} we present results from an
exploration of the effects introduced by using different values for
the axial-vector masses, $M_A$ (dipole) and $\widetilde{M}_A$
(monopole). The average value for $M_A$ taken from charged-current
neutrino-nucleus scattering reactions is $M_A=1.032\pm0.036$ GeV
(see~\cite{Ahrens}). However, recent measurements of quasielastic
neutrino -$^{12}$C cross sections obtained by MiniBooNE
collaboration~\cite{MiniBooNE} have shown an important discrepancy
with several model predictions unless a considerably larger value of
the axial-vector mass $M_A$ is used. Although no definitive
conclusions are yet in hand, and the MiniBooNE analysis may be taken
more as an indication of incompleteness of the theoretical
descriptions of the data, rather than as a strict indication for a
larger axial-vector mass, a detailed study of modeling versus
experiment for elastic PV electron-proton scattering may be able to
shed light on the reliability of the different values of the
axial-vector mass assuming not only the dipole shape but also the
monopole one.

In Fig.~\ref{fig:Axial-dependence} we show the absolute value of the
PV asymmetry and the separate axial-vector contribution $|{\cal
A}_A|$. The scattering angle is $\theta_e=170^o$, that is,
backward-angle scattering where the contribution of the axial-vector
term is maximized. We represent as shadowed areas the range of
values obtained for the asymmetry when different axial-vector masses
are considered for both functional dependencies, dipole (blue area)
and monopole (red). In the case of the standard dipole shape given
through $G_D^A(Q^2)$, the values of the axial-vector mass selected
are $M_A=1.032$ GeV (standard value) that corresponds to the lower
limit and $M_A=1.35$ GeV (upper limit in the region) which is
representative of what would bring modeling of $\nu - ^{12}$C cross
sections into agreement with data. Likewise, assuming a monopole
dependence $G_M^A(Q^2)$ (red region), the value
$\widetilde{M}_A=0.5$ GeV corresponds to the lower limit and
$\widetilde{M}_A=1.0$ GeV to the upper one. The strategy here is to
explore the differences that might occur when a monopole form is
employed rather than the conventional dipole form, making sure that
a similar range of values is being considered. Choosing this
(relatively generous) range for $\widetilde{M}_A$ assures that the
central values being parametrized are similar for the two choices of
parametrization, as seen in the figure, and that one could obtain
agreement with the neutrino experiments (given that the nuclear
physics issues there are being dealt with appropriately) with either
choice. Note that (by choice) the shadowed regions overlap. In fact,
a larger axial-vector mass $M_A$ in the dipole form leads to an
asymmetry in accord with results corresponding to a smaller monopole
axial-vector mass $\widetilde{M}_A$. Ultimately, results for the PV
asymmetry and the different descriptions of $G_A$, {\it i.e.,}
dipole versus monopole and values of the axial-vector mass should be
consistent with the analysis of neutrino scattering reactions, given
that nuclear physics issues in the latter can be resolved.

To conclude this analysis, the significant effects introduced in the
axial-vector contribution ${\cal A}_A$ by the particular description
of the axial-vector form factor lead to variations in the PV
asymmetry of the order of $\sim$5--6$\%$ for dipole and
$\sim$10--11$\%$ for monopole descriptions at $|Q^2|=1$ (GeV/c)$^2$.
In both cases they are significantly larger than the ones attached
to the particular description of the EM nucleon form factors,
namely, about $\sim$0.7$\%$ for $\theta_e=170^o$. However, these
variations are similar to what is found when we compare the fit of
Bernauer {\it et al.}~\cite{Bernauer} (see previous section) with
the GKex or any other description, namely,  $\sim$4.3$\%$ for
$\theta_e=170^o$ at $|Q^2|=0.8$ (GeV/c)$^2$. This means that for any
smaller value of the scattering angle the uncertainty associated
with using the Bernauer {\it et al.} fit versus any other EM
description is larger than the uncertainty linked to the particular
$G^e_A$ description. Again, recall that the relative contribution
of the axial-vector term diminishes with decreasing values of the
scattering angle.

\subsection{Dependence on nucleon strangeness} \label{sec:strangedep}

In this section we study the dependence of ${\cal A}^{PV}$ on the
strangeness content in the proton. We restrict our attention to the
electric and magnetic sectors, as the axial-vector strangeness does
not significantly modify the asymmetry. The functional dependence of
$G_{E,M}^{(s)}$ on $Q^2$ is taken as the standard dipole shape in
Eqs.~(\ref{GEs},\ref{GMs}), as well as through a monopole function
that is determined by a monopole vector mass (see Sect.~\ref{sec:strange} for
details). The static values and the parameters that characterize the
electric and magnetic strange form factors, namely, $\mu_s$ and
$\rho_s$, can be determined in principle from the analysis of data
and their comparison with theory. Hence, in what follows we present
a systematic study of the PV asymmetry with regards to the nucleon
strangeness content in the electric and magnetic channels. Our
theoretical predictions are compared with data taken at very
different kinematics.

In order to illustrate which are the most favorable kinematical
situations with regard to effects introduced by $\overline{s}s$ in
the electric and magnetic distributions, we rewrite the PV asymmetry
isolating the strangeness contributions in $\widetilde{G}_{E,M}$.
Introducing the explicit expressions for the coupling constants (at
tree level) and the electroweak form factors in
Eqs.~(\ref{WGEMpn},\ref{WGApn}), we can write:
\ba
& & \frac{2{\cal A}^{PV}}{{\cal A}_0} = -(1-4\sin^2\theta_W) \nonumber \\
&+& \frac{\varepsilon G_E^p\left(G_E^n+G_E^{(s)}\right)+\tau G_M^p\left(G_M^n+G_M^{(s)}\right)
+ \sqrt{\tau(1+\tau)(1-\varepsilon^2)} (1-4\sin^2\theta_W) G_M^p G_A^{e,p}}
{\varepsilon\left(G_E^p\right)^2+\tau\left(G_M^p\right)^2} \label{EMss} \, . \nonumber \\
\ea
In the backward-angle scattering limit, $\theta_e\rightarrow
180^o \Longrightarrow \varepsilon\rightarrow 0$, and hence the term
$G_E^{(s)}$ does not enter. In this situation the following
asymmetry results:
\be
\left[\frac{2{\cal A}^{PV}}{{\cal
A}_0}\right]_{\theta_e=180} =
-(1-4\sin^2\theta_W)\left[1 - \sqrt{1+\frac{1}{\tau}}\frac{G_A^{e,p}}{G_M^p}
\right] +\frac{G_M^n+G_M^{(s)}}{G_M^p} \, ,\label{backlimit}
\ee
where the entire
dependence on strangeness is given through the magnetic term
$G_M^{(s)}$ (as already discussed, strangeness in the axial-vector
form factor does not significantly modify ${\cal A}^{PV}$, and hence
its contribution does not alter the general discussion that
follows). From the above expressions it is clear that the basic
objective in ${\cal A}^{PV}$ measurements at backward-angle
scattering is the determination of the strange magnetic form factor.
This corresponds to the SAMPLE experiment performed at
MIT-Bates~\cite{SAMPLE2}. However, note that a precise determination
of $G_M^{(s)}$ requires good knowledge of the axial-vector form
factor. In some previous
work~\cite{BeckMcKeown,MusDon92,Donnelly1992,Thomas,Thomas2} the
correlation between the isovector axial-vector form factor
$G_A^{(1)}$ and the strange magnetic one $G_M^{(s)}$ has been
explored.

In the limiting forward-angle scattering kinematics $\theta_e
\rightarrow 0^o$, we simply have $\varepsilon\rightarrow 1$; thus,
the axial-vector contribution is zero and the PV asymmetry (at tree
level) reduces to \be \left[\frac{2{\cal A}^{PV}}{{\cal
A}_0}\right]_{\theta_e =0} =
-(1-4\sin^2\theta_W)+\frac{G_E^p\left(G_E^n+G_E^{(s)}\right)+\tau
G_M^p\left(G_M^n+G_M^{(s)}\right)}
{\left(G_E^p\right)^2+\tau\left(G_M^p\right)^2} \label{aforward} \,
, \ee which depends not only on the magnetic strange content, but
also on electric strangeness.

In addition to the TPE contributions linked to the
$\gamma\gamma$-box (commented on at the beginning of Section~IV) that
are believed to be responsible for the discrepancy between the
Rosenbluth and polarization transfer data~\cite{A-S,QAA2011}, one
should also consider the role played by the $\gamma Z$-box. A
detailed study on this topic, including also the $WW$ and $ZZ$-box
diagrams, has recently been presented in~\cite{GH2009,SBMT2010,GHM2011}. In those
works the $\gamma Z$-box contribution is examined making use of
dispersion relations and it is found to have a significant energy
dependence. Although this could have some impact on the $Q^2$-dependence and
strangeness extraction, some caution should be
exercised when drawing conclusions that are too strong from the results obtained for the strangeness
content of the nucleon, including its specific $Q^2$-dependence. 

From the analysis of the experimental data taken at $|Q^2|=0.1$
(GeV/c)$^2$ for different targets, H$_2$, D$_2$, $^4$He, and
considering forward and backward scattering angles, a correlation
diagram in the $G_E^{(s)}-G_M^{(s)}$ plane was
obtained~\cite{Lhuillier}. This compiles all data corresponding to
SAMPLE, HAPPEX, G0 and PVA4 and provides ellipses showing the $68\%$
and $95\%$ confidence level constraints on the vector strange form
factors. It is important to point out that these results only apply
to $|Q^2|=0.1$ (GeV/c)$^2$ where the effects introduced by different
descriptions of the EM and axial-vector nucleon form factors are
almost negligible (see previous sections). Our interest in this work
is to present a global analysis of the asymmetry for all
$Q^2$-values used in the experiments, showing its sensitivity to the
vector strangeness content and the consistency between data and
theoretical descriptions in the whole $Q^2$-range.

Within this general framework and because of the present lack of
knowledge on the strange form factors $G_{E,M}^{(s)}$ and their
specific dependence with $Q^2$, our main task in what follows is to
evaluate ${\cal A}^{PV}$ for different values of the magnetic and
electric strangeness parameters, $\rho_s$ and $\mu_s$, assuming not
only the standard dipole ($G^V_D(Q^2)$), but also the monopole
($G_M^V(Q^2)$) functional dependence. The values selected for the
vector masses are $M_V=0.84$ GeV, the standard value for EM form
factors, and $\widetilde{M}_V=1.02$ GeV which corresponds to using
the mass of the $\phi$ meson to explore what happens when a higher
mass is employed. The values chosen for $\widetilde{M}_A$ correspond
to using a range that yields results similar to what occurs when
using a dipole axial-vector form factor. Again, the objective of
this study is to see what happens to the PV asymmetry when the
functional $Q^2$ dependence is changed (see the motivations
presented in the previous section.)

In the figures that follow we show the evolution of
${\cal A}^{PV}$ with $|Q^2|$ for several scattering angles and
different strengths for the strangeness content. 
%
\begin{figure}[htbp]
    \centering
        \includegraphics[width=.75\textwidth,angle=270]{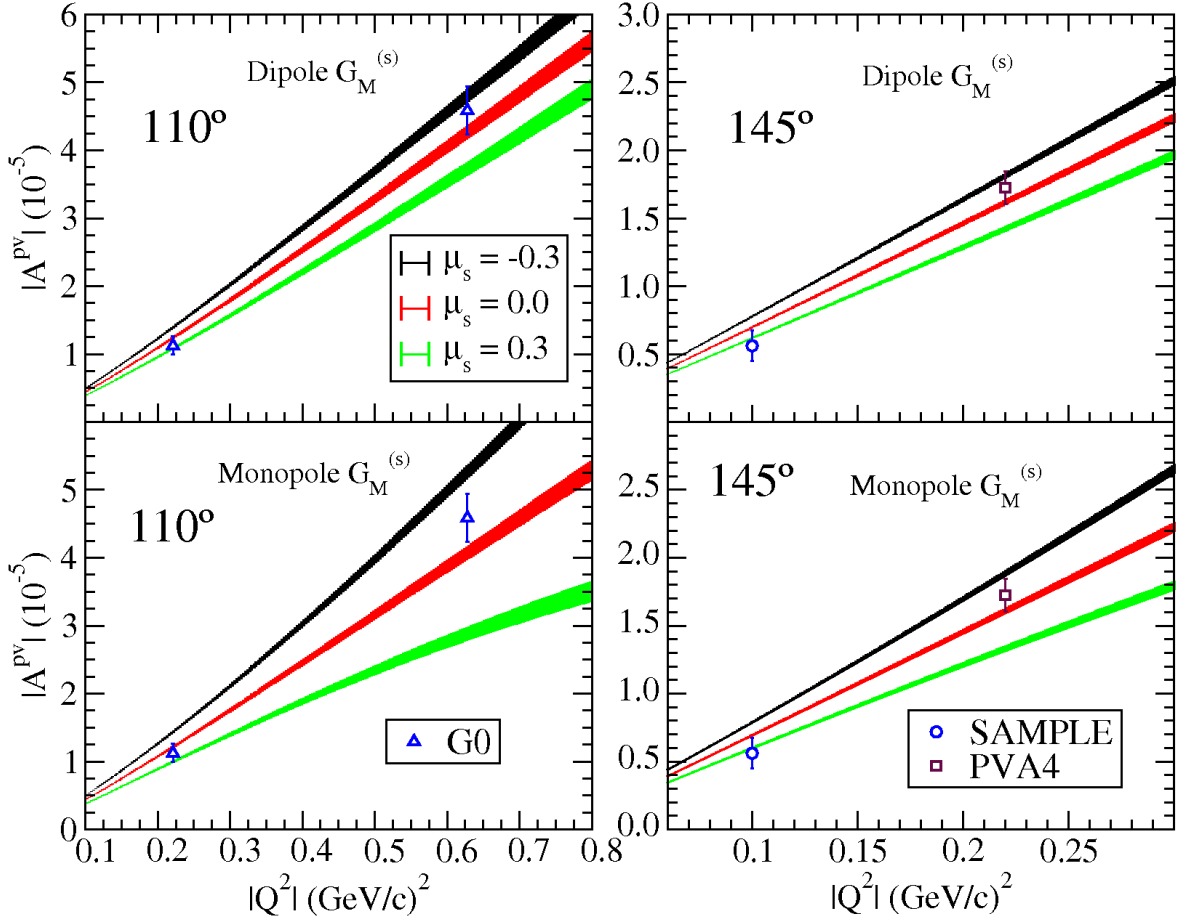}
    \caption{(Color online) PV asymmetry for backward scattering angles:
          $\theta_e=110^o$ (left panels) and $\theta_e=145^o$ (right panels).
      Results in upper panels refer to the use of the standard dipole shape in $G_M^{(s)}$,
          whereas the monopole form has been considered in the lower panels. Three different
          values for the magnetic strange parameter $\mu_s$ have been considered, namely,
          $-0.3$ (upper curves), 0 (middle curves) and $0.3$ (lower curves).
          In each case the shadowed area represents the uncertainty linked to the
          particular description of the WNC axial-vector form factor, specifically,
          the dipole functional dependence on $Q^2$ has been chosen and the corresponding
          values for the axial-vector mass: $M_A=1.03$ and $M_A=1.35$ GeV.
          The data are taken
          from~\cite{SAMPLE2,PVA4back,G0back}.}
    \label{fig:A-backward}
\end{figure}

All data error bars presented in this work represent the total
experimental error computed by adding in quadrature the statistical
and systematic errors. For the particular case of the G0 experiment,
the ``extra'' global systematic error has simply been added to the
experimental one obtained by adding in quadrature the statistical
and systematic errors. This follows from the general
analysis presented in~\cite{G0} where the error bars in the data include
the statistical uncertainty and statistical plus point-to-point
systematic uncertainties added in quadrature. The
global systematic uncertainty in that work comes partially from the
measurement and partially from the uncertainties in the evaluation of the
``no-vector-strange'' asymmetry (see~\cite{G0} for details). 

Results in Fig.~\ref{fig:A-backward} correspond to the asymmetry
${\cal A}^{PV}$ evaluated at backward angles, $\theta_e=110^o$ (left
panels) and $\theta_e=145^o$ (right panels). As already shown, in
this kinematical situation the contribution of $G_E^{(s)}$ can be
neglected. Hence the analysis of strangeness is focused on the
purely magnetic sector. The results in the upper panels have been
evaluated using the standard dipole shape for $G_M^{(s)}$, {\it
i.e.,} given through $G_D^V(Q^2)=(1+|Q^2|/M_V^2)^{-2}$ with
$M_V=0.84$ GeV. On the contrary, the lower panels present results
obtained with the monopole function
$G_M^V(Q^2)=(1+|Q^2|/\widetilde{M}_V^2)^{-1}$ with
$\widetilde{M}_V=1.02$ GeV. In all cases the GKex prescription
has been considered for the EM nucleon form factors and three
different values for the static strange magnetic content $\mu_s$
have been used: $-0.3$ (black), $0$ (red) and $0.3$ (green). These
values are consistent with some previous
work~\cite{Musolf1994,Amaro,Donnelly1992}.

The shadowed area for each $\mu_s$-value represents the region
spanned by the uncertainty linked to the description of the WNC
axial-vector form factor $G_A^e$ using the dipole function
as discussed in Fig.~\ref{fig:Axial-dependence}. The lower limit
corresponds to $M_A=1.032$ MeV and the upper one to $M_A=1.35$ MeV.
Note that when using the monopole shape for $G_A^e$ the
associated uncertainties increase to a few percent (see previous
section).

The comparison with available data, also provided in
Fig.~\ref{fig:A-backward}, shows some difficulties in reproducing
data from different experiments by using a single value for the
magnetic strangeness parameter $\mu_s$. This comment applies to both
backward angles no matter which particular functional dependence on
$Q^2$ is assumed for $G_M^{(s)}$. In this sense it is crucial to
point out that the shadowed regions corresponding to the three
$\mu_s$-values considered are clearly separated for all $Q^2$
(particularly true for $\theta_e=145^o$ and $\theta_e=110^o$ with
the monopole shape).

In order to clarify the above statements, let us start by discussing
the kinematics where $\theta_e=110^o$ (left panels). As observed,
the data taken from the G0 experiment at $|Q^2|\sim 0.2$ and $\sim
0.6$ (GeV/c)$^2$ are not in accord with theoretical results
evaluated with the same magnetic strangeness content. Whereas the
point at $|Q^2|\sim 0.2$ (GeV/c)$^2$ is consistent with positive and
close-to-zero $\mu_s$ (green and red regions), on the contrary data
at $|Q^2|\sim 0.6$ (GeV/c)$^2$ are fitted by results obtained with
negative or close-to-zero $\mu_s$-values. This situation is even
more transparent for $\theta_e=145^o$ (right panels) where the
dispersion linked to the axial-vector form factor is smaller due to
the lower range in $|Q^2|$ considered, and the regions for different
$\mu_s$ are widely separated. Note that the SAMPLE experiment
($|Q^2|=0.1$ (GeV/c)$^2$) agrees on average with $\mu_s=0.3$, but
PVA4 ($|Q^2|\sim 0.2$ (GeV/c)$^2$) shows consistency with negative
static strangeness. This conclusion, that applies to both functional
dependencies in $G_M^{(s)}(Q^2)$, dipole (top panel) and monopole
(bottom), should be taken with some caution due to the error bars shown by data.

In spite of the previous general discussion and the potentially
different slopes shown by theory and data as functions of $|Q^2|$,
the data error bands allow one to conclude that $\mu_s=0$ seems to
be the case where theory and data fit the best. As discussed
previously, backward-angle measurements of ${\cal A}^{PV}$ should be
considered as a means to isolate the contribution of $G_M^{(s)}$.
Results in Fig.~\ref{fig:A-backward} show the significant
sensitivity of the asymmetry to variations of the magnetic
strangeness, in fact, much more important than effects introduced by
other ingredients, the axial-vector and the EM form factors,
including the fit of Bernauer {\it et al.}. However, comparison
between theory and data taken at different $Q^2$-values (G0, SAMPLE
and PVA4) presents some discrepancies that still need further
investigation. From our general study using the entire set of
different descriptions of the EM, axial-vector and strangeness form
factors, it is not obvious that one is in a position to select a
specific $\mu_s$-value that provides a successful description of all of
the different $Q^2$-data measured. Results in
Fig.~\ref{fig:A-backward} show that the slope of $|{\cal A}^{PV}|$
increases with decreasing (larger) values of $\mu_s$ ($M_A$).


\begin{figure}[htbp]
    \centering
        \includegraphics[width=.78\textwidth,angle=270]{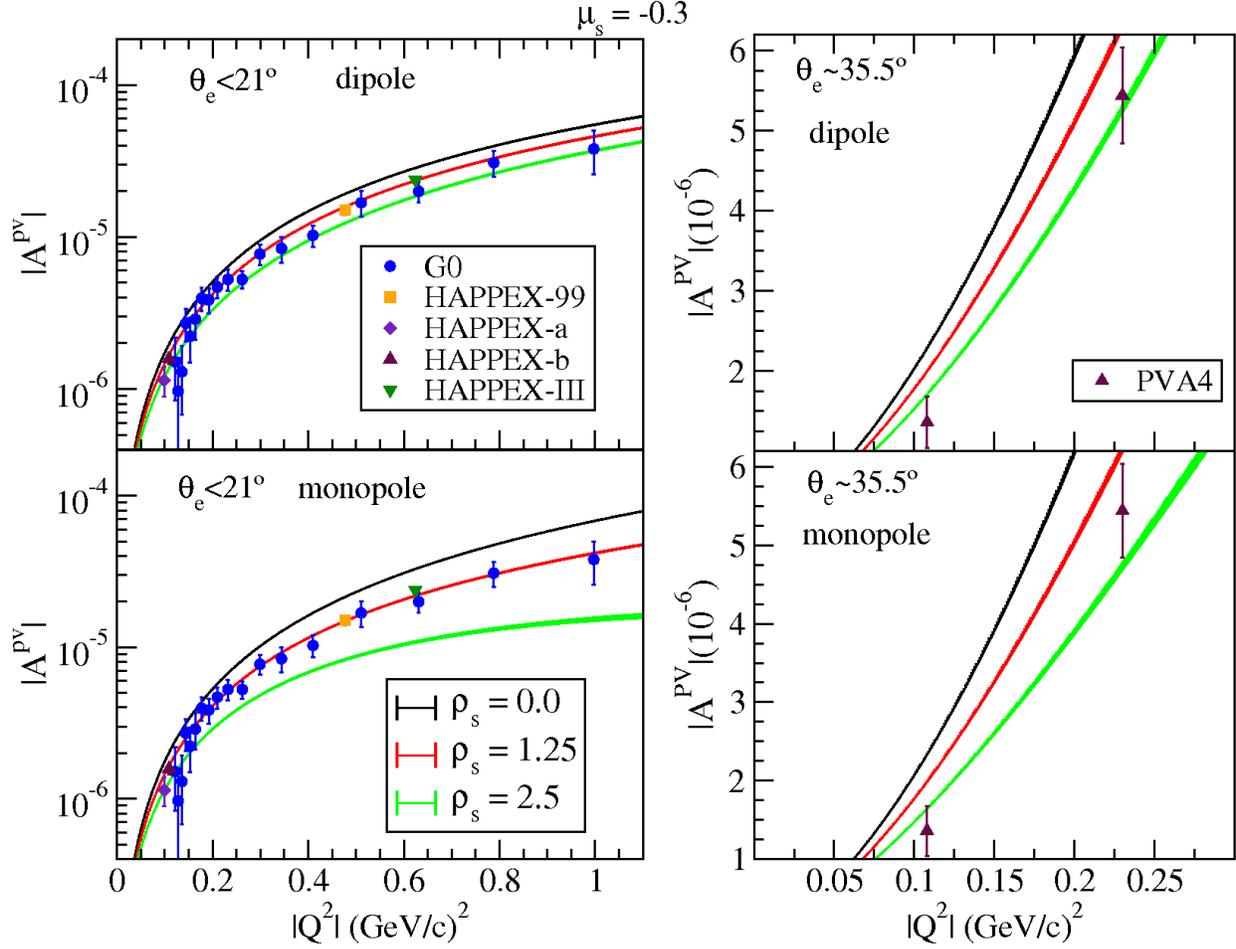}
    \caption{(Color online) PV asymmetry evaluated at forward scattering angles compared with
          experimental data. The GKex prescription for the EM nucleon form factors has been
          used. The top panels correspond to the electric and magnetic strange form factors given
          through the standard dipole shape (see text for details), and the bottom panels to a
          monopole description. The width of the various curves incorporates the total uncertainty
          linked to the WNC axial-vector form factor description (see text for details). The value of the
          static magnetic strange parameter is fixed to $\mu_s=-0.3$ and results are presented for
        three values of the electric strange content given through $\rho_s$: upper curves $\leftrightarrow$ $\rho_s=0$,
        middle curves $\leftrightarrow$ $\rho_s=1.25$ and lower curves $\leftrightarrow$ $\rho_s=2.5$. The data are taken
        from~\cite{G0,Happex-99,Happex-a,Happex-b,PVA4,PVA4-2,Happex-III}.}
    \label{fig:mus-1}
\end{figure}

\begin{figure}[htbp]
    \centering
        \includegraphics[width=.78\textwidth,angle=270]{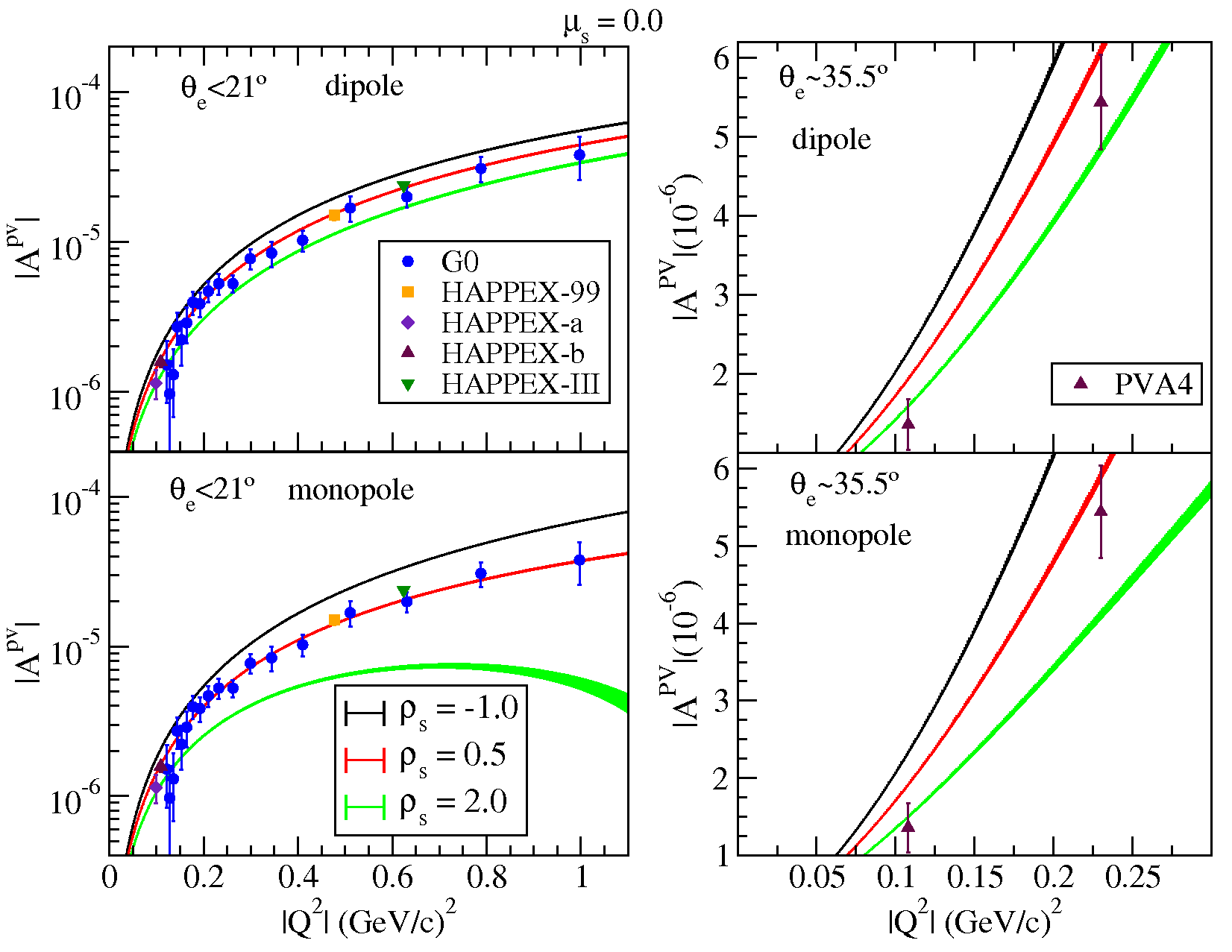}
    \caption{(Color online) As for Fig.~\ref{fig:mus-1}, except now for $\mu_s = 0.0$ and with upper
    curves $\leftrightarrow$ $\rho_s=-1.0$,
        middle curves $\leftrightarrow$ $\rho_s=0.5$ and lower curves $\leftrightarrow$ $\rho_s=2.0$.}
    \label{fig:mus-2}
\end{figure}

\begin{figure}[htbp]
    \centering
        \includegraphics[width=.78\textwidth,angle=270]{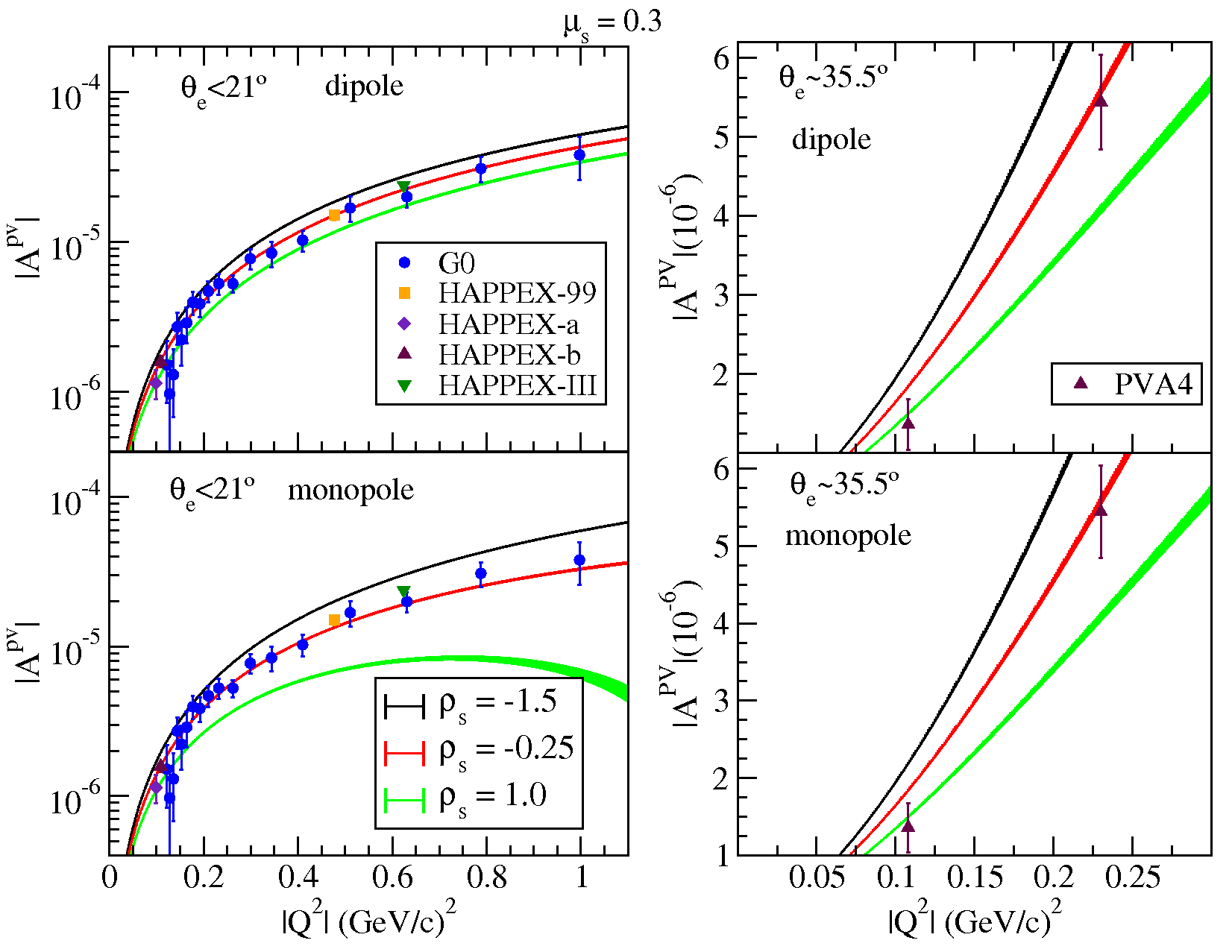}
    \caption{(Color online) As for Fig.~\ref{fig:mus-1}, except now for  $\mu_s = 0.3$ and with upper
    curves $\leftrightarrow$ $\rho_s=-1.5$,
        middle curves $\leftrightarrow$ $\rho_s=-0.25$ and lower curves $\leftrightarrow$ $\rho_s=1.0$.}
    \label{fig:mus-3}
\end{figure}

Our interest in the preceding discussion (with the results shown in
Fig.~\ref{fig:A-backward}) has been centered on a direct comparison
between data taken at different $|Q^2|$-values and theoretical
results evaluated assuming several prescriptions for the EM and WNC
nucleon form factors. Particular emphasis has been placed on the
roles played by the electric and magnetic strangeness content in the
nucleon and their impact on the PV asymmetry; not only the global
change in ${\cal A}^{PV}$ but also the specific behavior with $Q^2$
are investigated. This analysis allows one to get strong constraints
on the potential amount of strangeness contained in the
nucleon. Moreover, this study is also appropriate to dismiss
specific descriptions of the nucleon form factors in conjunction
with particular values of strangeness and/or $Q^2$-dependence. A
direct comparison between theory and data is also helpful to
determine clearly the dominant ingredients in ${\cal A}^{PV}$ for
different kinematics. This explains the analysis already presented
for backward scattering (Fig.~\ref{fig:A-backward}) and the detailed
study that follows for kinematics at forward-angles
(Figs.~\ref{fig:mus-1}--\ref{fig:mus-3}). However, in order to make
the discussion clearer, it is important to connect all of these
results with a global analysis of the world data. This is presented
in Section~IV.E, and complements the results shown in this section,
in order to help the reader to obtain a general and convincing
understanding of the ingredients and constraints connected with the
PV asymmetry and the nucleon form factors.

The analysis of forward-angle scattering kinematics is presented in
Figs.~\ref{fig:mus-1}--\ref{fig:mus-3} where, as in the previous
case, the EM nucleon form factors provided by GKex model have been
used. Each graph corresponds to results evaluated for specific
magnetic strangeness content: $\mu_s=-0.3$ (Fig.~\ref{fig:mus-1}),
$\mu_s=0$ (Fig.~\ref{fig:mus-2}) and $\mu_s=0.3$
(Fig.~\ref{fig:mus-3}). In all cases, theoretical results are
compared with data taken at different kinematics. In order to make
the discussion that follows easier, the data have been separated
into two basic categories: i) forward-angle scattering,
{\it i.e.,} $\theta_e<21^o$ (panels on the left), and ii) larger
scattering angles, $\theta_e\sim 35.5^o$ (right panels). In case i),
the extreme values of the scattering angle for data are
$\theta_e=6^o$ for HAPPEX-a and $21^o$ for G0 at $|Q^2|\cong1$ (GeV/c)$^2$.
Hence, the theoretical PV asymmetries have been computed at
$\theta_e=15^o$ since it corresponds approximately to the central
value and differences between theoretical asymmetries evaluated at
$\theta_e=15^o$ and $6^o$ and between $15^o$ and $21^o$ are small (less than 3$\%$).
The difference between top and bottom panels is linked to the functional
dependence assumed for the strangeness electric and magnetic form
factors on $|Q^2|$, namely, dipole shape (top panels) and monopole
one (bottom). In each case three values for the static electric
strangeness content $\rho_s$ have been considered. These are
indicated by the three color curves, green, red and black. The widths
of the curves take into account the uncertainty introduced by the
description of the axial-vector form factor as discussed in the
previous section. As expected, the width increases for
larger scattering angles; compare for instance
Fig.~\ref{fig:A-backward} with the right-hand panels in
Figs.~\ref{fig:mus-1}-\ref{fig:mus-3}. Notice that the relative
contribution of the WNC axial-vector form factor tends to vanish as
$\theta_e$ approaches zero (see Fig.~\ref{fig:Apv-componentes}).

As observed, the specific values selected for $\rho_s$ are different
in each figure in order to provide a region containing all data
taken at different $|Q^2|$-values. The dependence of the PV
asymmetry on the electric strangeness content $\rho_s$ is clearly
observable in all cases, particularly if the monopole shape is
assumed. Let us note how the different curves depart for increasing
$|Q^2|$-values (bottom panels). This clear separation may help in
disentangling which specific choice of electric strangeness is best
suited to the behavior of the data. This analysis should be coherent
with the study applied to backward-angle kinematics where the focus
is on the magnetic strangeness content. However, as already discussed in
that situation, the analysis of data at forward-angle scattering and
its comparison with theory also introduces some ambiguities that
need to be clarified.

Let us start with the case of smaller angles, {\it i.e.,}
$\theta_e<21^o$, and assuming the standard dipole form for the
strange nucleon form factors (left-top panels in
Figs.~\ref{fig:mus-1}-\ref{fig:mus-3}). As observed, the results for
the three $\mu_s$-values are very similar and most of the data are
located inside the region defined by the two $\rho_s$ limit values
considered in each graph. In particular, the results corresponding
to the intermediate values: $\rho_s=1.25$ (Fig.~\ref{fig:mus-1}),
 $\rho_s=0.5$ (Fig.~\ref{fig:mus-2}) and  $\rho_s=-0.25$
(Fig.~\ref{fig:mus-3}) fit nicely the behavior of the data with
$|Q^2|$. This outcome also applies to results provided with the
monopole functional dependence (left-bottom panels). However, in
this case the differences introduced by the limit $\rho_s$-values
are much larger, clearly over- or under-estimating the data for all
transferred momenta. This means that any variation in the static parameters
$\rho_s$ and/or $\mu_s$ may have a much stronger impact on the PV asymmetry
when a monopole shape for the vector strangeness is assumed.
This outcome is clearly shown in Section~IV.E where a global
analysis of world data is presented.

Results for larger scattering angles ($\theta_e\sim
35.5^o$) are presented in the right-hand panels. Assuming the
standard dipole parameterization for the strange form factors (upper
panels), we observe that the accordance between data and theory gets
better for the area between the green and red curves, that is, using
the larger values of $\rho_s$. On the contrary, the remaining value
(black curve) overestimates the data. This result holds for the
three magnetic strangeness cases considered and it introduces some
ambiguity in connection with the previous case, {\it i.e.,}
scattering angle values $\theta_e<21^o$. As shown, given a specific
$\mu_s$ and assuming the standard dipole shape for the functional
$|Q^2|$ dependence, the electric strangeness that provides the best
accord with data at $\theta_e<21^o$ on the contrary overestimates
the behavior of data at larger $\theta_e\sim 35.5^o$.

The use of a monopole shape in the form factors (right-bottom
panels) does not introduce significant differences with respect to the
previous case. This is consistent with the low-$Q^2$ region ($|Q^2|\leq 0.3$ (GeV/c)$^2$)
explored in this situation.

It is important to point out that theoretical results obtained for
$\mu_s=-0.3$ and $0.3$ are not in accord with all data measured at
different $|Q^2|$ for backward-angle kinematics (see
Fig.~\ref{fig:A-backward}). On the contrary, results evaluated for
$\mu_s=0$ (no magnetic strangeness) agree with data within the
experimental error bands, although being at the extreme. As already
stated, these general conclusions are consistent with the global
analysis presented in Section~IV.E where the PV asymmetry is
analyzed taking the characterization of the electric and magnetic
strangeness content of the nucleon as free parameters and leading to
a better appreciation of the range of values permitted by analyses
of the world data.

\subsection{Radiative Corrections}\label{sec:radcorr}

\begin{figure}[htbp]
    \centering
        \includegraphics[width=.6\textwidth,angle=270]{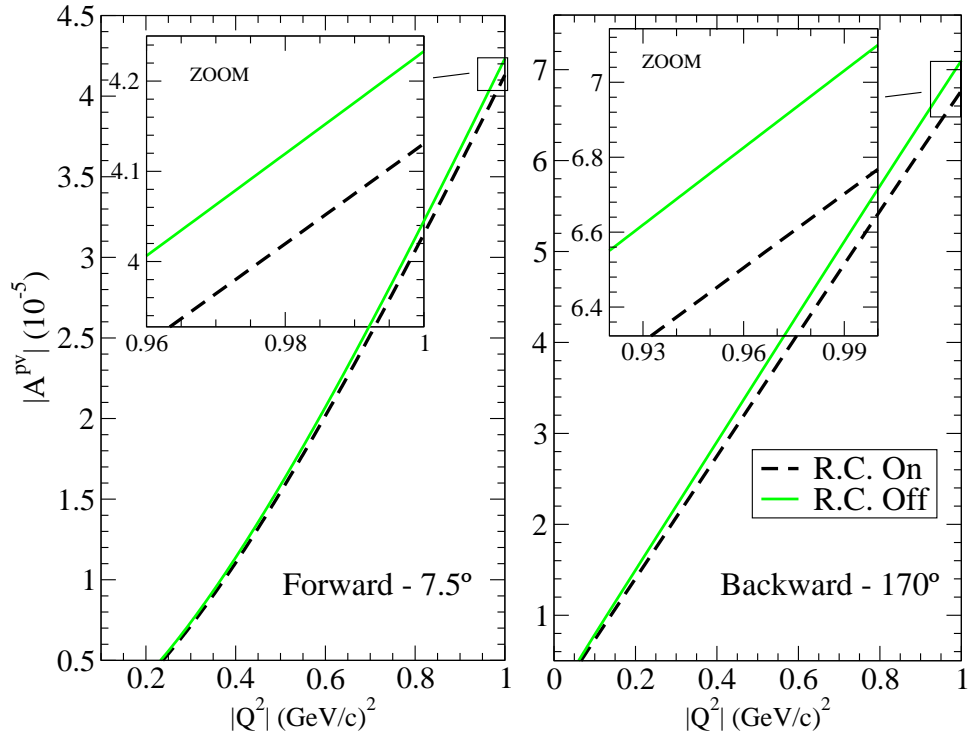}
    \caption{(Color online) PV asymmetry evaluated at forward, $\theta_e=7.5^o$ (left panel)
          and backward, $\theta_e=170^o$ (right) kinematics. In both cases, results obtained with
        the electroweak form factors evaluated at tree level (see Eqs.~(\ref{WGEMpn},~\ref{WGApn}))
        are compared with the ones
        incorporating radiative corrections as given in Eqs.~(\ref{radiative},~\ref{GAnew}).}
    \label{fig:Correc-Radiat}
\end{figure}

The effects introduced by radiative corrections in the PV asymmetry
are illustrated in Fig.~\ref{fig:Correc-Radiat}. Here the electric
and magnetic strange form factors have been fixed to $\rho_s=-0.25$,
$\mu_s=0.3$, and the dipole functional dependence with the
axial-vector mass $M_A=1.032$ GeV has been chosen. Dashed lines
correspond to results obtained at tree level, whereas the green
solid lines incorporate the effects introduced by the radiative
corrections at the level of the electric, magnetic and axial-vector
form factors. The analysis of these results shows that radiative
corrections are more important at backward-angle kinematics where
the difference amounts to $\sim$5$\%$ at $|Q^2|=1$ (GeV/c)$^2$,
falling to $\sim$2$\%$ at forward scattering angles. Concerning the
separate contribution of the radiative corrections in the
electric/magnetic (E/M) sector versus the axial-vector one, one
notes that at forward-angle kinematics the entire dependence comes
from the E/M sector. Indeed, the axial-vector contribution is
negligible for $\theta_e\rightarrow 0$. On the contrary, for
backward angles the radiative correction contribution in the
axial-vector form factor amounts to $\sim$3.5$\%$ at $|Q^2|=1$
(GeV/c)$^2$, hence being more important than the purely magnetic one
($\sim$1.5$\%$).

\subsection{Global analysis of ${\cal A}^{PV}$}\label{sec:fit}

As shown in previous sections, the result for the PV asymmetry may
depend significantly on the various ingredients that enter in its
evaluation, strangeness content in the nucleon, axial and vector
masses, $M_A$ and $M_V$, functional dependence of the weak form
factors with $Q^2$, {\it etc.} Moreover, a global comparison of
theory with available experimental data taken at different
kinematics leads to some problems and inconsistencies that need
further investigation. Thus, for completeness, we present in this
section a combined analysis of the world data on ${\cal A}^{PV}$
aiming to extract specific information on the WNC form factors.

To proceed we introduce the function \be
\chi^2=\sum_{j=1}^{28}\left(\frac{{\cal A}^{exp}_j-{\cal
A}_j^{th}}{\Delta {\cal A}^{exp}_j}\right)^2 \, , \ee where ${\cal
A}^{exp}_j$ refers to the experimental PV asymmetry, with
$\Delta{\cal A}^{exp}_j$ being its uncertainty and ${\cal A}_j^{th}$
the asymmetry given by the model. The sum runs over all available
experimental data (28) taken at different
kinematics~\cite{Happex-99,Happex-a,Happex-b,Happex-III,PVA4,PVA4-2,PVA4back,G0back,G0}.
To estimate the quality of the fit we have developed a code that
minimizes the value of $\chi^2$ as a function of the parameters
taken for the model and finds the absolute minimum. Special care has
been devoted to prevent the algorithm from being trapped in local
minima. It is important to point out that for all but the G0
experiment $\Delta{\cal A}_j^{exp}$ has been computed by adding in
quadrature the statistical and systematic errors. In the case of G0
experiment, as mentioned before, the ``extra'' global systematic
error has been added to the error already obtained by summing in
quadrature the statistical and systematic ones (see, however,
below).

To make contact with the discussion presented in previous sections,
we have selected as parameters of the model the electric and
magnetic strangeness content, $\rho_s$ and $\mu_s$, respectively.
The influence of the axial dipole mass $M_A$, and the vector dipole
(monopole) mass $M_V$ ($\widetilde{M}_V$) in the system has also
been studied.

\begin{figure}[htbp]
     \centering
         \includegraphics[height=16cm,width=12cm,angle=270]{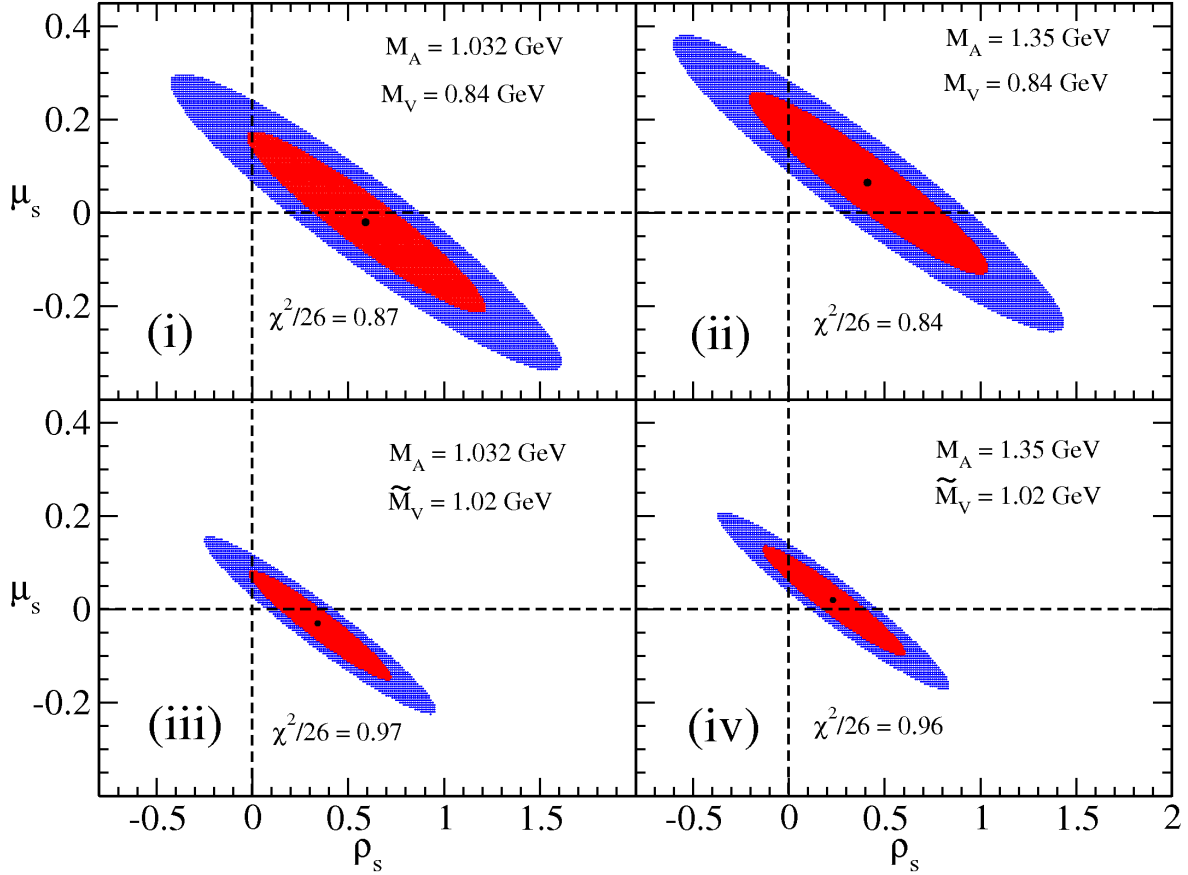}
     \caption{(Color online) World data constraint in the $\mu_s-\rho_s$ plane. The red
      (dark) and blue ellipses represent 68.27\% ($1\sigma\Rightarrow\Delta\chi^2 = 2.30$)
      and 95.45\% ($2\sigma\Rightarrow\Delta\chi^2 = 6.18$) confidence contours around the
      point of maximum likelihood (black), where $\Delta\chi^2$ is the increase over
      the minimum value of $\chi^2$. Zero $\mu_s$ and $\rho_s$ (dashed black lines)
      are indicated for reference. The GKex model has been used for the EM form factor description.
      Each panel corresponds to different values of the dipole axial mass, $M_A$, and dipole
      (monopole) vector strange mass, $M_V$ ($\widetilde{M}_V$). The corresponding
      reduced $\chi^2$ divided by the number of degrees of freedom for the system
      is also given in each panel (see text for details).}
     \label{fig:mus-rhos-GKex}
 \end{figure}

In Fig.~\ref{fig:mus-rhos-GKex}, we present the world data
constraint on $\mu_s$ and $\rho_s$ parameters as result of our
analysis. The ellipses are the projections of $\chi^2$ function on
the $\mu_s-\rho_s$ plane and represent the $1\sigma$ (inner) and
$2\sigma$ (outer) confidence level. The GKex model has been used for
the EM form factors and radiative corrections have also been
included. In the upper-left (lower-left) panel, a dipole (monopole)
vector strange mass, $M_V=0.84$ GeV ($\widetilde{M}_V=1.02$ GeV) and
standard dipole axial mass, $M_A=1.032$ GeV, have been assumed,
namely, situation-(i) (situation-(iii)). In the upper-right
(lower-right) panel, the same as in the left panels is assumed, but
now using the extreme value for the dipole axial mass, $M_A=1.35$
GeV, namely, situation (ii) (situation (iv)). The reduced $\chi^2$
is shown in each case. This is calculated as the ratio of $\chi^2$
and the number of degrees of freedom for the system, that is, the
number of experimental points minus the number of parameters. Again
we note that for the G0 experiment we have used a conservative
estimate where the global uncertainty is added to the other errors
computed in quadrature. If instead one were to make more stringent
assumptions, such as by entirely ignoring the global error (which
may be warranted to some extent, since some theoretical
uncertainties are apparently incorporated in the quantity), then the
error ellipses become a bit smaller, although the central values
change very little. Indeed, the new high-$Q^2$ HAPPEX point with its
very tight uncertainty dominates the analysis.

\begin{figure}[htbp]
     \centering
         \includegraphics[height=16cm,width=12cm,angle=270]{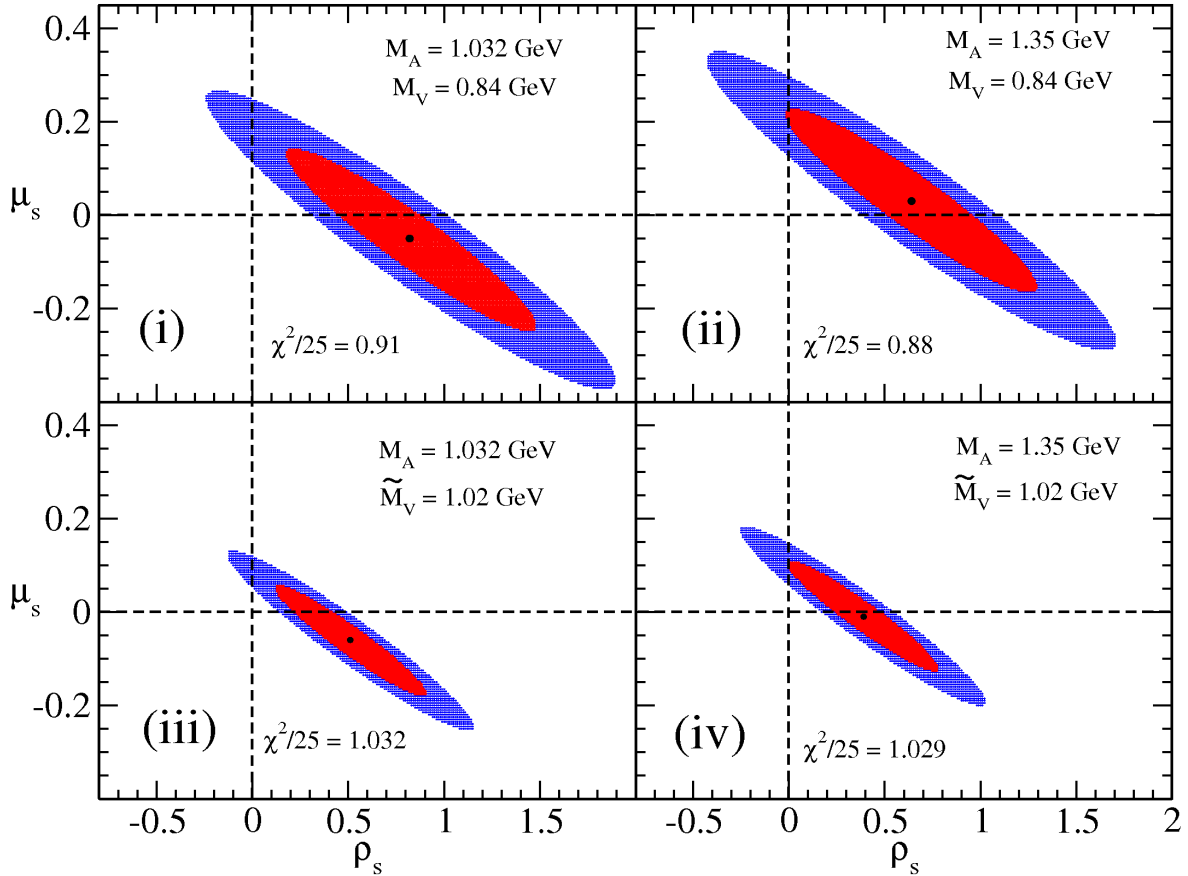}
     \caption{(Color online) As for Fig.\ref{fig:mus-rhos-GKex},
       except now the Bernauer {\it et al.} fit
      has been used for EM form factors of the proton.}
     \label{fig:mus-rhos-Bernauer}
 \end{figure}

Fig.~\ref{fig:mus-rhos-Bernauer} represents the same results as in
Fig.~\ref{fig:mus-rhos-GKex}, but in this case, the fit of Bernauer
{\it et al.}~\cite{Bernauer} has been used to the description of the
EM form factor of the proton. Note that here we have not included
the G0 point at $|Q^2|\sim1$ (GeV/c)$^2$ because the fit of Bernauer
{\it et al.} should not be used beyond $|Q^2|\sim0.9$ (GeV/c)$^2$.

Comparing results in Figs.~\ref{fig:mus-rhos-GKex} and
\ref{fig:mus-rhos-Bernauer} we observe on general grounds a similar
behavior for the confidence ellipses in the different situations. In
particular, note the significant reduction in the area of the
ellipses in the case of the monopole description for the vector
strange form factors (situations (iii) and (iv)). This is in accord
with Figs.~\ref{fig:mus-1}--\ref{fig:mus-3} in which the dispersion
in the asymmetry curves is larger when a monopole description is
employed than when the form factor has a dipole shape. Note that in
those figures $\rho_s$ was fixed for the monopole and dipole
descriptions, while for the ellipse here $\rho_s$ is a free
parameter. On the other hand, results obtained with GKex
(Fig.~\ref{fig:mus-rhos-GKex}) compared with the ones corresponding
to the fit of Bernauer {\it et al.}
(Fig.~\ref{fig:mus-rhos-Bernauer}), show the confidence ellipses in
the $\mu_s-\rho_s$ plane slightly shifted towards decreasing values
of $\rho_s$ (left) and increasing $\mu_s$ (upper). However, while
most of the $1\sigma$ ellipses are consistent with strictly positive
$\rho_s$-values (only in situations (ii) and (iv) in
Fig.~\ref{fig:mus-rhos-GKex} is the value $\rho_s=0$ clearly
contained into the $1\sigma$ confidence region), zero strangeness in
the magnetic sector is on the contrary supported by the analysis of
data in all situations. The analysis of the $2\sigma$ confidence ellipses
also shows consistency with negative values for $\rho_s$.

\begin{figure}[htbp]
     \centering
         \includegraphics[width=.4\textwidth,angle=270]{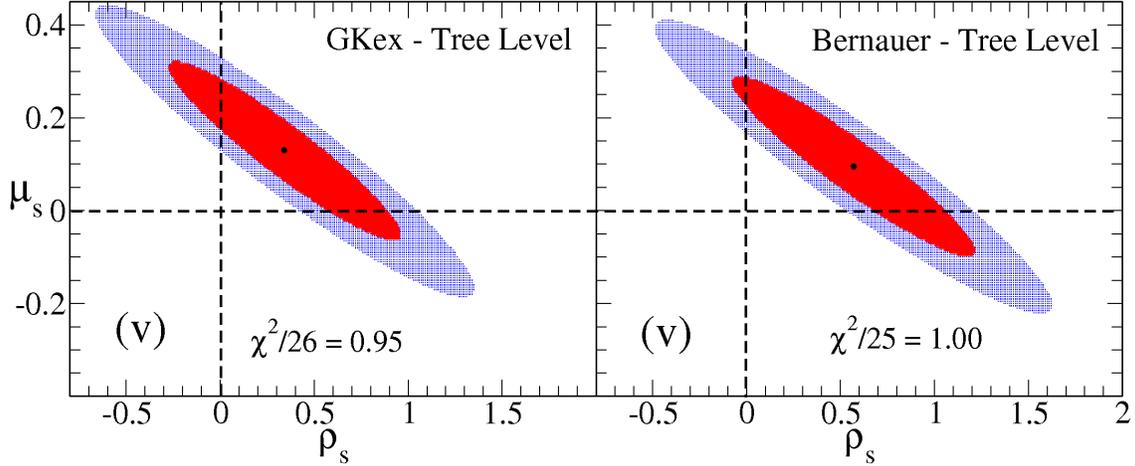}
     \caption{(Color online) Same general comments as for Fig.\ref{fig:mus-rhos-GKex}.
          In this case both panels correspond to a tree level calculation. A vector
          dipole mass, $M_V=0.84$ GeV, and standard axial mass, $M_A=1.032$ GeV,
          have been assumed. In the left-hand panel the GKex model has been assumed, while
          the fit of Bernauer {\it et al.} has been assumed for the right-hand one.}
     \label{fig:mus-rhos-RC}
\end{figure}

The effect on the PV asymmetry due to radiative corrections was
studied in section~\ref{sec:radcorr}. For completeness, in
Fig.~\ref{fig:mus-rhos-RC} we represent the results when tree level
is assumed. The standard dipole masses for both panels have been
assumed, $M_V=0.84$ GeV and $M_A=1.032$ GeV. Comparing these results
with the ones corresponding to situation (i) in
Figs.~\ref{fig:mus-rhos-GKex} and \ref{fig:mus-rhos-Bernauer}, one
observes that the confidence ellipses are significantly shifted to
larger (smaller) values of $\mu_s$ ($\rho_s$). Although zero
electric/magnetic strangeness is within the $1\sigma$ confidence
region (Fig.~\ref{fig:mus-rhos-RC}) for the two EM form factor
descriptions considered, notice that the central points in the
ellipses correspond to positive values of $\rho_s$ and $\mu_s$.

In Table~\ref{table:Chi2summary} we present a summary of all of the
results shown in this section. The values presented for the
parameters $\mu_s$ and $\rho_s$ correspond to the point of maximum
likelihood (central values of the parameters). Their errors have
been computed fixing one parameter at its central value and moving
the other parameter along the line out to the limit of the $1\sigma$
contour.

{%
\begin{table}
\newcommand{\mc}[3]{\multicolumn{#1}{#2}{#3}}
\begin{center}
\begin{tabular}{c|| c| c| c|| c| c| c||}
      \mc{4}{c}{ \ \ \ \ \ \  GKex} & \mc{3}{c}{  Bernauer}\\
\cline{2-7}
    & $\rho_s$\, & $\mu_s$\, & $\chi^2/26$\, & $\rho_s$\, & $\mu_s$\, & $\chi^2/25$\,\\
\cline{2-7}
(i)   \, & $0.59\pm0.21$ & $-0.020\pm0.065$ & 0.87 & $0.82\pm0.22$ & $-0.050\pm0.065$ & 0.91 \\
(ii)  \, & $0.41\pm0.21$ & $ 0.065\pm0.065$ & 0.84 & $0.64\pm0.22$ & $ 0.030\pm0.065$ & 0.88 \\
(iii) \, & $0.34\pm0.11$ & $-0.030\pm0.035$ & 0.97 & $0.51\pm0.11$ & $-0.060\pm0.035$ & 1.032\\
(iv)  \, & $0.23\pm0.11$ & $ 0.020\pm0.035$ & 0.96 & $0.39\pm0.11$ & $-0.010\pm0.035$ & 1.029\\
\cline{2-7}
(v)   \, & $0.34\pm0.21$ & $ 0.130\pm0.065$ & 0.95 & $0.57\pm0.21$ & $ 0.095\pm0.065$ & 1.00 \\
\cline{2-7}
\end{tabular}
\caption{Summary of the results shown in Fig~\ref{fig:mus-rhos-GKex}
    and Fig~\ref{fig:mus-rhos-Bernauer}. Situation (v) corresponds to
    Fig.~\ref{fig:mus-rhos-RC}.}\label{table:Chi2summary}
\end{center}
\end{table}
}%

From this analysis we conclude that the specific values of the
electric and magnetic strangeness parameters show some dependence
on the particular description of the EM form factors,
namely, GKex versus the fit of Bernauer {\it et al.} The former
gives rise to smaller central values for $\rho_s$, although they
overlap with the ones obtained with the fit of Bernauer {\it et al.}
when error bars are included. However, in all of the situations
considered the analysis of data is consistent with positive or zero
$\rho_s$. The particular value selected for the axial mass in
addition to the monopole/dipole description of the functional
dependence of $G_{E,M}^{(s)}$ with $Q^2$ introduces important
effects; note the significant reduction in $\rho_s$ when moving from
situation (i) to (iv).

Concerning the magnetic strangeness parameter $\mu_s$, the use of
the fit of Bernauer {\it et al.} leads to smaller values, although
the results are very close to zero strangeness. The modifications
introduced using monopole/dipole form factors and the specific axial
mass are also visible, leading to changes in the sign of $\mu_s$.
However, the specific values of $\mu_s$ are close to zero in all
cases. Notice that $\mu_s=0$ is contained in the $1\sigma$
confidence region in all situations.

The results obtained at tree level (Fig.~\ref{fig:mus-rhos-RC}),
also summarized in Table~\ref{table:Chi2summary} as situation (v),
show some significant aspects that should be mentioned. The
confidence ellipses are shifted towards smaller $\rho_s$ and larger
$\mu_s$. However, whereas the value of $\rho_s$ is still positive
but smaller than the ones corresponding to situation (i) in the
previous cases, the central value of $\mu_s$ is clearly shifted into
the positive region. Note that $\rho_s=0$ and/or $\mu_s=0$ are
contained in the $1\sigma$ confidence region, in contrast with
Figs.~\ref{fig:mus-rhos-GKex} and \ref{fig:mus-rhos-Bernauer}
(situation (i)) where $\rho_s=0$ is outside the $1\sigma$ confidence
region. However, the analysis of results extended to the $2\sigma$ region 
also supports zero $\rho_s$ strangeness.

To conclude, a common feature that emerges from all of the results shown in
Figs.~\ref{fig:mus-rhos-GKex}--\ref{fig:mus-rhos-RC} is that either
$\rho_s=0$ or $\mu_s=0$ are contained within the $95\%$ confidence
level. Moreover, no strong statement on the specific sign of
$\rho_s$ and/or $\mu_s$ can be established. In fact, from
Figs.~\ref{fig:mus-rhos-GKex}--\ref{fig:mus-rhos-RC} we observe that
the only situation that is 
outside the $1\sigma$ confidence contour corresponds to both $\rho_s$ and $\mu_s$ taking on
negative values. Moreover, the special case of no strangeness, that
is, $\rho_s=\mu_s=0$, is also outside the $1\sigma$ contour in most
of the cases. However, some caution should be drawn for these specific
$\rho_s,\mu_s$-values due to the analysis of the extended $2\sigma$ confidence region.

%

Although not shown here, for completeness we have also
analyzed the results obtained by using the weak mixing angle as
given in~\cite{Erler2}. In this case, the confidence ellipses favor
smaller values of $\mu_s$, while the values of $\rho_s$
permitted by the analysis of world data do not change significantly, that is, the ellipses simply move to more negative $\mu_s$-values. In
fact, the point of maximum likelihood in the magnetic strangeness
goes from a value close to zero to $\mu_s\sim [-0.1,-0.2]$ and 
the case of no strangeness, $\rho_s=\mu_s=0$, is within the
$1\sigma$ confidence contour.

\subsection{Q-weak Experiment}\label{sec:qweak}

In this section we analyze in detail the potential impact the
variations we are considering might have on the Q-weak
experiment~\cite{Qweak1,Qweak2,Qweak3,NPA805,hallaweb,Young}. As is
well known, the main aim of Q-weak is to search for new physics at
the TeV scale via the measurement of the proton's weak charge at
$|Q^2|=0.028$ (GeV/c)$^2$, an incident energy of $1.16$ GeV and
$\theta_e\sim8^o$. Some basic objectives of the experiment are to
determine the weak charge of the proton to $4\%$, to extract
$\sin^2\theta_W$ to $0.3\%$ and to set limits on PV new physics at
an energy scale of $\sim$2-3 TeV. However, in spite of the low
$|Q^2|$ value considered, the hadronic form factors still play a
role and make the extracting the asymmetry and its associated error
not entirely clean. Any variations we have discussed in the
preceding sections may have some impact on the results. In what
follows we briefly evaluate how the interpretation of the Q-weak
result might be affected when different options for the description
of the nucleon form factors are considered.

The general expression for the Q-weak term can be written as
\be
Q_W^p=1-4\sin^2\theta_W= \frac{-2G^2{\cal A}^{PV}/{\cal
A}_0+B_{pn}+B_{ps}}{G^2(1+R_V^p)- \sqrt{\tau(1+\tau)(1-\epsilon^2)} G_M^p G_A^{e,p}} \, .
\label{qweak}
\ee

In the limit of extreme forward-angle kinematics,
{\it i.e.,} $\theta_e\rightarrow 0$, the axial-vector term does not
contribute and the previous expression reduces to
\be
Q_W^p=\frac{-2G^2{\cal A}^{PV}/{\cal
A}_0+B_{pn}+B_{ps}}{G^2(1+R_V^p)} \, . \label{qweaklimit}
\ee
As before, we have introduced
$G^2=\epsilon(G_E^p)^2+\tau(G_M^n)^2$ and the functions
\ba
B_{pn}&=&(\epsilon G_E^pG_E^n+\tau G_M^pG_M^n)(1+R_V^n) \\ \nonumber
B_{ps}&=&(\epsilon G_E^pG_E^s+\tau G_M^pG_M^s)(1+R_V^{(0)}) \, .
\ea

It is important to point out that our interest in this section is
not to determine a precise value for the weak charge of the proton.
On the contrary, it is to analyze how, as for the weak
mixing angle, such a quantity
may change due to the uncertainties associated with
the description of the nucleon's EM and WNC form factors, with
special emphasis being placed on the potential electric/magnetic strangeness
content in the nucleon.

Thus, the general procedure we adopt in this section is as
follows: the value for the PV asymmetry at $|Q^2|=0.028$ (GeV/c)$^2$
is taken from the global fit to all electroweak data presented
in~\cite{PRL99}. This analysis leads to ${\cal A}^{PV}=-1.91\times
10^{-7}$. Hence, the value of $Q_W^p$ is computed through
Eq.~(\ref{qweak}) making use of different descriptions for the EM
and WNC nucleon form factors. The results obtained are shown in
Table~\ref{table-qweak-EM} and \ref{table-qweak-WNC}. All
calculations have been made considering the situation GKex-(i)
defined in the previous section as reference (see
Table~\ref{table:Chi2summary}), {\it i.e.,} $M_A=1.032$ GeV,
$M_V=0.84$ GeV, $\rho_s=0.59\pm0.21$ and $\mu_s=-0.020\pm0.065$.

In Table~\ref{table-qweak-EM} it is seen that the effects introduced
by the EM description of the form factors are very small, less than
$\sim$3$\%$ to $Q^p_W$, {\it i.e.,} less than $\sim$0.2$\%$ to
$\sin^2\theta_W$. Note that these differences emerge from the
comparison of results evaluated with the six prescriptions for the
EM form factors presented in Table~\ref{table-qweak-EM}. In fact,
the maximum discrepancy corresponds to the BHM-SC prescription
compared with Bernauer's fit.
\begin{table}
\begin{tabular}{|c|cccccc|}
\hline \hline
EM  & A-S & Kelly  &  BHM-SC  &  BHM-pQCD  & GKex & Bernauer\\
\hline
$Q_W^p$          & 0.05373 & 0.05337 & 0.05422 & 0.05405 & 0.05340 & 0.05257 \\
$\sin^2\theta_W$ & 0.23657 & 0.23666 & 0.23644 & 0.23649 & 0.23665 & 0.23686 \\
\hline \hline
\end{tabular}
\caption{Values for $Q_W^p$ and $\sin^2\theta_W$ obtained through
Eq.~(\ref{qweak}) for different descriptions of the EM nucleon form
factors (see text for details). Electric and magnetic strangeness
content of the nucleon have been fixed to the GKex-(i) situation
(Table~\ref{table:Chi2summary}): $\rho_s=0.59$ and $\mu_s=-0.020$.}
\label{table-qweak-EM}
\end{table}

Table~\ref{table-qweak-WNC} shows the sensitivity of $Q_W^p$ and
$\sin^2\theta_W$ to variations of the strangeness parameters,
$\rho_s$ and $\mu_s$. As stated above, the PV asymmetry is taken as
${\cal A}^{PV}=-1.91\times 10^{-7}$, and the GKex prescription is
employed for the EM form factors. The specific values of the
strangeness parameters presented in Table~\ref{table-qweak-WNC}
correspond to the maximum and minimum values permitted by the
($1\sigma$) errors in GKex-(i), considered as a reference case. The
last column in the table shows the percentage deviation between the
two values presented for $Q_W^p$ (likewise for $\sin^2\theta_W$).
\begin{table}
\begin{tabular}{|c|cc|c|}
\hline \hline
E-strange  ($\mu_s=-0.020$)&  $\rho_s=0.80$  & $\rho_s=0.38$ & \hspace{0.5cm}(\%)\hspace{0.5cm} \\ \hline
$Q_W^p$                & 0.05499         & 0.05182  & 5.8      \\
$\sin^2\theta_W$           & 0.23625         & 0.23704  & 0.33     \\
\hline \hline
M-strange  ($\rho_s=0.59$) &  $\mu_s=-0.085$  &  $\mu_s=0.045$ &\hspace{0.5cm}(\%)\hspace{0.5cm} \\ \hline
$Q_W^p$                &  0.05203         &  0.05478 & 5.0      \\
$\sin^2\theta_W$           &  0.23699         &  0.23630 & 0.3      \\
\hline \hline
\end{tabular}
\caption{Values for $Q_W^p$ and $\sin^2\theta_W$ obtained from Eq.(\ref{qweak}) for different descriptions of the WNC nucleon
form factors (see text for details). The last column corresponds to
    the difference between the cases considered.}
\label{table-qweak-WNC}
\end{table}

Notice that the deviations linked to the magnetic and electric
strangeness are similar or even a little bit larger than the
expected uncertainty of the Q-weak experiment. Hence a word of
caution should be expressed on the analysis and interpretation of
results derived from the Q-weak experiment.

Before concluding, we recall once more that the aim of this study is
focused on the role played by the description of the EM and WNC form
factors, particularly concerning the strangeness content, in the
determination of the weak charge and/or the weak mixing angle. Thus,
we have applied our global analysis as presented in previous
sections to the particular kinematics corresponding to the Q-weak
experiment, assuming for the PV asymmetry its value taken from an
extrapolation to the forward angle using all current
data~\cite{PRL99}. Note that this study differs from the work shown
in~\cite{GHM2011} (see also~\cite{GH2009,SBMT2010}) where the focus is placed on exploring the
theoretical uncertainties associated with contributions of hadronic
intermediate states. The authors of \cite{GHM2011} reexamine the
$\gamma Z$-box contribution in the framework of dispersion relations
and provide an estimate of the absolute size and
uncertainty of the $\gamma Z$ correction to the PV asymmetry in the
forward-angle limit. This analysis is applied to the Q-weak
experiment, providing a correction to ${\cal A}^{PV}$ equivalent to
a shift in the proton weak charge.

From all of this discussion it is clear that the precision
measurement of the asymmetry at Jefferson Lab (Q-weak experiment)
may help us in resolving some of the basic uncertainties that are
still present in the study of PV elastic electron-proton scattering,
and should help to deepen our understanding of the internal
structure of the nucleon. Following these general ideas, and still
waiting for the final analysis of the Q-weak experiment, in next
section we extend our global study by incorporating the {\sl
``hypothetical''} Q-weak asymmetry value and examine its effect on
the nucleon structure.

\subsection{Additional analysis: Neutral weak effective couplings}

In this section we extend our previous analysis of strangeness in
order to improve our knowledge of the WNC effective coupling
constants. Thus, we provide an estimate of the electroweak
parameters ($C_{1u}$, $C_{1d}$) consistent with the global analysis
of world data presented in Section~IV.E. Following the work
in~\cite{PRL99} we evaluate the confidence level contour ellipses
displayed in the $C_{1u}+C_{1d}$ versus $C_{1u}-C_{1d}$ plane. To
clarify what is done here, we explain in some detail the general procedure we
have considered.

We start with the general expression for the PV asymmetry in terms
of the nucleon form factors and electroweak couplings, \be {\cal
A}^{PV}=\frac{{\cal A}_0}{2}\left[ \frac{a_A\left(\varepsilon
G_E^N\widetilde{G}_E^p+\tau G_M^p\widetilde{G}_M^p\right) -
a_V\sqrt{1-\varepsilon^2}\sqrt{\tau(1+\tau)}G_M^p G_A^{e,p}}
{\varepsilon(G_E^p)^2+\tau (G_M^p)^2} \right] \, , \ee and set the
leptonic coupling constants to their values in the Standard Model at
tree level, {\it i.e.,} $a_A=-1$ and $a_V=-1+\sin^2\theta_W$.
Coherently with our study in previous sections and also following
the general arguments introduced in~\cite{Liu}, the weak mixing
angle is fixed to $\sin^2\theta_W=0.23122$. 
Thus, we choose as free parameters in the analysis the values of the weak
neutral couplings, $\xi_V^p$ and $\xi_V^n$, that enter in the WNC
form factors through
\begin{eqnarray}
 \widetilde{G}_{E,M}^p = \xi_V^p G_{E,M}^p + \xi_V^nG_{E,M}^n + \xi_V^{(s)}G_{E,M}^{(s)} \, .
\end{eqnarray}
The strangeness contribution is taken as $\xi_V^{(s)}=-(1+R_V^{(0)})$ with $R_V^{(0)}=-0.0123$.

\begin{figure}[htbp]
    \centering
        \includegraphics[width=0.6\textwidth,angle=270]{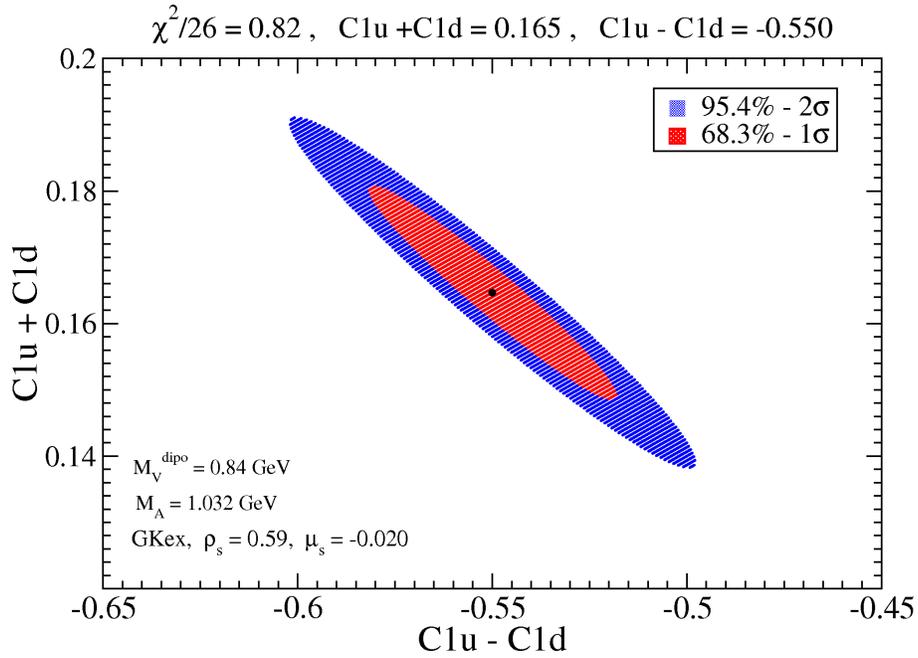}
    \caption{(Color online) World data constraint in the $C_{1u}+C_{1d}$ versus $C_{1u}-C_{1d}$ plane.
        The red and blue
    ellipses represent 68.27\% (1$\sigma$) and 95.45\% (2$\sigma$) confidence contours around
    the point of maximum likelihood (black dot). The value of the minimum $\chi^2$ and the
        point of maximum likelihood are indicated in the top legend.}
    \label{fig:C1u-C1d}
\end{figure}

In what follows we present a global analysis of the world data on
${\cal A}^{PV}$ by showing the $1\sigma$ (inner) and $2\sigma$
(outer) confidence level ellipses in the $C_{1u}+C_{1d}$ versus
$C_{1u}-C_{1d}$ plane. Results have been obtained by using the GKex
description for the EM form factors and fixing the electric and
magnetic strange parameters to the point of maximum likelihood using
GKex-(i), {\it i.e.,} $\rho_s=0.59$, $\mu_s=-0.020$ (see
Table~\ref{table:Chi2summary} and upper-left panel in
Fig.~\ref{fig:mus-rhos-GKex}). Dipole vector strange and axial
masses have been assumed: $M_V=0.84$ GeV, $M_A=1.032$ GeV. As
observed, for the specific GKex-(i) situation the $1\sigma$ contour
region spreads to the range $(-0.52,\, -0.58)$ for $C_{1u}-C_{1d}$
and $(0.148,\, 0.18)$ for $C_{1u}+C_{1d}$. In the $2\sigma$ case the
range is extended to $(-0.5,\, -0.6)$ [$C_{1u}-C_{1d}$ axis] and
$(0.138,\, 0.19)$ [$C_{1u}+C_{1d}$]. The value of the point of
maximum likelihood, also presented as a label in
Fig.~\ref{fig:C1u-C1d}, is $C_{1u}+C_{1d}=0.165$ and
$C_{1u}-C_{1d}=-0.550$. 

As already mentioned, the use of a different value for the weak mixing angle,
for instance $\sin^2\theta_W=0.23867$ (see~\cite{Erler2}), leads to
differences in the confidence level ellipses. In particular, the new
contour moves to the right by $\sim$5-6\% (larger values of
$C_{1u}-C_{1d}$) and decreases ($\sim$4-5\%) along the
$C_{1u}+C_{1d}$ axis.

%
%

\begin{figure}[htbp]
    \centering
        \includegraphics[width=0.6\textwidth,angle=270]{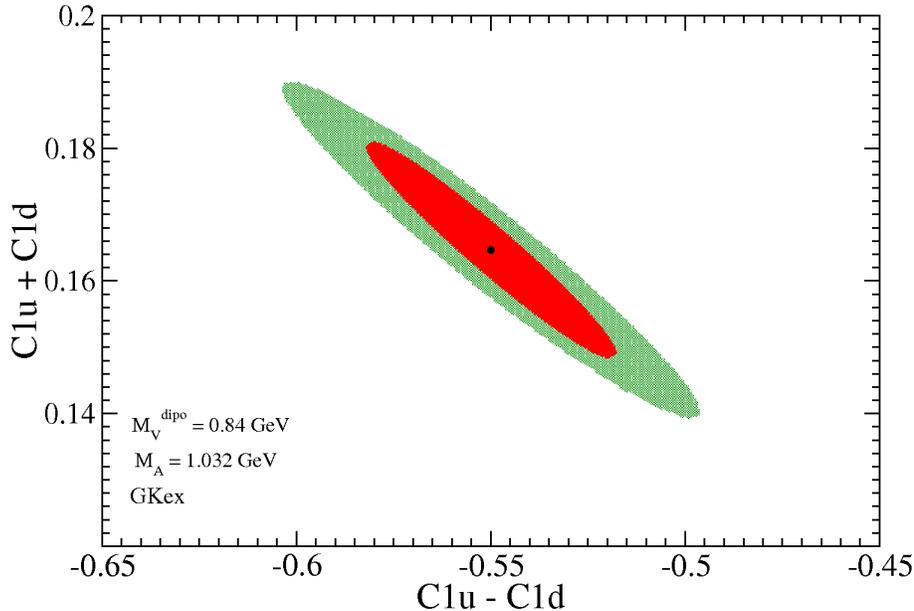}
    \caption{(Color online) Green area represents the superposition of the full
          set of 1$\sigma$-ellipses (see text for
          details). 
          The red area is the original $1\sigma$-ellipse.}
    \label{fig:C1u-C1d-green}
\end{figure}

We complete the study by analyzing the dependence of the confidence
level ellipses on the particular strangeness content in the
nucleon. Thus, we redo the global analysis producing new contour
ellipses that take into account the variation in strangeness, that
is, the information contained in
Figs.~\ref{fig:mus-rhos-GKex}-\ref{fig:mus-rhos-RC}. In particular,
as considered in Fig.~\ref{fig:C1u-C1d}, we restrict ourselves to
the GKex-(i) situation and let the electric and magnetic strange
parameters ($\rho_s,\mu_s$) to float around the boundary contour
corresponding to the $1\sigma$ region in the upper-left panel in
Fig.~\ref{fig:mus-rhos-GKex} (red region). Each point in the
($\rho_s,\mu_s$) contour leads to a specific confidence level
ellipse in the $C_{1u}/C_{1d}$ plane. In
Fig.~\ref{fig:C1u-C1d-green} we show the global region (green area)
produced by the superposition of a full set of ellipses, each one
corresponding to a specific ($\rho_s,\mu_s$) value. As reference, we
compare this global contour with the original one presented in
Fig.~\ref{fig:C1u-C1d} (red area) where the centroid: $\rho_s=0.59$,
$\mu_s=-0.02$, was selected (see discussion above). Notice that only
the $1\sigma$ confidence level ellipses are presented in
Fig.~\ref{fig:C1u-C1d-green}. 

\begin{figure}[htbp]
    \centering
        \includegraphics[width=0.45\textwidth,angle=270]{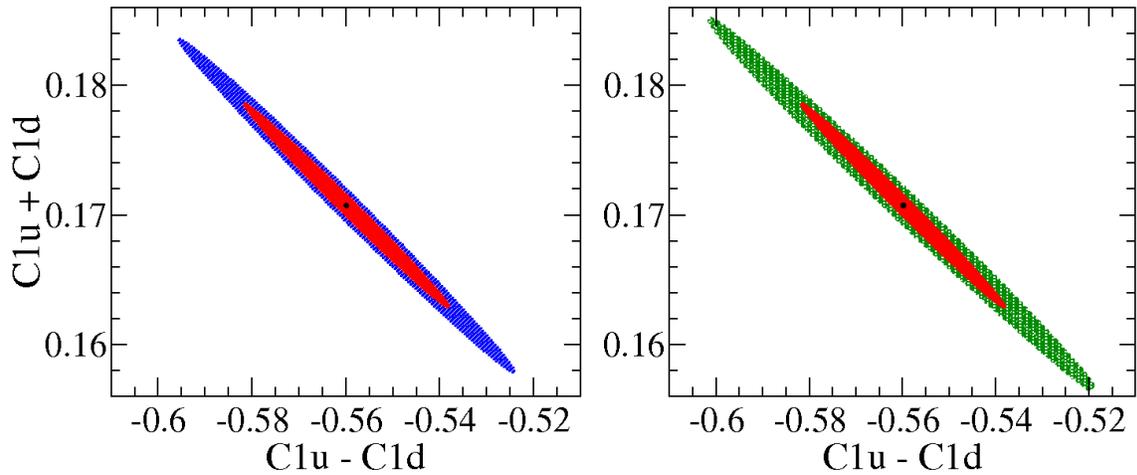}
    \caption{(Color online) As for Fig.~\ref{fig:C1u-C1d} (left panel) and
          Fig.~\ref{fig:C1u-C1d-green} (right panel), except that now the hypothetical
          $Q_{weak}$ result is included in the global analysis.}
    \label{fig:C1u-C1d-Qw}
\end{figure}

Next we investigate how the world data constraint in the
$C_{1u}/C_{1d}$ plane changes when introducing the {\sl
``hypothetical''} Q-weak measurement. To make the discussion
that follows clear, we consider the specific situation shown in
Fig.~\ref{fig:C1u-C1d}, and analyze how the contour changes by
adding to the global fit the value ${\cal A}^{PV}=-1.91\times
10^{-7}$ with an error of $\sim$2.2\% (expected Q-weak
result). The new results are presented in Fig.~\ref{fig:C1u-C1d-Qw}
(left panel) that shows the $1\sigma$ and $2\sigma$ contours to be
reduced very significantly; the areas in Fig.~\ref{fig:C1u-C1d-Qw} are about $\sim 85\%$ smaller than the ones shown
in Fig.~\ref{fig:C1u-C1d}. To conclude, we apply the
same analysis to results in Fig.~\ref{fig:C1u-C1d-green}, {\it
i.e.,} letting the strange parameters float along the $1\sigma$
level confidence contour in Fig.~\ref{fig:mus-rhos-GKex} (upper-left
panel), and add again to the global fit the {\it ``hypothetical''}
Q-weak result. The new ellipses (linked to
Fig.~\ref{fig:C1u-C1d-green}) are shown in the right panel of
Fig.~\ref{fig:C1u-C1d-Qw}. As in the previous case, note the strong
reduction in the area enclosed by the ellipses.

Finally, we note that a new experiment is planned to be run at Mainz
making use of the MESA accelerator facility~\cite{MESA}. Its main
aim is to provide a very precise measurement for the weak charge of
the proton and weak mixing angle in the low-energy regime. The
energy of the electron beam is 137 MeV and the scattering angle
$20^0 \pm 10^0$. The transferred momentum is fixed to $|Q^2|=0.0022$
(GeV/c)$^2$. As shown in~\cite{GHM2011}, the impact of the $\gamma
Z$-box correction is expected to be very small for these kinematics, and
hence this result should complement the Q-weak experiment, at the same
time providing a test for the Standard Model while also leading to a
deeper understanding of the internal structure of the nucleon.

\subsection{Projections for Higher-Energy Experiments}\label{sec:expt}

\begin{figure}[htbp]
    \centering
        \includegraphics[width=.6\textwidth,angle=270]{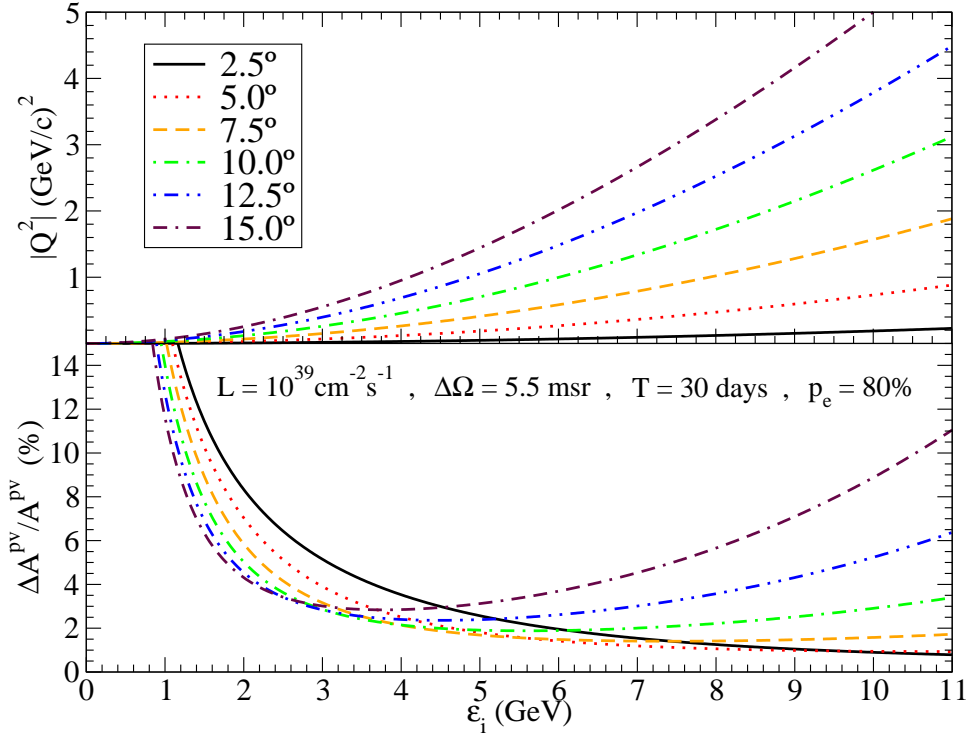}
    \caption{(Color online). Fractional precision $\Delta {\cal A}^{PV}/{\cal A}^{PV}$ as a function
of the incident electron energy $\varepsilon_i$ (bottom panel). The
top panels show the transferred four-momentum versus $\varepsilon_i$
for different scattering angles.}
    \label{fig:q2-ei-Aerror}
\end{figure}

To obtain some idea about the likelihood that high precision elastic e-p PV 
experiments may be feasible at higher energies we include here a brief section projecting what fractional precision could be attained. 
It should be noted that at present, even with JLab being upgraded to 11 GeV at the high currents needed for such studies, 
there are no approved experiments of this type; however, for completeness the following has been included. The results 
presented below are characterized using
the standard figure of merit, \be {\cal F} =
({\cal A}^{PV})^2\ \sigma^{PC} \, , \ee where the asymmetry ${\cal
A}^{PV}$ and the parity-conserving differential cross section
$\sigma^{PC}$ are evaluated within our theoretical model. This
expression governs the level of statistical precision that can be
achieved for given experimental conditions. The fractional precision
in the asymmetry, assuming that systematic errors are under control
is given by \be \frac{\Delta {\cal A}^{PV}}{{\cal A}^{PV}} =
\frac{1}{p_e\sqrt{\Delta\Omega\ T\ {\cal L}\ {\cal F}}} \, , \ee
where $p_e$ represents the incident electron polarization,
$\Delta\Omega$ is the detector solid angle, $T$ is the runtime and
${\cal L}$ is the luminosity. In this work we choose specific
``canonical'' values for all of these (see
Fig.~\ref{fig:q2-ei-Aerror}). For more details on these choices the
reader can look at \cite{Musolf1994, hallaweb}. The resulting
fractional precision is shown in the bottom panel of
Fig.~\ref{fig:q2-ei-Aerror} for a large variety of scattering angles
at forward kinematics. As observed, for a wide range of kinematical
conditions, {\it i.e.,} values of the energy $\varepsilon_i$, the
precision expected for our ``canonical'' conditions lies in the
vicinity of $2\%$, being even smaller ($\sim$1$\%$) for high
energies and very forward scattering angles. Notice that this
situation corresponds to the lowest values of the transferred
momentum $|Q^2|$ (see top panel). This is consistent with the fact
that at very high $|Q^2|$ it is hard to avoid uncertainties coming
from ingredients beyond the elastic scattering regime.

As an example of using these projections, consider the 7.5$^o$ case
(dashed lines): at 11 GeV, the upgraded energy of the JLab facility
for experiments in Halls A and C, one could reach a fractional error
of better than 2\% at momentum transfers reaching almost 2
(GeV/c)$^2$.

\section{Conclusions}\label{sec:concl}

This work presents a systematic study of elastic electron-nucleon
scattering including parity violation (PV) in its description, {\it
i.e.,} not only the dominant EM interaction is considered but also
the role played by the weak interaction. The basic goal of the
present study is to deepen our knowledge of the hadronic structure involved
with a special focus on the analysis of strangeness and axial-vector
content in the electroweak nucleon form factors.

Within the general framework of the Electroweak Theory of the
Standard Model, we have summarized the general formalism needed in
the description of the above-mentioned scattering process. Hadronic
response functions and the parity-violating asymmetry or helicity
asymmetry (denoted simply as PV asymmetry, ${\cal A}^{PV}$) have
been evaluated. The latter is defined as the ratio between the
difference and the sum of the electron scattering cross sections for
positive and negative incoming electron helicities. The helicity
asymmetry being different from zero is a clear signature of the
presence of the weak interaction, and thus its measurement allows
one to gain insight into the electroweak structure of the nucleon.
In recent years significant efforts from both the experimental and
theoretical points of view have been devoted to this problem. New
experiments have been devised and performed for a variety of
kinematical situations. Data reported at backward and forward
scattering angles are compared in this work with theoretical
modeling showing the role played by the various ingredients that enter into
the description of the reaction mechanism.

One of the basic ingredients in the PV asymmetry comes from the EM
structure of the nucleon. Our present knowledge about the EM nucleon
form factors, particularly in the case of the proton, is rather
precise through measurements of elastic (parity-conserving) electron-nucleon scattering.
These have involved techniques based on the Rosenbluth separation,
as well as the use of nucleon polarization measurements. Although a
proper description of data is provided by different models, still
some ambiguities emerge from the analysis of different experiments.
In particular, the behavior of the nucleon form factors with $Q^2$
(transferred four-momentum) may differ significantly with the
particular prescription considered. In this work we have
investigated a wide selection of models for the nucleon's EM
structure and have analyzed the impact of these choices on the
helicity asymmetry. A precise description of the EM structure of the
nucleon is essential for the analysis of the PV asymmetry.

A systematic study of the PV asymmetry including its dependence on
the various nucleon form factors has been presented in Sect.
\ref{sec:analysis}. In addition to the purely EM nucleon structure,
the effects introduced by the WNC axial-vector form factor have been
analyzed at depth. Functional dependencies on $Q^2$ based on dipole
and monopole shapes have been assumed using different values for the
axial-vector mass in both cases. The strangeness content in the
axial-vector form factor does not introduce significant effects in
the PV asymmetry. From our study, when applied to backward-angle
kinematics where the axial-vector contribution has its largest
impact, the uncertainty in ${\cal A}^{PV}$ linked to the description
of the axial-vector form factor is of the order of $\sim$5--6$\%$
for dipole and $\sim$10--11$\%$ for monopole descriptions at
$|Q^2|=1$ (GeV/c)$^2$. This uncertainty has been evaluated assuming
the axial-vector mass to span the ranges $M_A=1.032-1.35$ GeV
(dipole) and $\widetilde{M}_A=0.5-1$ GeV (monopole).

One of the main objectives in studies of PV electron scattering
concerns the role of the strange quark in the electric and magnetic
sectors ${\cal A}_E$ and ${\cal A}_M$ which do not involve the
axial-vector form factor. Thus, the helicity asymmetry has been
evaluated using different approaches to describing the strange form
factors $G_{E,M}^{(s)}(Q^2)$. The specific strangeness content is
given by the static strangeness parameters $\rho_s$ and $\mu_s$ in
the electric and magnetic sectors, respectively, whereas the
specific dependence on $Q^2$ is taken as dipole (with the vector
mass fixed to $M_V=0.84$ GeV) and monopole ($\widetilde{M}_V=1.02$
GeV). The analysis performed includes backward and forward
scattering angles spanning the kinematics involved in the
experiments.

Backward-angle measurements of ${\cal A}^{PV}$ are meant to isolate
the contribution of $G_M^{(s)}$ (the electric sector is severely
reduced there). From our modeling and comparisons with data some
significant discrepancies emerge that need further investigation.
Although being cautious because of the error bands, the behavior of
data versus $|Q^2|$ presents a positive slope which is larger than
the one resulting in the theoretical calculations. This applies to
both dipole and monopole functional dependencies for the strange and
axial-vector form factors, and shows the difficulty in reproducing
all data with a specific value for the static strange parameter
$\mu_s$. Only $\mu_s$-values close to zero overall reproduce the data at
different $Q^2$ (just touching the extreme error bands). However,
data at small $|Q^2|=0.1$ (GeV/c)$^2$ (SAMPLE) agree better with
results for positive $\mu_s$, whereas for $|Q^2|=0.6$ (G0) the best
agreement emerges for slightly negative $\mu_s$. Finally, data
measured at $|Q^2|\sim 0.2$ (GeV/c)$^2$ for G0 ($\theta_e=110^o$)
and SAMPLE ($\theta_e=145^o$) also present some difficulties when
compared with theory. Whereas the former implies an intermediate
value of $\mu_s$ within the range $[0,0.3]$, the latter is
consistent with $\mu_s\in [-0.3,0]$.

Forward scattering kinematics, where a wide selection of data taken
at different $Q^2$ are available, has been also studied in detail in
Sect.~IV. Comparison between theory and data shows in general good
accordance concerning the behavior with $Q^2$. However, some
significant differences which need to be clarified also emerge. In
particular, assuming the dipole shape for the strange form factors
with a fixed value for the magnetic strangeness parameter $\mu_s$,
it is difficult to reproduce data taken at forward angles, $\theta_e<21^o$, and at
$\theta_e\sim 35.5^o$ using a single value for the
electric strangeness content $\rho_s$. This is, for instance, the
case of $\mu_s=0$, {\it i.e.,} no magnetic strangeness, where data
at $\theta_e<21^o$ are in accord with calculations for $\rho_s=0.5$,
whereas data for larger scattering angles, $\theta_e\sim 35.5^o$,
are located within the area between the curves corresponding to $\rho_s=0.5$ and larger $\rho_s=2$.
Similar comments apply to the other $\mu_s$ values selected, $\pm 0.3$.
Although the spread of results corresponding to different $\rho_s$
is not very large, further investigation is needed in order to settle
the reason for this discrepancy. When assuming a monopole shape for the
strange form factors the spread between the curves calculated for
different $\rho_s$ is much higher,  the different cases being
clearly separated, even if uncertainties linked to the description
of the axial-vector form factor are included.
Discussion of the role of $\rho_s$ and comparison with data follows similar trends to the
ones already applied to the dipole description.

Summarizing, the general discussion presented in previous paragraphs
clearly indicates that further studies and investigations are needed
before definite conclusions on the strangeness content in the
nucleon can be drawn. Not only the specific values of the
strangeness content given through the parameters $\mu_s$ and
$\rho_s$ should be reviewed, but also the specific functional
dependence with $Q^2$ has to be explored in depth. Moreover, the
role played by the WNC axial-vector form factor is also crucial in
understanding the results for the PV asymmetry and its comparison
with data. Contrary to some previous
work~\cite{Young,Lhuillier,Thomas,Thomas2} where the focus was
placed on the analysis of specific data taken at fixed $|Q^2|$, here
our interest has been to provide a general and coherent description
of all data measured at different transferred momenta. We also have
estimated the amount of uncertainty in ${\cal A}^{PV}$ linked to
different descriptions of the electroweak form factors, namely to
their strengths and $Q^2$-dependencies. Radiative corrections in the
electric, magnetic and axial-vector form factors have been also
analyzed and their impact on the asymmetry evaluated.

Following these general discussions a global analysis of the
asymmetry ${\cal A}^{PV}$ has been performed by presenting the world
data constraint on the electric and magnetic strangeness parameters.
Ellipses showing the $1\sigma$ and $2\sigma$ confidence regions in
the $\rho_s$-$\mu_s$ plane have been presented in different
situations, using GKex and the fit of Bernauer {\it et al.} for the
EM form factors, as well as assuming monopole/dipole functional
dependencies for $G_{E,M}^{(s)}$ and different values for the
axial/vector masses. From this global analysis consistency of world
data with positive values of $\rho_s$ emerges, although the specific
central value of $\rho_s$ depends on the particular situation
considered. Nevertheless, the $1\sigma$ confidence ellipses are
located in most of the cases in the positive $\rho_s$-region (only
situations (ii) and (iv) in Fig.~\ref{fig:mus-rhos-GKex} and (v) in
Fig.~\ref{fig:mus-rhos-RC} touch negative $\rho_s$-values at the
extreme). The $2\sigma$ confidence level extends the validity of
$\rho_s$ to slightly negative values in all cases. Concerning the magnetic
sector, zero strangeness, $\mu_s=0$, is
located inside the $1\sigma$ confidence region in all situations. In
fact, the central values obtained using GKex
(Fig.~\ref{fig:mus-rhos-GKex}) and the fit of Bernauer {\it et al.}
(Fig.~\ref{fig:mus-rhos-RC}) are very close to $\mu_s=0$ (within the
error bands). Only when radiative corrections are neglected
(Fig.~\ref{fig:mus-rhos-RC}), does the point of maximum likelihood
for $\mu_s$ clearly become positive, although zero strangeness is
still contained inside the $1\sigma$ confidence region. In general,
we conclude that the magnitude of strangeness effects are
constrained to be quite small. However, the analysis of the
$1\sigma$ and $2\sigma$ confidence ellipses shows that
the case of no strangeness,
{\it i.e.,} $\rho_s=\mu_s=0$, is excluded by most of the fits, as is
the region where the signs are both negative.

Additionally, some 
considerations concerning the likelihood of future high precision PV 
experiments have been presented, together with discussions of the
kinematical conditions under which the precision is expected to be
maximum. This is likely to be the situation for the planned MESA experiment~\cite{MESA}
where effects coming from $\gamma Z$-box corrections~\cite{GH2009,SBMT2010,GHM2011} and
isospin violations~\cite{Kubis2006,Viviani2009} are expected to be very small.

Finally, the potential impact the variations considered in this
work might have on interpretations of the Q-weak experiment has
been also discussed. From this study, we conclude that a rough
variation in the parameters of the model consistent with available
data ($1\sigma$ confidence region) may lead to the proton's weak
charge determined by $\sim$5-6$\%$, {\it i.e.,} $\sin^2\theta_W$
extracted to $\sim$0.3$\%$. This is similar to the basic objectives
pursued by the Q-weak experiment.

\section*{Acknowledgments}
This work was
partially supported by DGI (Spain): FIS2011-28738-C02-01, by the Junta de
Andaluc\'{\i}a, the Spanish Consolider-Ingenio
2000 programmed CPAN (CSD2007-00042), and part (TWD) by U.S.
Department of Energy under cooperative agreement DE-FC02-94ER40818.
R.G.J. acknowledges support from the Ministerio de Educaci\'on.
We thank K.~Kumar for helpful discussions.

%

\end{document}